\makeatletter\let\latex@xfloat\@xfloat\makeatother

\documentclass[numbers,sort&compress]{elsarticle}
\usepackage{etex}
\usepackage{lineno}
\usepackage[normalem]{ulem}
\usepackage{soul}
\usepackage[bookmarks,bookmarksopen,bookmarksdepth=2]{hyperref}
\hypersetup{colorlinks=true,citecolor=blue,linkcolor=blue}
\modulolinenumbers[5]

\usepackage{tikz}
\usepackage{pgfplots}
\usetikzlibrary{shapes.misc}
\usetikzlibrary{patterns,decorations.pathreplacing}

\tikzset{cross/.style={cross out, draw=black, minimum size=2*(#1-\pgflinewidth), inner sep=0pt, outer sep=0pt},
cross/.default={1pt}}
\usepackage{gensymb}
\usepackage{indentfirst}
\usepackage{fancyhdr}
\usepackage{setspace}
\usepackage{dsfont}
\usepackage{amsfonts,amssymb,amsmath,mathtools,amsthm,ulem,cancel}
\usepackage{graphicx,color,geometry,epstopdf}
\usepackage{tikz}
\usepackage{pifont}
\usepackage{epstopdf}
\usepackage{caption}
\usepackage{multirow}
\usepackage[export]{adjustbox}
\usepackage{subfig}
\usepackage{cleveref}
\usepackage{subfloat}

\graphicspath{{figures/}}

\newcommand{\verteq}{\rotatebox{90}{$\,=$}}
\newcommand{\equalto}[2]{\underset{\scriptstyle\overset{\mkern4mu\verteq}{#2}}{#1}}
\definecolor{commenti}{rgb}{0.13,0.55,0.13}
\definecolor{stringhe}{rgb}{0.63,0.125,0.94}
\usepackage[font=footnotesize]{caption}

\usepackage{listings}
\usepackage{keyval}

\newcommand{\om}{\omega}
\newcommand{\la}{\lambda}
\newcommand{\sta}{\star}

\newcommand{\vare}{\varepsilon}

\definecolor{my-color}{RGB}{201,2,2}
\definecolor{Lou-color}{RGB}{0,0,255}
\definecolor{Shahriar-color}{RGB}{0,252,10}
\begin{document}

\begin{frontmatter}

\title{Interfacial Dynamics of Thin Viscoelastic Films and Drops}

\author{Valeria Barra}
\author{Shahriar Afkhami\corref{correspondingauthor}}
\cortext[correspondingauthor]{Corresponding author}
\ead{shahriar.afkhami@njit.edu}

\author{Lou Kondic\corref{}}

\address{Department of Mathematical Sciences,
New Jersey Institute of Technology,
Newark, NJ, 07102, USA}

\begin{abstract}
We present a computational investigation of thin viscoelastic films and drops on a solid
substrate subject to the van der Waals interaction force, in two spatial dimensions. The governing equations
are obtained within a long-wave approximation of the Navier-Stokes equations with Jeffreys
model for viscoelastic stresses. We investigate the effects of viscoelasticity, Newtonian
viscosity, and the substrate slippage on the dynamics of thin viscoelastic films.
We also study the effects of viscoelasticity on drops that spread or recede on a prewetted substrate. For dewetting films, the
numerical results show the presence of multiple secondary droplets for
higher values of elasticity, consistently with experimental findings. For drops, we find that
elastic effects lead to deviations from the Cox-Voinov law for partially wetting
fluids. In general, elastic effects enhance spreading, and suppress retraction, compared to
Newtonian ones.
\end{abstract}


\end{frontmatter}

\section{Introduction}

Thin liquid films play a central role in many real life applications and therefore are
studied widely theoretically, numerically, and experimentally. Thin polymer films, in particular, are of special
importance due to their presence in a broad variety of applications, for example,
in the food, chemical, and pharmaceutical industries, as well as in materials science. Polymeric liquids are one example of a wider class of viscoelastic liquids, constituted by a Newtonian (viscous) solvent and a non-Newtonian (polymeric) solute. In general, viscoelastic films combine characteristics of viscous fluids with features typical of elastic matter. The interface between the liquid and the surrounding fluid (usually a gaseous phase) is a free and deformable boundary, and therefore thin liquid films can display a variety of dynamics and interfacial instabilities. As widely presented in the literature, see, for instance,~\cite{Oron, Myers,Sharma}, these instabilities can lead to film breakup, dewetting the substrate. The understanding of the instability mechanisms relevant to thin polymer films has thus motivated many theoretical and experimental studies, see, e.g.,~\cite{Bird,B-W,Renardy,Safran}. Perhaps one of the first experimental works on this matter has been
carried out by Reiter \cite{Reiter}, where the influence of the film thickness on
the interfacial instability of polymer films of nanometer size is examined. His study shows that, when dewetting occurs, a rim can form ahead of the dewetting edge and subsequently decay into drops on the substrate.
Since his work, the investigation of thin polymer film morphologies at the nanoscale
has been a major focus of many studies, see, e.g.,~\cite{SarkarSharma,GreenGanesan,GabrieleEtAl,SharmaReiter,Herminghaus1,Herminghaus2}. Additional works focused on the stability, the dynamics, and the morphology of the fluid interface due to rheological properties \cite{DeGennes,Christensen,GabrieleEtAl2,Shaqfeh,WuChou}; these investigations are carried out with the goal of understanding whether the effects related to viscoelasticity, slippage, surface heterogeneities, or forces of electrohydrodynamic origin play a key role in the development of surface instabilities.

Despite numerous works focusing on polymeric films, very few studies consider numerical simulations of the interface of thin layers of viscoelastic fluids dewetting a solid substrate,
see, e.g.,~\cite{VilminRaphael,TomarEtAl}.
In particular, Vilmin and Rapha{\"e}l develop a model based on a simplified dewetting
geometry of the film, neglecting the surface tension \cite{VilminRaphael}. They demonstrate that the friction force and the residual stresses, due to the film viscoelasticity, can have an opposing influence on the dewetting dynamics.
They show that these residual stresses can accelerate the onset of the dewetting,
followed by a slow, quasi-exponential, growth of the hole.
Although their model is useful to explain the main features of the dynamics of the evolving rim,
it is unable to provide a detailed description of the dewetting process and a
quantitative investigation of the final morphological structures. An earlier study of
Tomar et al.~\cite{TomarEtAl} uses the lubrication model derived by Rauscher et al.~\cite{Rauscher2005}
for thin viscoelastic films of Jeffreys type, although without including the substrate slippage. Using both linear stability analysis and nonlinear
simulations, they show that viscoelasticity does not have a major influence on the dewetting dynamics. Their numerical solutions suggest that
the length scale of instability in the nonlinear regime is unaltered by the viscoelasticity.

In this work, we present a detailed description of numerical solutions of the nonlinear governing
equation based on the long-wave (lubrication) model developed by Rauscher et al.~\cite{Rauscher2005}
for thin viscoelastic films, with Jeffreys constitutive model for viscoelastic stresses \cite{Jeffreys,Bird}. In this model, viscoelastic stresses are described with a Newtonian contribution (due to the solvent)
plus a polymer contribution that is governed by the linear Maxwell model \cite{Renardy}. To model the film breakup and the consequent dewetting process, as well as to impose the contact angle, we include the van der Waals attraction/repulsion interaction force. This force introduces an equilibrium film on the solid substrate, leading to a prewetted (often called precursor) layer in nominally dry regions. In particular, we focus on the emerging length scales due to the instability of
a viscoelastic film at dewetting stage, that to date have not been reported in the literature.
Unlike the previous numerical studies introduced, we consider the effect of transitioning from no-slip to weak slip
on the initial instability development and the dewetting dynamics.
A surprising finding is that the resulting morphologies are influenced by viscoelasticity and slippage.
In fact, we show the formation of not only main drops, as previously demonstrated in \cite{TomarEtAl},
but also of multiple satellite droplets that are completely absent for Newtonian films. These secondary droplets are comparable with those found experimentally (see, e.g.~\cite{SharmaReiter,Reiter} or \cite{GreenGanesan}, where they are called ``nanodroplets''), but, to the best of our knowledge, have not been found in previous computational studies of the evolution of viscoelastic films.

The first part of our investigations concerns the spontaneous dewetting of a thin viscoelastic film, initially at rest, due to van der Waals interactions, in two spatial dimensions. Consistently with \cite{Blossey2006}, we find that in the linear regime, the critical and most unstable wavenumbers are neither dependent on the viscoelastic parameters, nor on the slip length, but only on the van der Waals interactions with the substrate. We then provide numerical simulations of the evolution of the interface in the nonlinear regime. In this regime, we find that the instability and the final configuration of the fluid in primary and secondary droplets are affected by the viscoelastic parameters and the slippage of the substrate. We show how a larger viscoelasticity induces the formation of secondary droplets, and how the slip at the substrate prevents them from forming. We thus provide, for the first time, numerical simulations leading to novel morphologies
for thin viscoelastic dewetting films.

Finally, we focus on the spreading/receding of viscoelastic planar drops on a solid substrate. We study viscoelastic drops that spread/recede spontaneously due to the imbalance between the initial and the equilibrium contact angles. The theoretical and experimental studies of spreading or retracting drops, both for Newtonian and viscoelastic fluids, are numerous (see, e.g.,~\cite{DeGennesEtAl,DeGennes2,Kyle2,KimEtAl,KimEtAl2,WeiEtAl,WeiEtAl2,BartoloEtAl}). Surprisingly, only a few studies report computational results for viscoelastic dynamic contact lines, see \cite{YueFeng,WangEtAl,Izbassarov}. Yue and Feng \cite{YueFeng} use a phase-field model to simulate the displacement flow of Oldroyd-B fluids in a channel formed by parallel plates. They show that viscoelastic stresses close to the contact line region affect the bending of the interface. Also, Wang et al.~\cite{WangEtAl} use an axisymmetric formulation to describe the spreading of viscoelastic drops, comparing the Giesekus (shear-thinning) and the Oldroyd-B models. They show that the spreading speed depends on the viscoelastic relaxation time. Most recently, Izbassarov and Muradoglu \cite{Izbassarov} study the effects of viscoelasticity on drop impact and spreading on a solid substrate. They investigate the spreading rate of viscoelastic drops, using the FENE-–CR model, and find that viscoelastic effects enhance the spreading speed.

In the present work, we consider partial wetting by accounting for van der Waals interactions between the solid and the fluid.
Although our approach is developed strictly for configurations characterized by small interfacial slopes, we expect that it still provides reasonably accurate results for the situations such that the contact angle is not small (see, e.g.,~\cite{Kyle2} for a discussion of this topic). Our numerical results show that viscoelasticity enhances the spreading in the early stage of wetting by smoothing the interface in the contact line region. Similar considerations are drawn by Wang et al.~\cite{WangEtAl}, and Izbassarov and Muradoglu \cite{Izbassarov}. Finally, the study of the advancing dynamic contact angle allows us to determine that the Cox-Voinov law \cite{Cox1986,Voinov} holds for the viscous Newtonian fluid, but not for the viscoelastic counterpart. For retracting drops, we find that the interface of a viscoelastic drop provides more resistance to the motion, causing the drop to retract slower, consistently with the experimental study \cite{BartoloEtAl}. Our results regarding receding viscoelastic drops show a deviation from the Cox-Voinov law as well.

It is appropriate to make a remark about the choice of the constitutive model. Although linear viscoelastic models, such as the Jeffreys model, are known to be valid only for flows with small displacement gradients \cite{Bird,Rauscher2005,TomarEtAl}, we expect a linear constitutive model to be sufficiently adequate to describe the viscoelastic behavior in the context of spontaneous wetting/dewetting processes. For more complex flows, one should incorporate more general viscoelastic models, such as the Oldroyd-B model \cite{Bird}. However, as noted also by Tomar et al.~\cite{TomarEtAl}, the nonlinear convective terms of the stress tensor in an Oldroyd-B model would not change the linear stability analysis, and therefore our results for the linear regime are valid for both linear and nonlinear viscoelastic models. Furthermore, our numerical simulations show how, in the final stage of the nonlinear evolution, dewetting viscoelastic films display a slow, viscous dynamics, for which a linear viscoelastic model is considered to be appropriate. Additionally, we have verified that displacement gradients (hence the shear rate) are not large even in the intermediate time of the dewetting process, in which viscoelastic fluids exhibit a non-Newtonian response to deformations. In summary, in the context of spontaneous wetting/dewetting processes driven by the van der Waals interaction force only, the assumption of small displacement gradients is not violated, and a linear viscoelastic model suffices to describe the effects of viscoelasticity.

The rest of this paper is organized as follows: In \S~\ref{Sec1}, we introduce the governing equations; In \S~\ref{Sec2}, we outline the numerical methods used to solve the nonlinear problem; In \S~\ref{sec:NumericalResults}, we present the linear stability analysis (LSA), and discuss the numerical results for both dewetting and wetting studies; In \S~\ref{Sec4}, we draw our conclusions; We finally report the derivation of the governing equations and their numerical discretization in Appendices A and B, respectively.

\section[Governing Equations]{Governing Equations}
\label{Sec1}

We consider an incompressible liquid, with constant density $\rho$,
surrounded by a gas phase assumed to be inviscid, dynamically passive, and of constant pressure.
The equations of conservation of mass and momentum, respectively, for the liquid phase then become
\begin{subequations}\label{Eq:N-S}
\begin{align}
\rho \left( \partial_t \textbf{u} + \textbf{u} \cdot \nabla \textbf{u}\right) &= - \nabla (p + \Pi) + \nabla \cdot \mathbf{ \tau} \ , \label{GovMomentum}\\
\nabla \cdot \textbf{u} &= 0 \, , \label{Incompressability}
\end{align}
\end{subequations}
\noindent where $\textbf{u} = (u(x,y,t),v(x,y,t))$ is the velocity field in the Cartesian $xy$-plane (as by convention, the $x$-axis is horizontal, and the $y$-axis is vertical), and $\nabla= (\partial_x, \partial_y)$; $\mathbf{ \tau}$ is the stress tensor, $p$ is the pressure, and $\Pi$ is the disjoining pressure induced by the van der Waals solid-liquid interaction force (we note that $\nabla \Pi =0$ except at the liquid-gas interface). This force is attractive (destabilizing) for thicker films and repulsive (stabilizing) for thin ones, leading naturally to the concept of equilibrium film thickness, defined below as $h_{\sta}$, at which repulsive and attractive forces balance each other \cite{Isreaelachvili}. We provide the form of the disjoining pressure used in this work in Appendix A, where the long-wave approximation of the system (\ref{Eq:N-S}) is described in detail.

To model the stresses, we use a generalization of the Maxwell model for viscoelastic liquids: the Jeffreys model. It describes the non-Newtonian nature of the stress tensor $\mathbf{\tau}$, interpolating between a purely elastic and a purely viscous behavior. The stress tensor according to Jeffreys model follows the constitutive equation
\begin{equation}
\label{Jeffreys}
\mathbf{\tau} + \lambda_1 \partial_t \mathbf{\tau} = \eta (\dot{\gamma} + \lambda_2 \partial_t \dot{\gamma}) \, ,
\end{equation}
\noindent where $\dot {\gamma}$ is the strain rate tensor, e.g.~$\dot{\gamma}_{12}= \partial_x v + \partial_y u$ (other components of $\dot{\gamma}$ are similarly expressed in terms of derivatives of $\mathbf{u}$), and $\eta$ is the shear viscosity coefficient. In Jeffreys model, the response to the deformation of a viscoelastic liquid is characterized by two time constants, $\la_1$ and $\la_2$, the \textit{relaxation time} and the \textit{retardation time}, respectively, related by
\begin{equation}
\label{L1&L2}
\lambda_2 = \la_1 \frac{\eta_s}{\eta_s + \eta_p} \, .
\end{equation}
\noindent Here $\eta_s$ and $\eta_p$ are the viscosity coefficients of the Newtonian solvent and the polymeric solute, respectively, such that $\eta = \eta_s + \eta_p$. Noting that the ratio $\eta_r=\eta_s/(\eta_s + \eta_p) \leq 1$, we have that $\la_1 \geq \la_2$ \cite{Bird,Rauscher2005,TomarEtAl}. We also observe that, within the Jeffreys model, we recover the Maxwell viscoelastic model when
$\lambda_2=0$, and a Newtonian fluid when $\lambda_1=\lambda_2$.

\begin{figure}[t]
\centering
\includegraphics[width=0.78\linewidth]{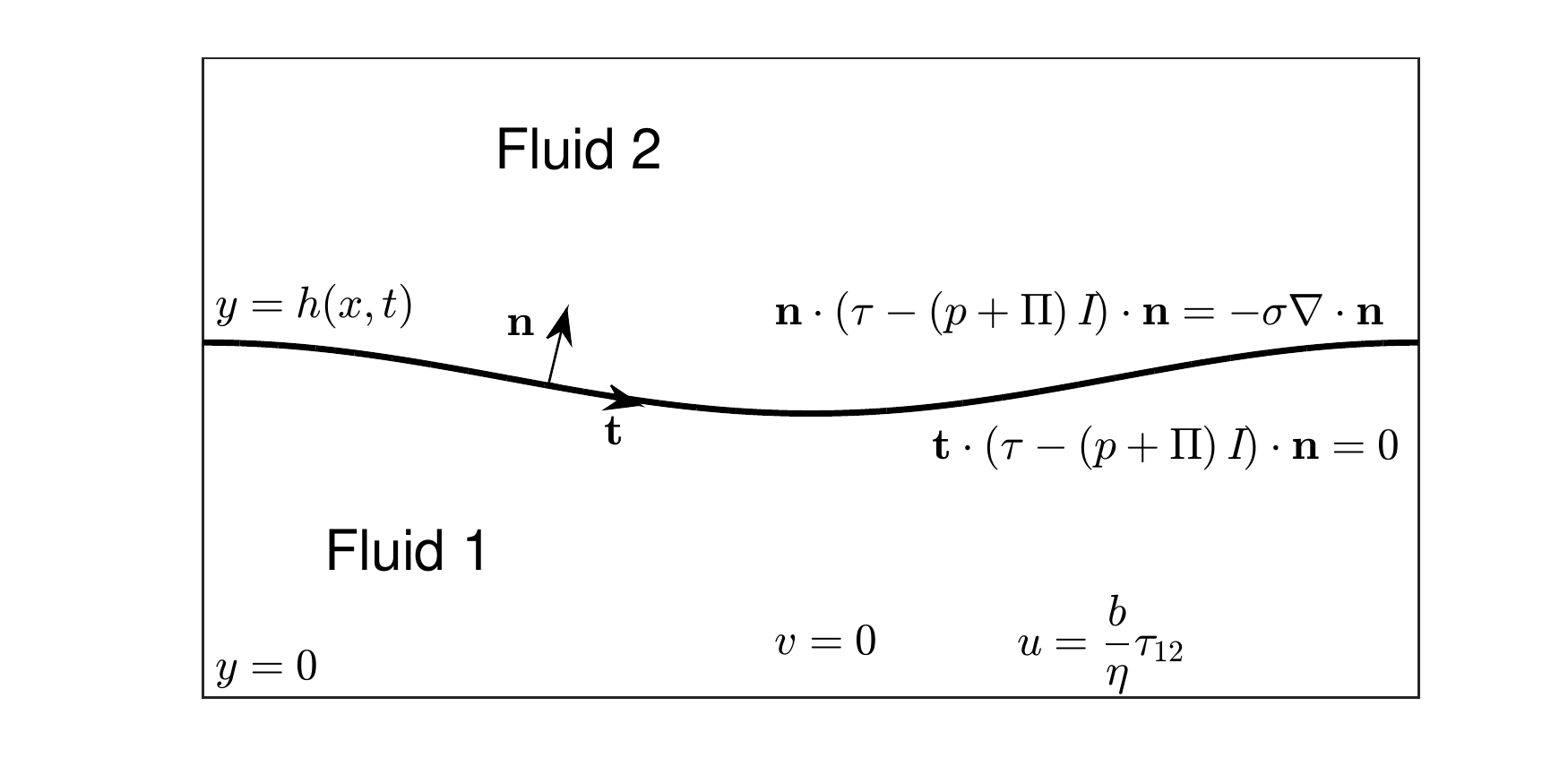}
\caption{Schematic of the fluid interface and boundary conditions. Fluid $1$ is the viscoelastic liquid and fluid $2$ is the ambient (passive) gas.}\label{fig:Setup}
\end{figure}

Figure \ref{fig:Setup} shows a schematic of the fluid interface, represented parametrically by the function $f(x,y,t)=y-h(x,t)=0$, and
the boundary conditions at the free surface ($y=h(x,t)$) and at the $x$-axis ($y=0$). At the latter, we apply the non-penetration and the Navier slip boundary conditions, with slip length coefficient denoted by $b \geq 0$ ($b=0$ implies no-slip). As discussed in \cite{Fetzer,Munch}, long-wave models for thin films can be derived in different slip regimes. In the present work, we will focus on the weak slip regime; for strong slip, a different system of governing equations is derived in \cite{Blossey2006}.
The interested reader can find the derivation of the governing equation for the evolving interface $h(x,t)$ in Appendix A, where all quantities and scalings are defined. We report here its final dimensionless form
\begin{align}
\label{GovEq2D}
&(1 + \la_2 \partial_t) h_t + \frac{\partial}{\partial x} \left\{  (\la_2 - \la_1 )\left( \frac{h^2}{2} Q - h R \right) h_t +\right. \nonumber\\
&\left. \qquad\left[(1 + \la_1 \partial_t) \frac{h^3}{3} + (1 + \la_2 \partial_t ) bh^2\right] \frac{\partial }{\partial x}\left(\frac{\partial^2 h }{\partial x^2} + \Pi(h) \right) \right\} =0 \, ,
\end{align}
\noindent where $Q=Q(h)$ and $R=R(h)$ satisfy, respectively,
\begin{subequations}\label{Eq:Q&R}
\begin{align}
(1 + \la_2 \partial_t) Q &= -\frac{\partial }{\partial x}\left(\frac{\partial^2 h }{\partial x^2} + \Pi(h) \right) \, , \label{Eq:Q}\\
(1 + \la_2 \partial_t) R &= -h\frac{\partial }{\partial x}\left(\frac{\partial^2 h }{\partial x^2} + \Pi(h)\right)\, . \label{Eq:R}
\end{align}
\end{subequations}
\noindent We note that in the absence of viscoelasticity (i.e.~with $\la_1=\la_2$), equations (\ref{GovEq2D}) and (\ref{Eq:Q&R}) reduce to the well-known long-wave formulation for viscous Newtonian films (see, e.g.,~\cite{Myers}).

\section[Numerical Methods]{Numerical Methods}\label{Sec2}

To numerically solve equations (\ref{GovEq2D}) and (\ref{Eq:Q&R}), we use the finite differences technique, described in detail in Appendix B. To simplify, we first consider a purely elastic (Maxwell) liquid, that is $\la_2=0$. Thus, the governing equations (\ref{GovEq2D}) and (\ref{Eq:Q&R}) reduce to a first order in time formulation given by
\begin{align}\label{Eq:DiscreteGovWithL2=0}
 h_t -  \frac{\partial}{\partial x} \left\{\la_1 \frac{h^2}{2} \left( h_{xxx} + \Pi'(h)h_x \right) h_t - \left[ \frac{h^3}{3} + bh^2 + \la_1 \frac{\partial}{\partial t} \left(\frac{h^3}{3}\right)\right] \left( h_{xxx} + \Pi'(h)h_x \right)  \right\} =0\, .
 \end{align}
\noindent To discretize equation (\ref{Eq:DiscreteGovWithL2=0}), we isolate the time derivatives from the spatial ones, so that we can apply an iterative scheme to find the approximation to the solution at the new time step. We do so by differentiating the spatial derivatives and, assuming the partial derivatives of $h(x,t)$ to be continuous, we obtain
{\setlength{\abovedisplayskip}{5pt}
  \setlength{\belowdisplayskip}{\abovedisplayskip}
  \setlength{\abovedisplayshortskip}{4pt}
  \setlength{\belowdisplayshortskip}{4pt}
\begin{align}\label{Eq:DiscretizeL2=0}
&\underbrace{ \frac{\partial}{ \partial x} \left[ \left( \frac{h^3}{3} + bh^2 \right) \left( h_{xxx} + \Pi'(h)h_x  \right)\right] }_{f(h)} + \underbrace{\left[1 -  \frac{1}{2}\la_1  \frac{\partial}{\partial x} \left[ h^2  \left( h_{xxx} + \Pi'(h)h_x  \right) \right]  \right]}_{g(h)} h_t + \nonumber \\
&\frac{\partial}{\partial t}\underbrace{ \left( \frac{\partial h}{\partial x}\right)}_{l(h)} \underbrace{\left\{ - \frac{1}{2} \la_1  \left[ h^2 \left( h_{xxx} + \Pi'(h)h_x  \right) \right] \right\}}_{m(h)} + \la_1 \frac{\partial}{\partial t}\underbrace{ \left[\frac{\partial}{ \partial x} \left(\frac{h^3}{3}  \left( h_{xxx}+ \Pi'(h)h_x  \right) \right) \right]}_{p(h)} =0 \, .
\end{align}
}
\noindent We can now differentiate the time derivatives and use the Crank-Nicolson scheme on the term $f(h)$ as
\begin{align}
\label{Eq:AllCircled}
g(h)_i^n \frac{h_i^{n+1} - h_i^n}{\Delta t} + m(h)_i^n \frac{l(h)_i^{n+1} - l(h)_i^n}{\Delta t}  + \la_1 \frac{p(h)_i^{n+1} - p(h)_i^n}{\Delta t}= -\frac{1}{2}\left[ f(h)_i^n + f(h)_i^{n+1}\right] \, .
\end{align}

\noindent The nonlinear terms $h^2$ and $h^3$ are computed at the cell-centers, as outlined in \cite{Kondic2003,Bertozzi}. After the linearization, we obtain a system of equations of the form $A_1\xi=B_1$, that we numerically solve for the correction term, $\xi$, using a direct method \cite{TeSheng}. The initial condition given for $h(x,0)$ is a known function that either describes the initial perturbation of the fluid interface for the film simulations or a circular cap for the drop simulations (see \S~\ref{sec:NumericalResults}).

For $\la_2 \neq 0$, the governing equation (\ref{GovEq2D}), after differentiating the spatial derivative, and isolating the time derivatives, can be recast to a second order in time equation given by
{
  \setlength{\abovedisplayskip}{5pt}
  \setlength{\belowdisplayskip}{\abovedisplayskip}
  \setlength{\abovedisplayshortskip}{4pt}
  \setlength{\belowdisplayshortskip}{4pt}
\begin{align}\label{Eq:DiscretizeL2neq0}
& \la_2h_{tt} +  \underbrace{\frac{\partial}{\partial x} \left[\left( \frac{h^3}{3}   + bh^2 \right) \left( h_{xxx} + \Pi'(h)h_x \right) \right] }_{f(h)}+ \underbrace{ \left\{1 + (\la_2 - \la_1 ) \left[ \frac{\partial}{\partial x} \left( \frac{h^2}{2} Q - h R \right) \right]\right\}}_{\widehat{g}(h)} h_t  + \nonumber \\
&\frac{\partial}{\partial t} \underbrace{\left(\frac{ \partial h}{\partial x}\right)}_{{l}(h)} \underbrace{(\la_2 - \la_1) \left( \frac{h^2}{2} Q - h R  \right)}_{\widehat{m}(h)} + \la_1 \frac{\partial}{ \partial t} \underbrace{\left[ \frac{\partial}{\partial x} \left( \frac{h^3}{3} \left( h_{xxx} + \Pi' (h) h_x \right) \right) \right]}_{p(h)} + \nonumber \\
& \la_2  \frac{\partial}{\partial t}\underbrace{ \left[ \frac{\partial}{\partial x}\left(bh^2 \left(  h_{xxx} + \Pi'(h)h_x \right)  \right) \right]  }_{q(h)} =0 \, ,
\end{align}
}
where now the discrete versions of the equations (\ref{Eq:Q}) and (\ref{Eq:R}) are:
\begin{subequations}
\begin{align}
\frac{Q^{n+1}_i - Q^n_i}{\Delta t} &= - \frac{Q_i^n}{\la_2} - \frac{1}{\la_2} \left( h_{xxx} + \Pi ' (h)h_x  \right)_i^n \, , \label{Eq:DiscretizedQ} \\
\frac{R^{n+1}_i - R^n_i}{\Delta t} &= - \frac{R_i^n}{\la_2} - \frac{1}{\la_2} h_i^n \left( h_{xxx} + \Pi ' (h)h_x  \right)_i^n  \, ,  \label{Eq:DiscretizedR}
\end{align}
\end{subequations}
\noindent that we simply solve by the forward Euler method with initial conditions $Q_i^0 = 0$ and $R_i^0 = 0$. Again, discretizing all terms and applying Crank-Nicolson scheme we obtain
\begin{align}\label{Eq:AllCircledL2}
&\la_2\frac{h_i^{n+1} -2h_i^n + h_i^{n-1}}{{\Delta t}^2} + \widehat{g}(h)_i^n \frac{h_i^{n+1} - h_i^n}{\Delta t} + \widehat{m}(h)_i^n \frac{l(h)^{n+1}_i - l(h)^n_i}{\Delta t} + \la_1 \frac{p(h)^{n+1}_i - p(h)^n_i}{\Delta t} + \nonumber \\
&\quad \la_2 \frac{q(h)^{n+1}_i- q(h)^n_i}{\Delta t}=  -\frac{1}{2}\left[ f(h)_i^n + f(h)_i^{n+1}\right] \, .
\end{align}
\noindent Similarly, we proceed by linearizing the nonlinear terms and solving the resulting system $A_2\xi=B_2$. We note that in this case the partial differential equation is second order in time. We therefore need a two-step method with a second initial condition, in addition to the prescribed $h(x,0)$. We use $h_t (x,0) = 0$, resulting from the assumption that the considered films and drops are initially at rest.

\section[Results]{Results and discussion}
\label{sec:NumericalResults}

\subsection[Linear stability analysis]{Linear stability analysis}\label{LSA}

To study the fluid response to a prescribed disturbance, we perform the linear stability analysis (LSA). We perturb a flat film of initial thickness $h_0$ by a Fourier mode of amplitude $\delta h_0$ (such that $\delta \ll 1$), with wavenumber $k$ and growth rate $\om$. Hence we let $h(x,t)= h_0 + \delta h_0 e^{ikx + \omega t}$. The dispersion relation $\omega = \omega(k)$ is
\begin{equation}\label{Eq:CompleteLSA}
\la_2 \om^2 + \left[1 + (k^4 - k^2 \Pi'(h_0) )\left( \la_1 \frac{h_0^3}{3} +\la_2 bh_0^2 \right)\right]\om + (k^4 - k^2 \Pi'(h_0))\left( \frac{h_0^3}{3} + bh_0^2 \right)=0 \, .
\end{equation}
\noindent Solving for the two roots of this quadratic equation, we obtain one strictly negative root, $\om_2$, and one root with varying sign,
$\om_1$. The latter one is positive (unstable) for $k^2 < \Pi'(h_0)$. We note that both the critical wavenumber, $k_c$, given by
$k_c^2= \Pi'(h_0)$, and the wavenumber of maximum growth, $k_{m}=k_c/\sqrt{2}$, do not depend on $\la_1$ and $\la_2$, nor on the slip length $b$ (as also discussed in \cite{Blossey2006}).
Moreover, we note that in the absence of retardation, i.e.~for $\la_2=0$, the dispersion relation for a purely elastic film leads to an unbounded growth rate for $\la_1= -3 / h_0^3(k^4 - k^2 \Pi '(h_0))$. However, for $\la_2 \neq 0$, the growth rate $\om$ is always finite. This observation about the unboundedness of the growth rate in purely elastic films has also been drawn by other authors, see, for instance,~\cite{TomarEtAl,WuChou}.
We also note that the maximum growth rate, $\om_m=\om(k_m)$, is an increasing function of $\la_1$ and $b$,
while a decreasing function of $\la_2$.
In \S~\ref{sec:ThinFilmsResults}, we will discuss in more details the effects of $\la_1$, $\la_2$, and $b$ on the dewetting dynamics.

\subsection[Dewetting of thin viscoelastic films]{Dewetting of thin viscoelastic films}\label{sec:ThinFilmsResults}

\begin{figure}[H!t1]
\captionsetup{type=figure}
\centering

\subfloat[]{\includegraphics[scale=0.35,valign=t,trim=0.15in 0in 0.3in 0in,clip=true]{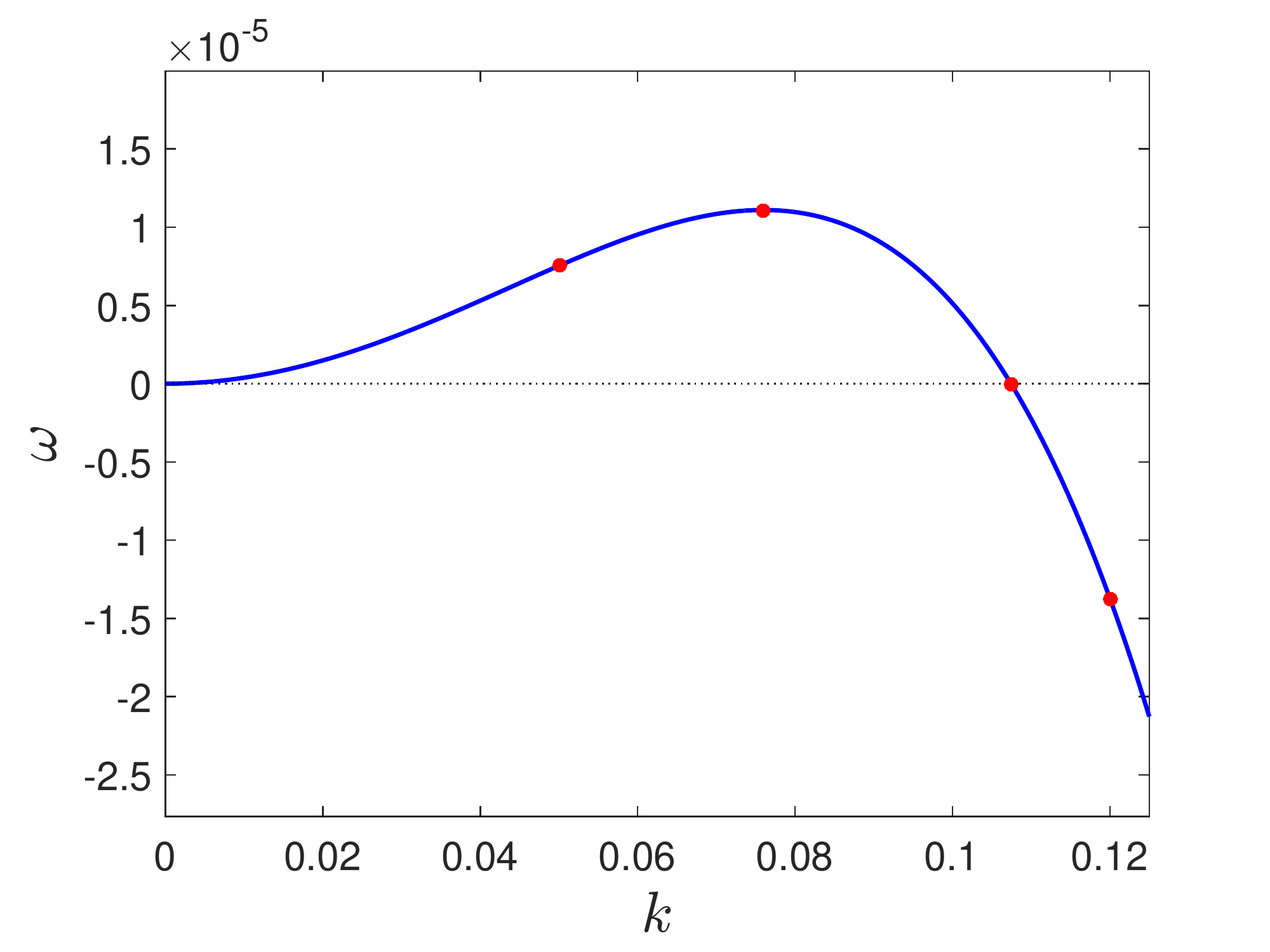}\label{fig:GrowthRateCheckL1=5,L2=0,B=0}}
\subfloat[]{\includegraphics[scale=0.35,valign=t,trim=0.09in 0in 0.3in 0in,clip=true]{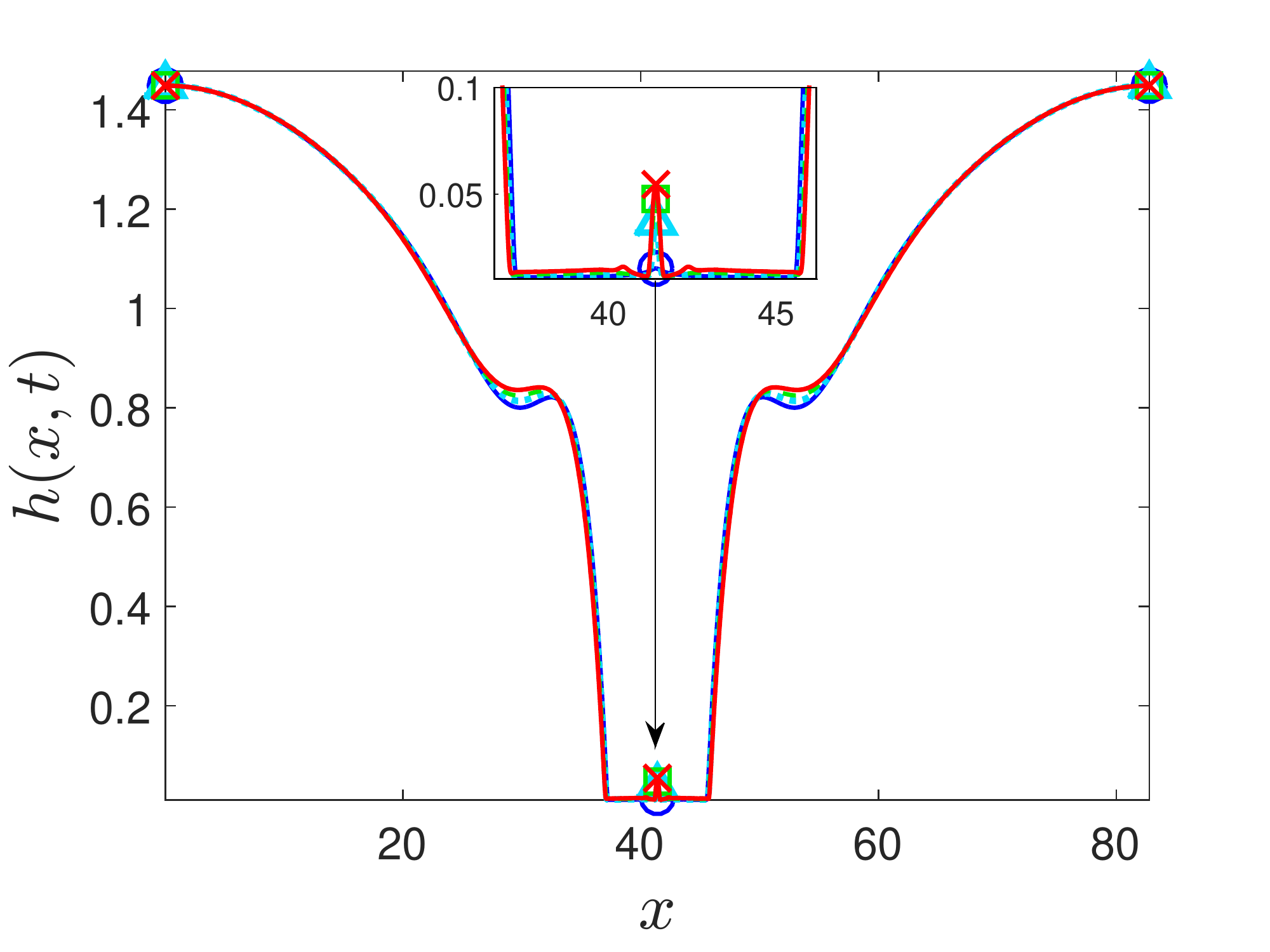}\label{fig:L1=0,2,4,6,t4}}\\
\subfloat[]{\includegraphics[scale=0.35,valign=t,trim=0.09in 0in 0.3in 0in,clip=true]{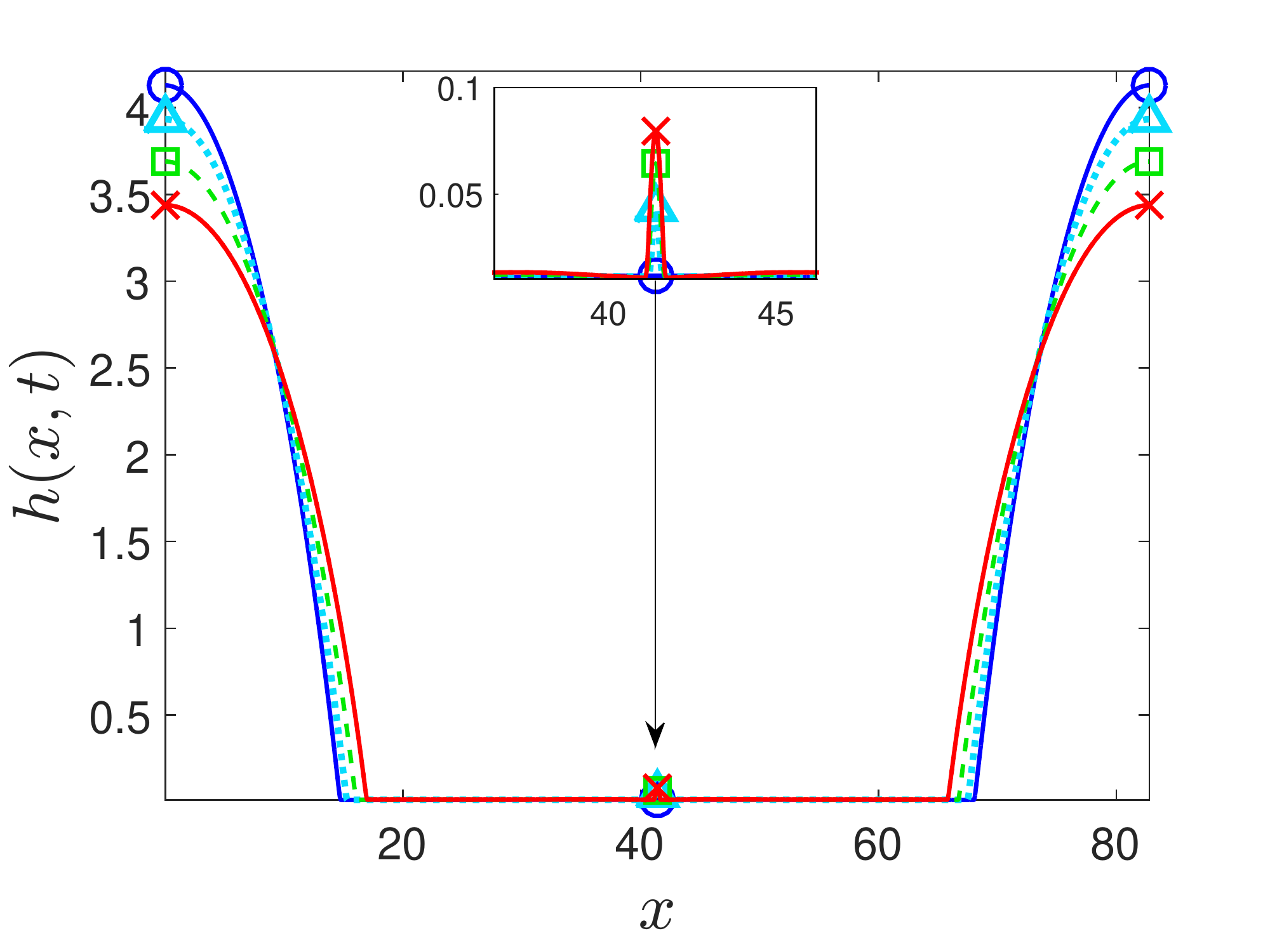}\label{fig:L1=0,2,4,6,t10}}
\subfloat[]{\includegraphics[scale=0.35,valign=t,trim=0.15in 0in 0.3in 0in,clip=true]{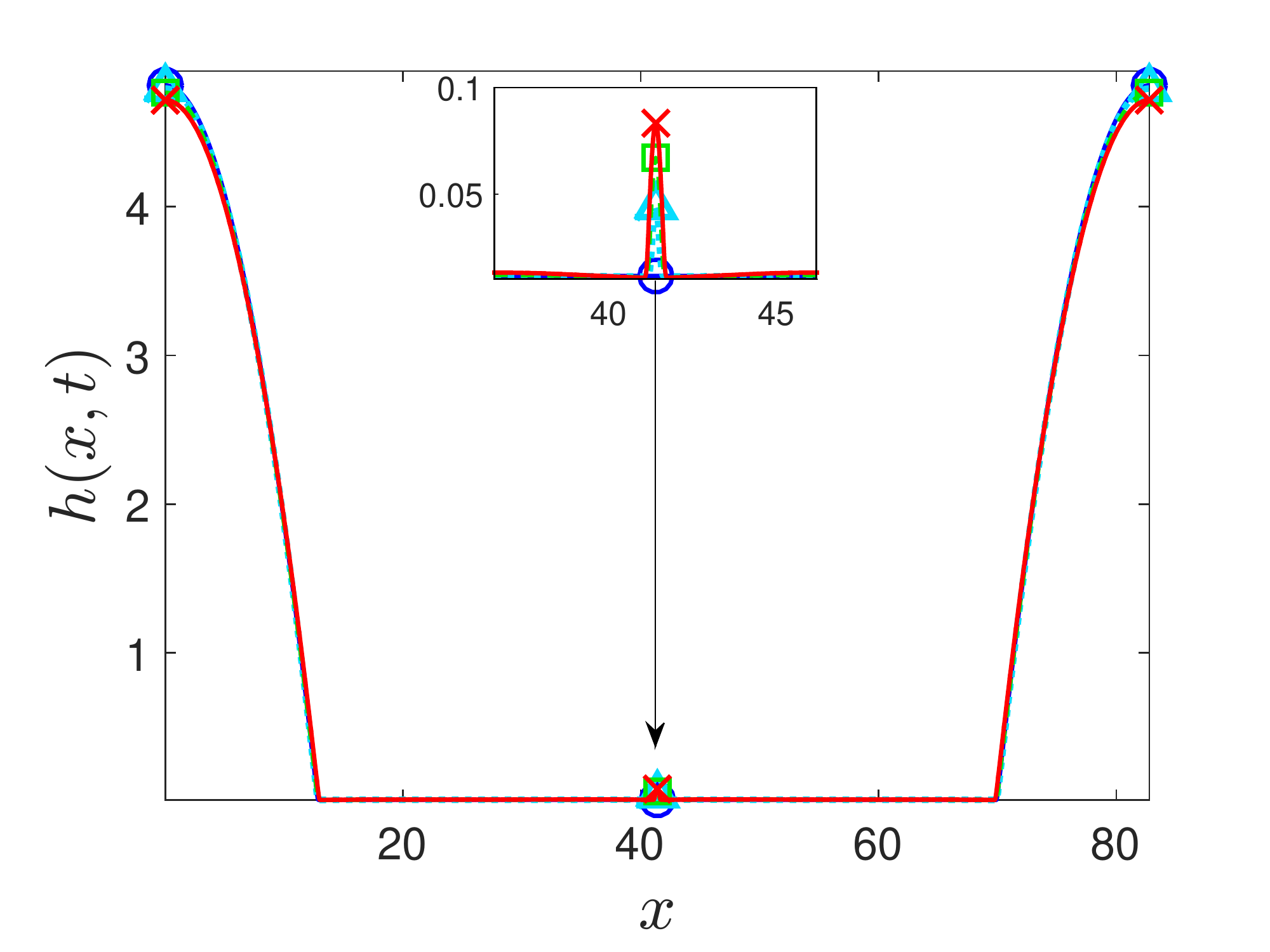}\label{fig:L1=0,2,4,6,t12}}
\caption{\protect\subref{fig:GrowthRateCheckL1=5,L2=0,B=0}~The comparison of the computed growth rate (red circles) with the prediction of the LSA (blue solid line) given by equation (\ref{Eq:CompleteLSA}), for $h_0=1$, $h_{\sta}=0.01$, $b=0$, $\la_2=0$, $\la_1=5$.
\protect\subref{fig:L1=0,2,4,6,t4}-\protect\subref{fig:L1=0,2,4,6,t12} Evolution of four distinct dewetting films with $\la_1=0$ (blue circles), $2$ (cyan triangles), $4$ (green squares), $6$ (red crosses), at three selected times. At $t=3.345 \times 10^5$ \protect\subref{fig:L1=0,2,4,6,t4}, the separation of the rims and the formation of a secondary droplet for values of $\la_1 \neq 0$ are shown. In \protect\subref{fig:L1=0,2,4,6,t10}, $t=3.37 \times 10^5$, and in \protect\subref{fig:L1=0,2,4,6,t12}, $t=3.38 \times 10^5$. The insets show a close-up of the region where a secondary droplet forms.}\label{fig:EvolutionL1=0,2,4,6}
\end{figure}

In this section, we present the numerical results for a dewetting thin film under the influence of the van der Waals interaction force. We perturb the initially flat fluid interface of thickness $h_0$ as described in \S~\ref{LSA}, with $k=k_m$ and $\delta=0.01$.
We choose the domain size to be equal to the wavelength of maximum growth, that is $\Lambda= 2 \pi / k_m$, unless noted otherwise. For unstable perturbations, the van der Waals interaction force is attractive, causing the liquid interface to
retract towards the substrate. When the fluid interface approaches the substrate, dewetting occurs, i.e.~a hole
(nominally a dry region) forms, leading to the formation of a rim that retracts and collects the liquid at the edge.
The system then gradually evolves toward an equilibrium state, corresponding to separate drops on the substrate
characterized by the equilibrium contact angle, $\theta_e$. We are mainly interested in the dynamics of the instabilities and the resulting morphologies, so we will only show results for unstable films. The set of parameters used for all simulations shown hereafter is: an initial normalized height of the fluid $h_0=1$, an equilibrium film thickness $h_{\sta}=0.01$, a constant contact angle $\theta_e=45^{\circ}$, a normalized surface tension $\sigma=1$, and a fixed grid size $\Delta x=5 \times 10^{-3}$, unless specified differently. All numerical results shown in this work are verified to be mesh-independent. We also validate our numerical investigations by comparing the growth rate for the early times with the LSA (equation (\ref{Eq:CompleteLSA})). Figure \subref*{fig:GrowthRateCheckL1=5,L2=0,B=0} shows the comparison of the computed growth rates for different wavenumbers (red circles) with the one given by the dispersion relation $\om_1(k)$ (blue solid curve), for a film with $\la_1=5$ and $\la_2=0$, when $b=0$. For these numerical simulations, we choose the domain size according to the wavelength that corresponds to the specified wavenumber. Although not shown here, comparisons between computed growth rates and the LSA are performed for all following results as well.

We begin our analysis with the simplest case of a purely elastic fluid, $\la_2=0$, and with no-slip at the substrate, $b=0$. Figures \subref*{fig:L1=0,2,4,6,t4}--\subref*{fig:L1=0,2,4,6,t12} show the evolution of four distinct films with different values of the relaxation time, $\la_1=0$ (blue circles), $2$ (cyan triangles), $4$ (green squares), and $6$ (red crosses), at three selected times. The interfacial dynamics can be divided into three phases. The initial regime corresponds to the short-time viscous response of the fluid, until the film separates in two retracting rims (figure \subref*{fig:L1=0,2,4,6,t4}). During intermediate times, the fastest dynamics occurs, and the liquid responds elastically: in this stage holes and retracting edges form (figure \subref*{fig:L1=0,2,4,6,t10}). In the last phase, the rims grow further in height, until the interface reaches its final configuration, attaining an equilibrium contact angle with the substrate (figure \subref*{fig:L1=0,2,4,6,t12}). During this third stage, the fluid shows a long-time Newtonian response again. These observations of the dewetting dynamics are in agreement with results in \cite{TomarEtAl,VilminRaphael}; moreover, we note that the shape of the dewetting front that forms a retracting rim is consistent with findings in \cite{Herminghaus1,Herminghaus2,TomarEtAl}. In addition, we observe that non-zero values of $\la_1$ not only slightly increase the speed of the breakup, but also allow for the formation of a satellite droplet, in contrast to the Newtonian film (with $\la_1=0$). In particular, in the inset in figure \subref*{fig:L1=0,2,4,6,t4}, we distinguish the formation of dips on the interface in the vicinity of the secondary droplet formed in the film with the highest value of $\la_1=6$. These oscillations disappear at a later time, as the interface around the secondary droplet flattens (figures \subref*{fig:L1=0,2,4,6,t10} and \subref*{fig:L1=0,2,4,6,t12}). To study analytically the observed oscillations, we perform a linear analysis, as the one presented in \cite{Rauscher2005,Munch}, for the
inner region of the growing hole. We find that the linearized solution, under quasi steady state conditions, does not depend on the viscoelastic parameters. Therefore the oscillations that the viscoelastic interface exhibits in the inner region of the dewetting hole cannot be analytically described with a linear analysis. In what follows, we also show that increasing the elasticity even further, provided a small retardation time (e.g.~$\la_2=0.01$) is also included, can lead to multiple strongly pronounced dips, and subsequently form numerous secondary droplets.

\begin{figure}[H!t1]
\captionsetup{type=figure}
\centering
\subfloat[]{\includegraphics[scale=0.265,valign=t,trim=0.07in 0in 0.35in 0in,clip=true]{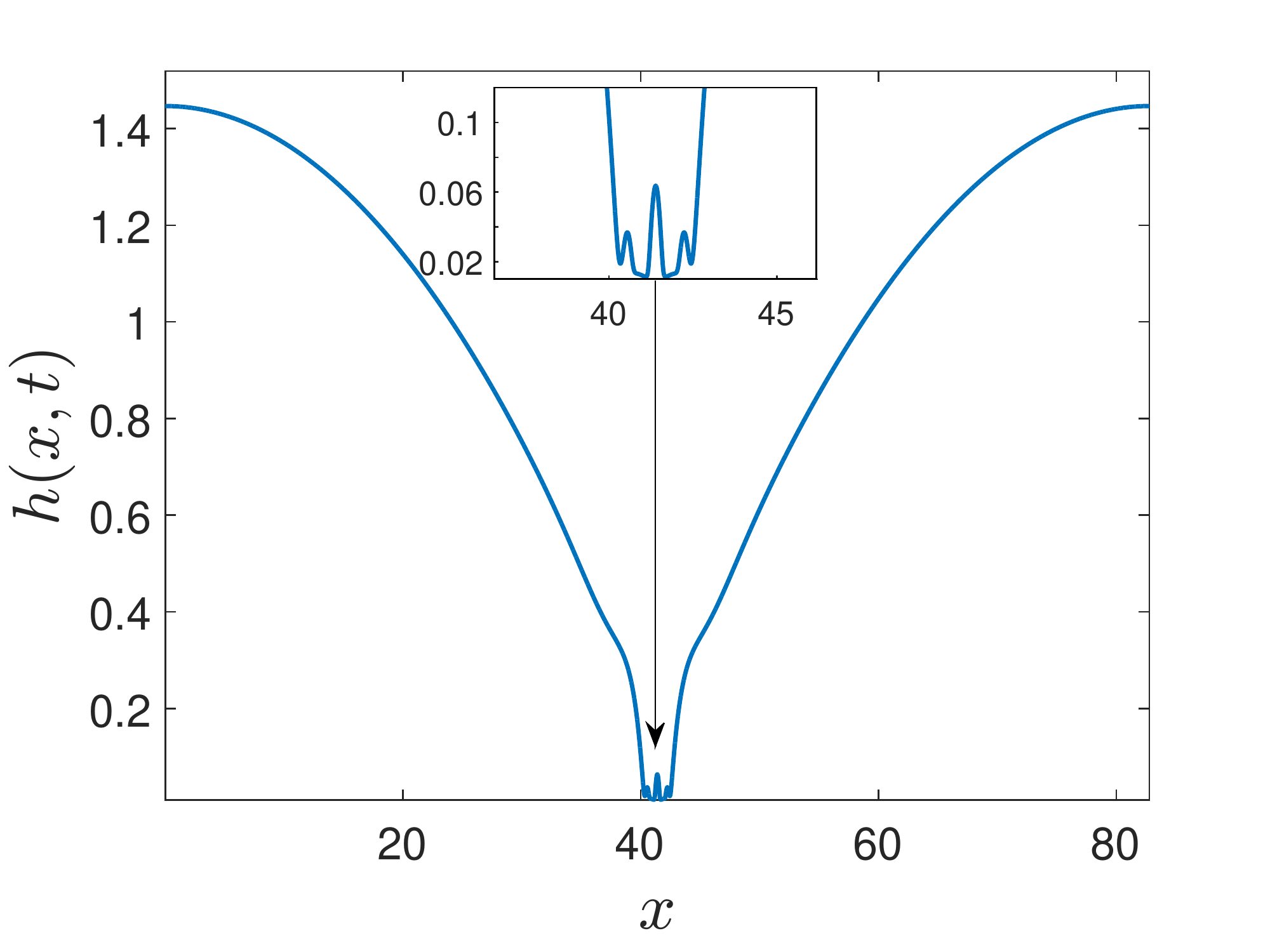}\label{fig:L1=10,L2=0,01,t2}}
\subfloat[]{\includegraphics[scale=0.265,valign=t,trim=0.07in 0in 0.35in 0in,clip=true]{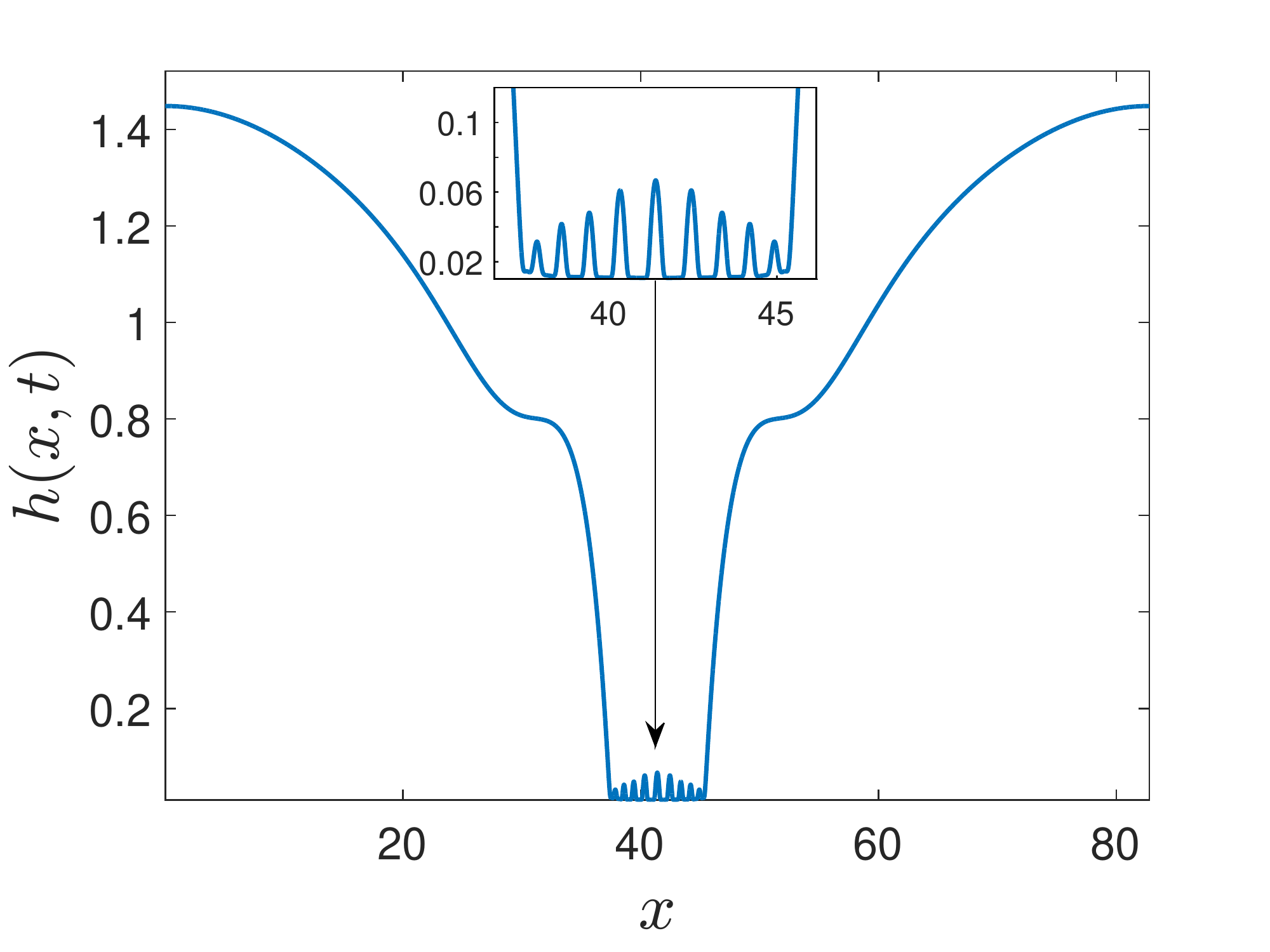}\label{fig:L1=10,L2=0,01,t6}}
\subfloat[]{\includegraphics[scale=0.265,valign=t,trim=0.07in 0in 0.35in 0in,clip=true]{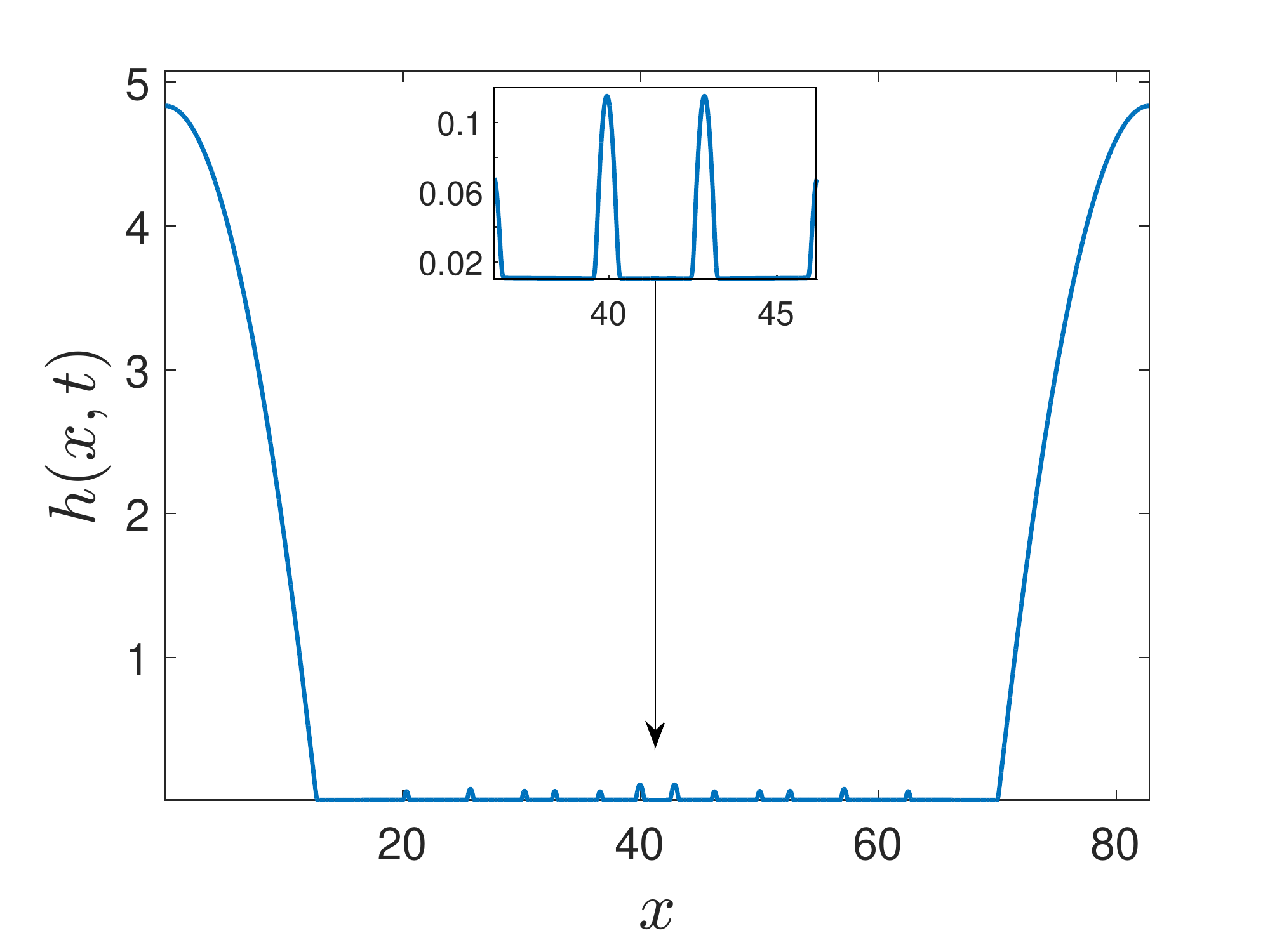}\label{fig:L1=10,L2=0,01,t20}}
\caption{Evolution of a viscoelastic film with $h_0=1$, $h_{\star}=0.01$, $b=0$, $\la_1=10$ and $\la_2=0.01$, at three selected times. \protect\subref{fig:L1=10,L2=0,01,t2} The separation of the two rims ($t=3.341 \times 10^5$). \protect\subref{fig:L1=10,L2=0,01,t6} The formation of oscillations that lead to the formation of secondary droplets ($t=3.345 \times 10^5$). The secondary droplets remain present until the final configuration shown in \protect\subref{fig:L1=10,L2=0,01,t20} ($t=4 \times 10^5$). In all three figures, the insets show a detailed close-up of the dewetting region.}\label{fig:L1=10,L2=0,01}
\end{figure}

Next, we take into account the viscosity of the Newtonian solvent by including $\la_2 \neq 0 $. As anticipated in \S~\ref{LSA}, $\la_2$ has the effect of slowing down the growth rate of the instability. We find that a slower dynamics provides a numerical advantage as well: By stabilizing the computations, hence avoiding the high Weissenberg number\footnote{Defined in Appendix A.} problem (see \cite{HaoPan} and references therein for a discussion on the computational challenges regarding this aspect), which otherwise can destabilize the numerical solutions. In fact, our simulation results show that when $\la_2=0$, for $\la_1>6$, unfeasibly small time steps would be required to overcome numerical instabilities, due to the rapid growth rate for purely elastic films. On the contrary, when viscoelastic films are considered, even a small contribution of the retardation time (e.g.~$\la_2=0.01$) allows simulations of films with a high Weissenberg number, that are yet numerically stable.

Figures \subref*{fig:L1=10,L2=0,01,t2}--\subref*{fig:L1=10,L2=0,01,t20} show the evolution of a viscoelastic dewetting film with $\la_1=10$, $\la_2=0.01$, at three selected times. In particular, in the inset of figure \subref*{fig:L1=10,L2=0,01,t2}, we observe the separation of the two rims (at time $t=3.341 \times 10^5$), and the formation of oscillations on the interface. These undulations lead to multiple secondary droplets that are shown in figure \subref*{fig:L1=10,L2=0,01,t6} ($t=3.345 \times 10^5$), and that remain present until the final configuration shown in figure \subref*{fig:L1=10,L2=0,01,t20} ($t=4 \times 10^5$). To our knowledge, secondary droplets of this nature have not been reported in numerical investigations of thin viscoelastic films, but their observation is consistent with experimental findings (see, for instance,~\cite{SharmaReiter,Reiter,GreenGanesan}). While we do not attempt a direct comparison with experiments in the present work, our results suggest that the emergence of secondary droplets in simulations is related to viscoelastic phenomena, in accordance with experimental observations. Our results therefore highlight the need for more refined numerical models for a comprehensive prediction of the instabilities in viscoelastic thin films.

\begin{figure}[H!t]
\captionsetup{type=figure}
\centering
\subfloat[]{\includegraphics[scale=0.265,valign=t,trim=0.07in 0in 0.35in 0in,clip=true]{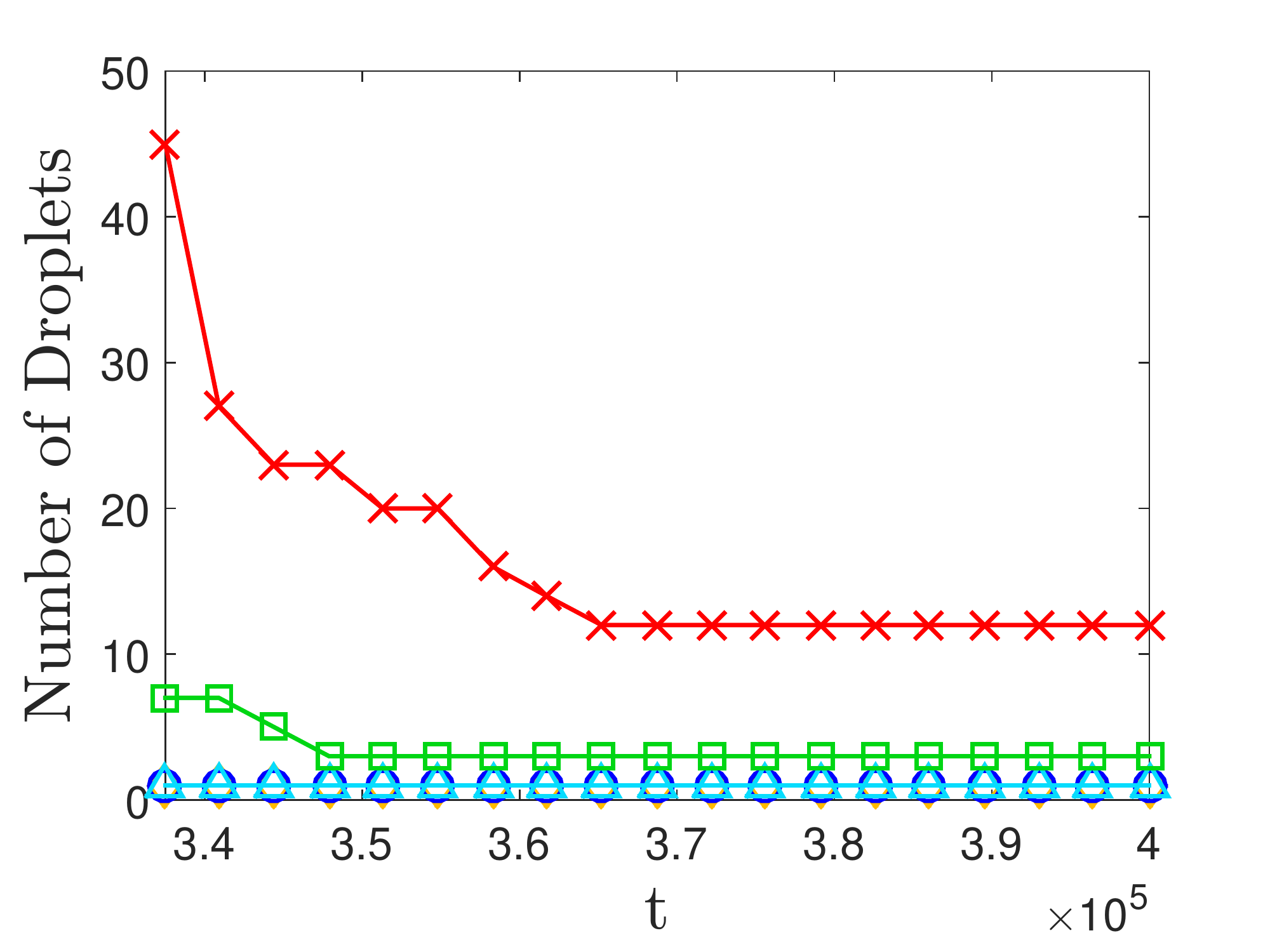}\label{fig:NumberDroplets,L2=0,01,L1=2,4,6,8,10,Dx=5x10^-3}}
\subfloat[]{\includegraphics[scale=0.265,valign=t,trim=0.07in 0in 0.35in 0in,clip=true]{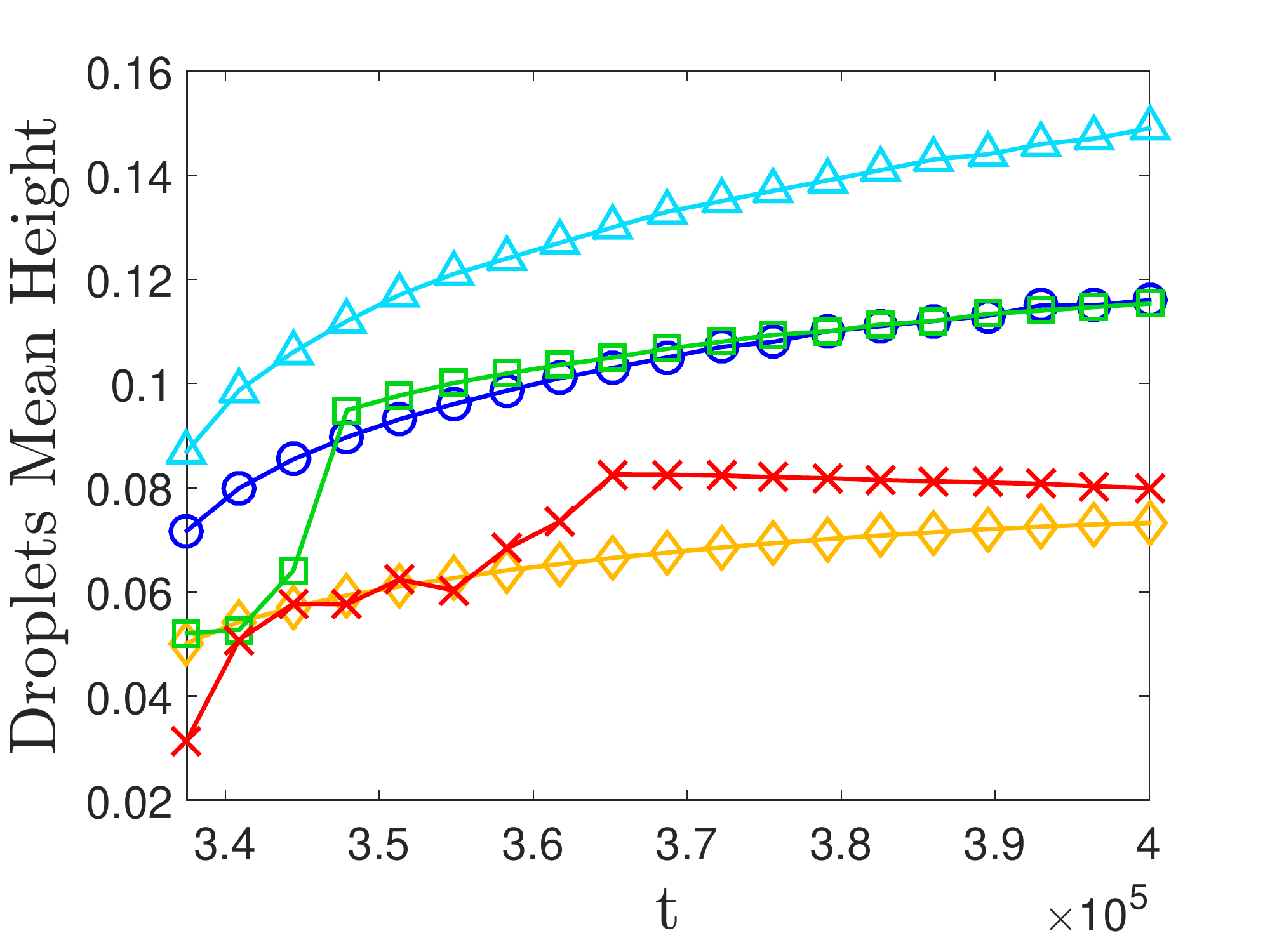}\label{fig:MeanHeightDroplets,L2=0,01,L1=2,4,6,8,10,Dx=5x10^-3}}
\subfloat[]{\includegraphics[scale=0.265,valign=t,trim=0.07in 0in 0.35in 0in,clip=true]{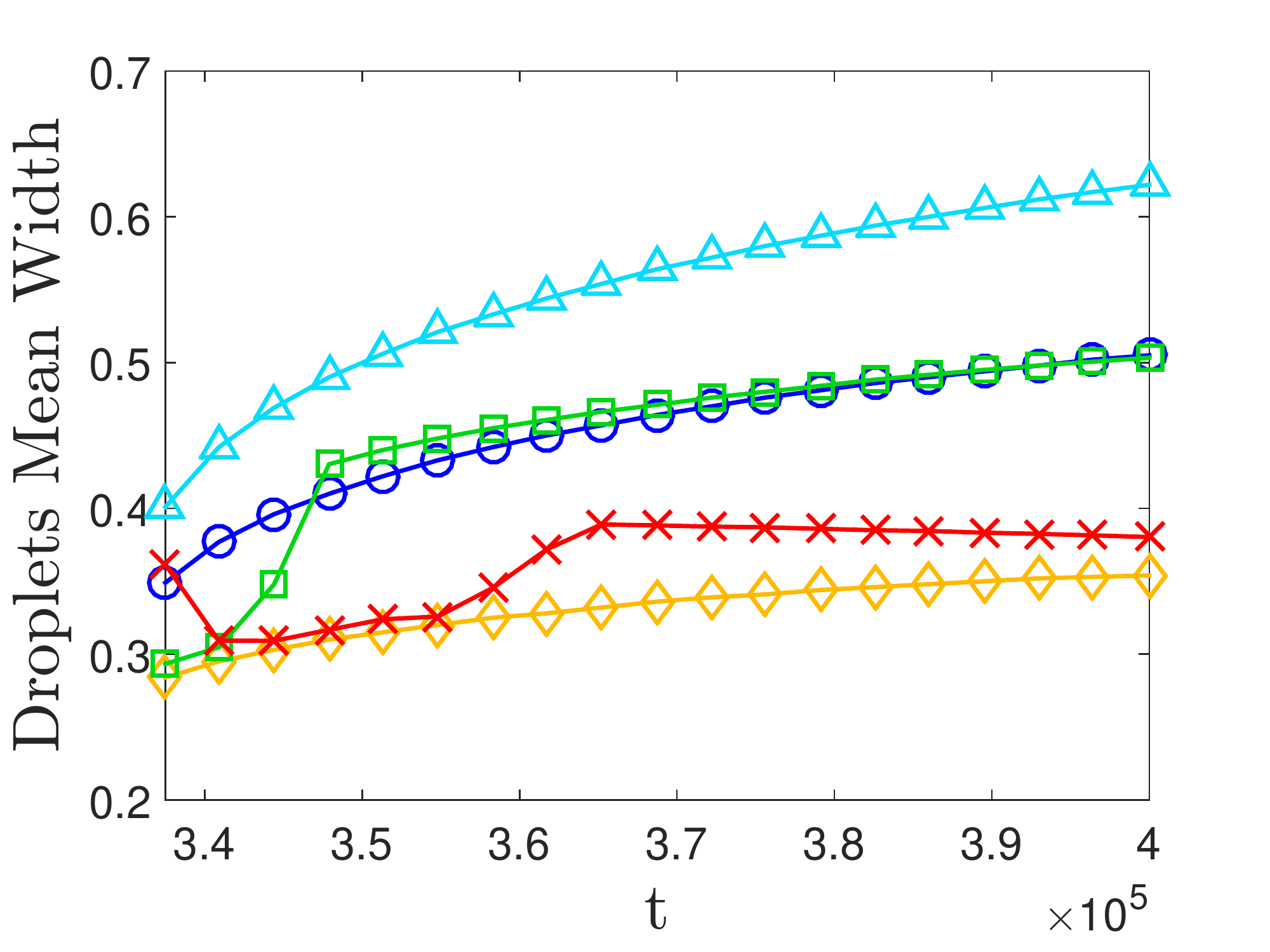}\label{fig:MeanWidthDroplets,L2=0,01,L1=2,4,6,8,10,Dx=5x10^-3}}
\caption{Evolution of the secondary droplets with $h_0=1$, $h_{\star}=0.01$, $b=0$, $\la_2=0.01$, and $\la_1=2$ (yellow diamonds), $4$ (blue circles), $6$ (cyan triangles), $8$ (green squares), and $10$ (red crosses). \protect\subref{fig:NumberDroplets,L2=0,01,L1=2,4,6,8,10,Dx=5x10^-3} The number of droplets versus time. For $\la_1=2,4,6$, only one secondary droplet is formed, so the lines overlap. For values of $\la_1>6$, there are multiple secondary droplets; due to coalescence, the number of secondary droplets decrease in time. \protect\subref{fig:MeanHeightDroplets,L2=0,01,L1=2,4,6,8,10,Dx=5x10^-3} The mean height of the secondary droplets versus time. \protect\subref{fig:MeanWidthDroplets,L2=0,01,L1=2,4,6,8,10,Dx=5x10^-3} The width (at half height) of the secondary droplets versus time.}\label{fig:DropletAnalyses}
\end{figure}

We can rheologically explain the presence of the secondary droplets by noting that a higher relaxation time $\la_1$ manifests a higher molecular weight of the polymers \cite{Larson}. Thus, in the presence of an extensional flow, as the one produced by the two separating rims, the chains of molecules are more stretched, and the elastic response to deformation is more visible. Similar considerations can be found in studies on beads-on-string structures of viscoelastic jets (see, for instance,~\cite{ChangEtAl,LiFontelos}). We note that both in our study and in the cited works on extensional flows of viscoelastic filaments, there is no strong correlation between the breakup time and the relaxation time. The latter mostly influences the formation of droplets, their migration and coalescence \cite{LiFontelos}. We also note that additional simulations have shown that $\lambda_2$ does not affect significantly the final configurations (results not shown for brevity). We moreover remark that a high Weissenberg number does not break the assumption of small shear rates, for which linear viscoelastic constitutive models, such as the one that we consider, are valid. This observation is confirmed numerically by analyzing the quantity $ \left|{\partial u}/{\partial x}\right| \sim \left| { h_t(x,t)}/{h(x,t)}\right|$ over the entire time of the evolution. In particular, for the times presented in figures \subref*{fig:L1=10,L2=0,01,t2}--\subref*{fig:L1=10,L2=0,01,t6}, $ \left|{\partial u}/{\partial x}\right|$ does not exceed the value of $10^{-2}$, and for the final stage of the evolution, shown in figure \subref*{fig:L1=10,L2=0,01,t20}, it has an order of magnitude of $10^{-5}$. Hence, we can confirm that for the flows considered, even for a high value of the dimensionless relaxation time, $\la_1 U/L$, which corresponds to a high Weissenberg number (see Appendix A), the assumption of small deformation gradients is not violated.

\begin{figure}[H!t1]
\captionsetup{type=figure}
\centering
\subfloat[]{\includegraphics[scale=0.265,valign=t,trim=0.07in 0in 0.35in 0in,clip=true]{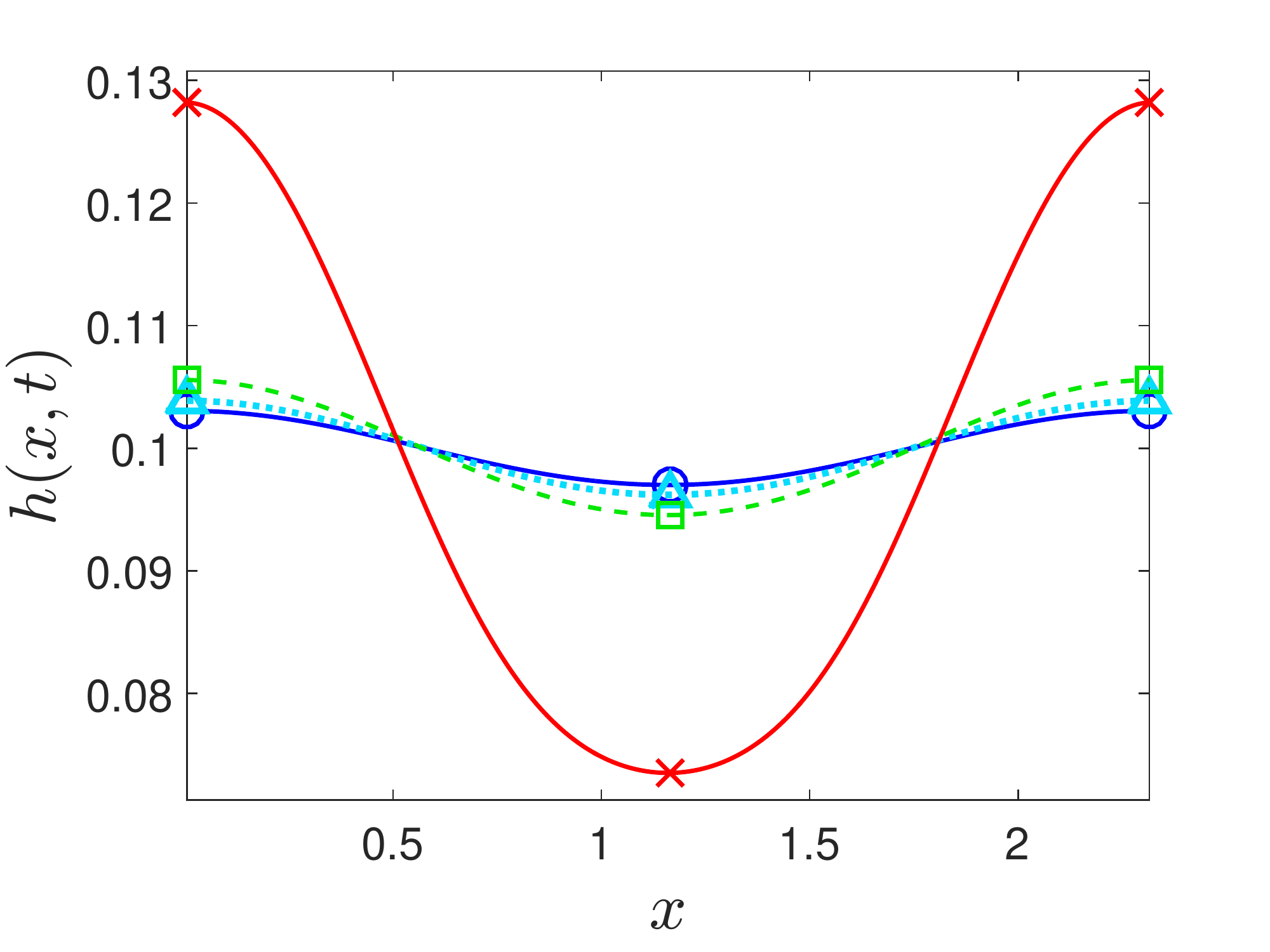}\label{fig:L2=0,B=0,L1=0,10,20,50,t2}}
\subfloat[]{\includegraphics[scale=0.265,valign=t,trim=0.07in 0in 0.35in 0in,clip=true]{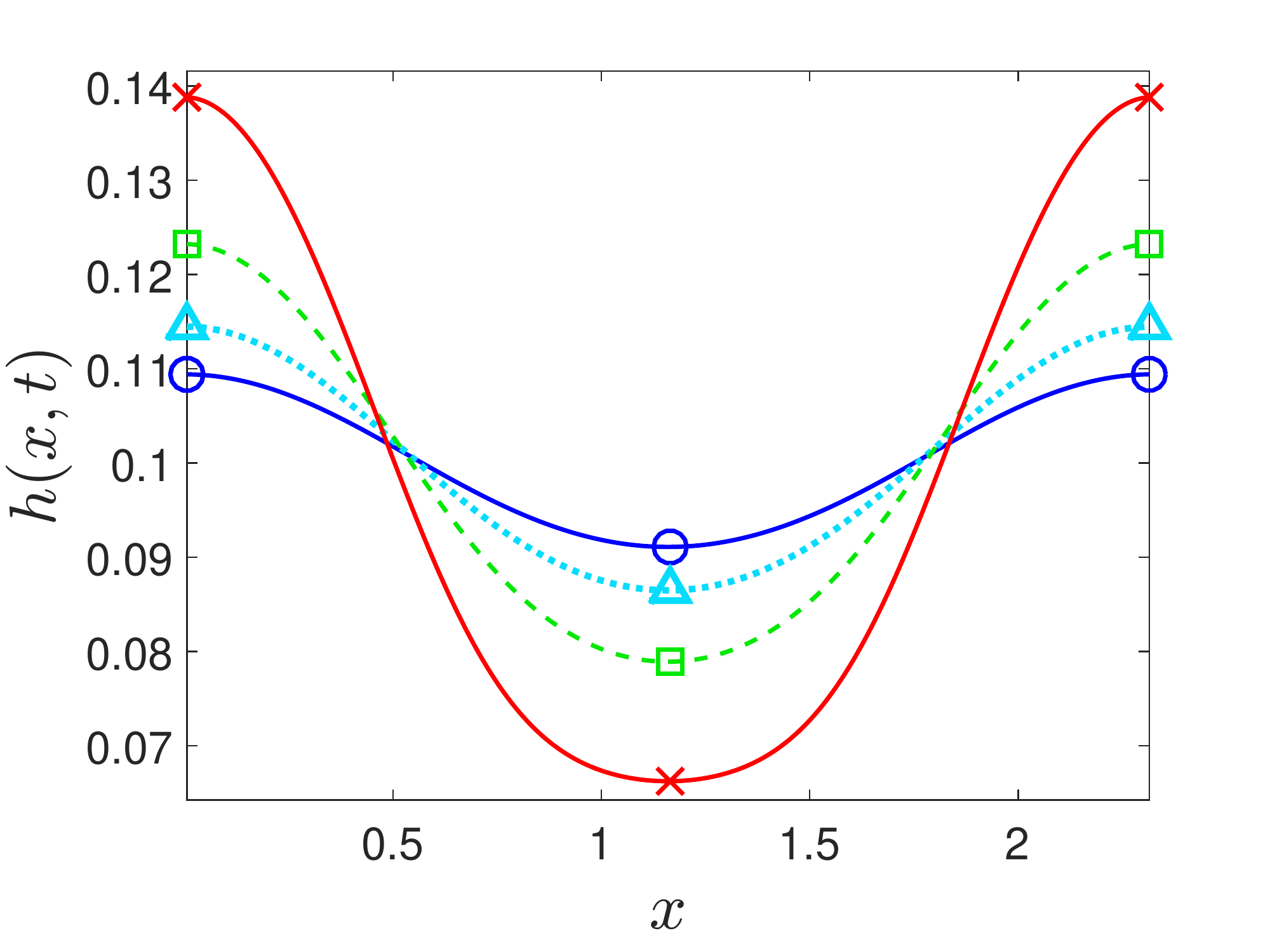}\label{fig:L2=0,B=0,L1=0,10,20,50,t3}}
\subfloat[]{\includegraphics[scale=0.265,valign=t,trim=0.07in 0in 0.35in 0in,clip=true]{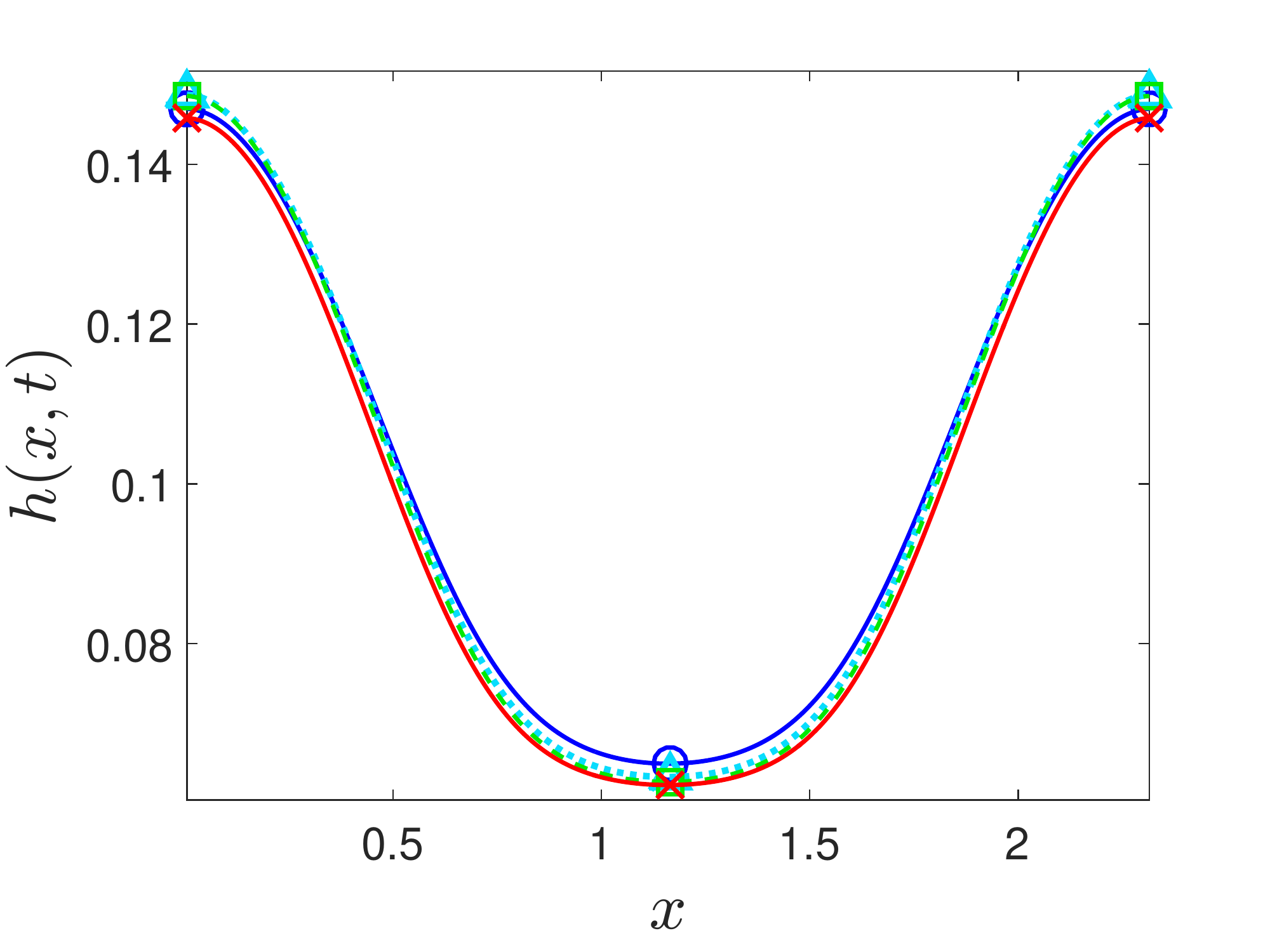}\label{fig:L2=0,B=0,L1=0,10,20,50,t5}}
\caption{Evolution of four distinct films with $h_0=0.1$ and $h_{\sta}=0.06$, $\la_2=0$, $b=0$, and $\la_1=0$ (blue circles), $10$ (cyan triangles), $20$ (green squares), and $50$ (red crosses), at $t=50$ \protect\subref{fig:L2=0,B=0,L1=0,10,20,50,t2},
$t=70$ \protect\subref{fig:L2=0,B=0,L1=0,10,20,50,t3} and $t=250$ \protect\subref{fig:L2=0,B=0,L1=0,10,20,50,t5}. For this set of parameters, the dynamics of the breakup is faster compared to the simulations considered in earlier figures.}
\end{figure}

We proceed by analyzing the influence of $\lambda_1$ on the characteristic length scales of the secondary droplets. The corresponding results are shown in figure \ref{fig:DropletAnalyses}, where we keep $\la_2=0.01$, and study the resulting morphologies for $\la_1=2,4,6,8,10$. Figure \subref*{fig:NumberDroplets,L2=0,01,L1=2,4,6,8,10,Dx=5x10^-3} shows the number of secondary droplets versus time. For $\la_1=2,4,6$, denoted by yellow diamonds, blue circles, and cyan triangles respectively, only one secondary droplet is formed. Whereas, for values of $\la_1=8,10$, denoted by green squares and red crosses respectively, multiple secondary droplets form. At later times and when $\la_1=8,10$, the secondary droplets can coalesce, resulting in a sudden change in their numbers, height, and width. From figure \subref*{fig:NumberDroplets,L2=0,01,L1=2,4,6,8,10,Dx=5x10^-3}, we also notice that the merging of secondary droplets is much more severe for $\la_1=10$, implying that the elastic force is responsible for the secondary droplets migration and coalescence. We note that the coarsening process of droplets under the influence of the disjoining pressure has been studied for Newtonian fluids \cite{GlasnerWitelski,BertozziWitelski}. Extending such analyses to include viscoelastic effects would be of interest, but is beyond the scope of the present work.

The results presented so far suggest that the relaxation time $\la_1$ significantly affects the final morphologies of dewetting films, but it has a weaker effect on the growth rate and the breakup time. We note that in the results presented above, the viscoelastic relaxation time, $\la_1$, is relatively short with respect to the breakup time. Hence, in the linear early-time regime of the evolution, the viscoelastic fluid shows a Newtonian response. It is therefore reasonable to raise the question of the effect of the liquid viscoelasticity when the breakup time is comparable to the relaxation time of the liquid, i.e.~when $\om_m^{-1} \approx \la_1$. Figures \ref{fig:L2=0,B=0,L1=0,10,20,50,t2}--\ref{fig:L2=0,B=0,L1=0,10,20,50,t5} present the evolution of four distinct films with $h_0=0.1$ and $h_{\sta}=0.06$, $\la_2=0$, $b=0$, and $\la_1=0$ (blue circles), $10$ (cyan triangles), $20$ (green squares), and $50$ (red crosses), at three selected times. For this set of parameters, the relaxation time and the breakup time are comparable. We see that in this regime, the influence of the elasticity parameter $\la_1$ on the time scale of the evolution of the dewetting prior to rupture is more pronounced, with respect to the cases analyzed earlier. We furthermore notice that for this set of parameters, the van der Waals interaction force, higher in this case, prevents formation of satellite droplets.

\begin{figure}[ht1]
\captionsetup{type=figure}
\centering
\subfloat[]{\includegraphics[scale=0.265,valign=t,trim=0.07in 0in 0.35in 0in,clip=true]{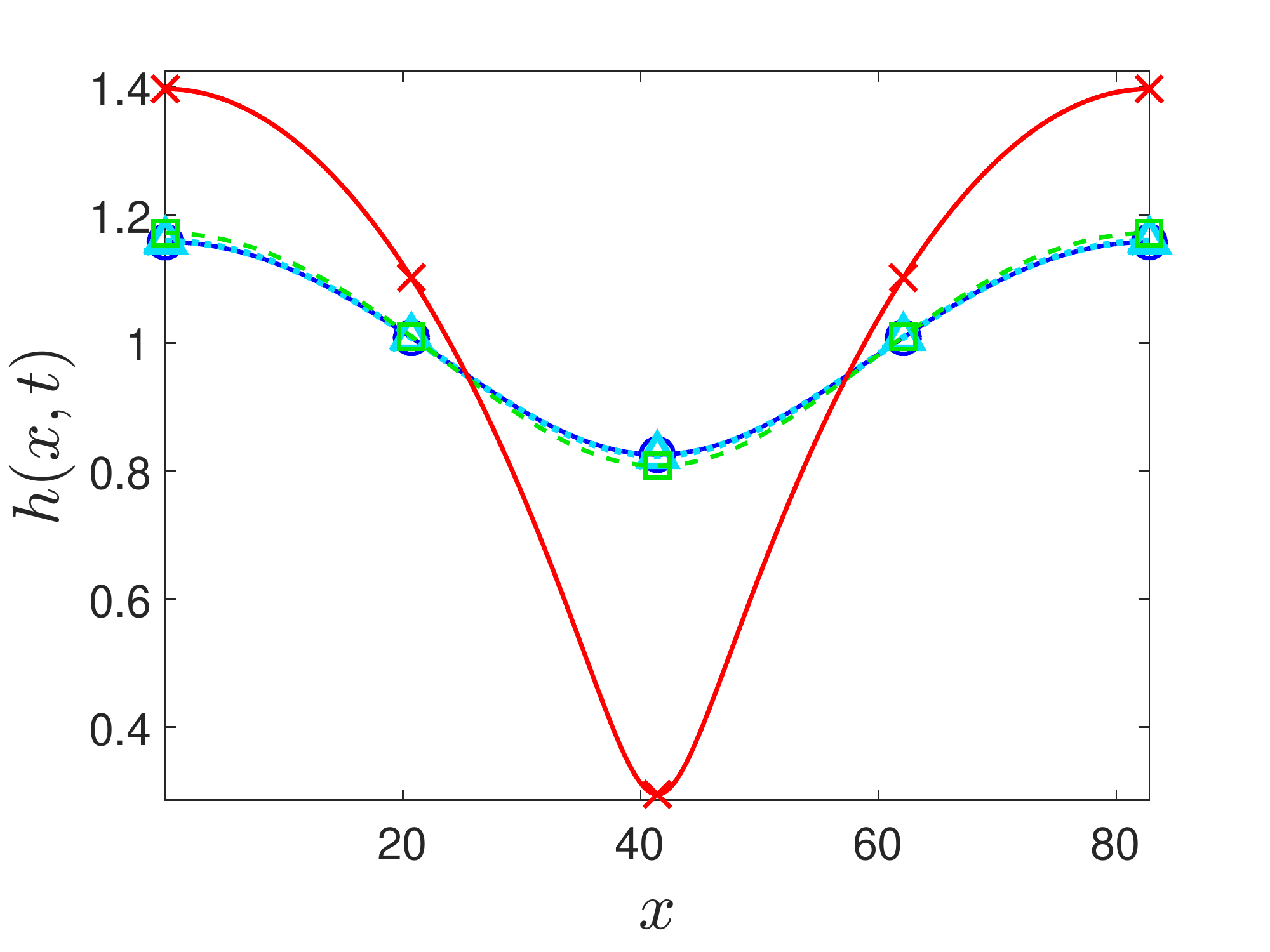}\label{fig:L1=10,L2=0,01,B=0VSB=0,1,t1}}
\subfloat[]{\includegraphics[scale=0.265,valign=t,trim=0.07in 0in 0.35in 0in,clip=true]{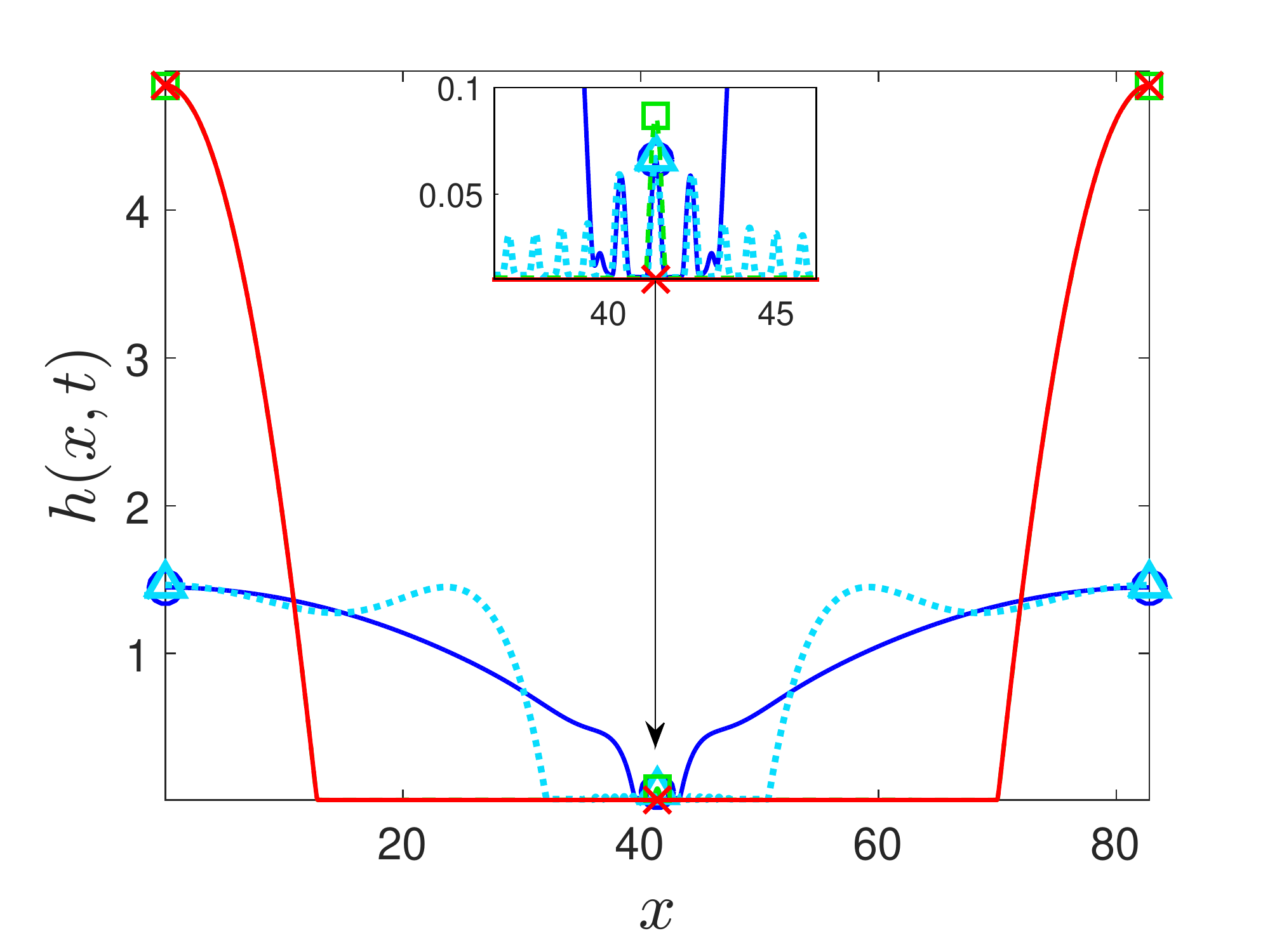}\label{fig:L1=10,L2=0,01,B=0VSB=0,1,t3}}
\subfloat[]{\includegraphics[scale=0.265,valign=t,trim=0.07in 0in 0.35in 0in,clip=true]{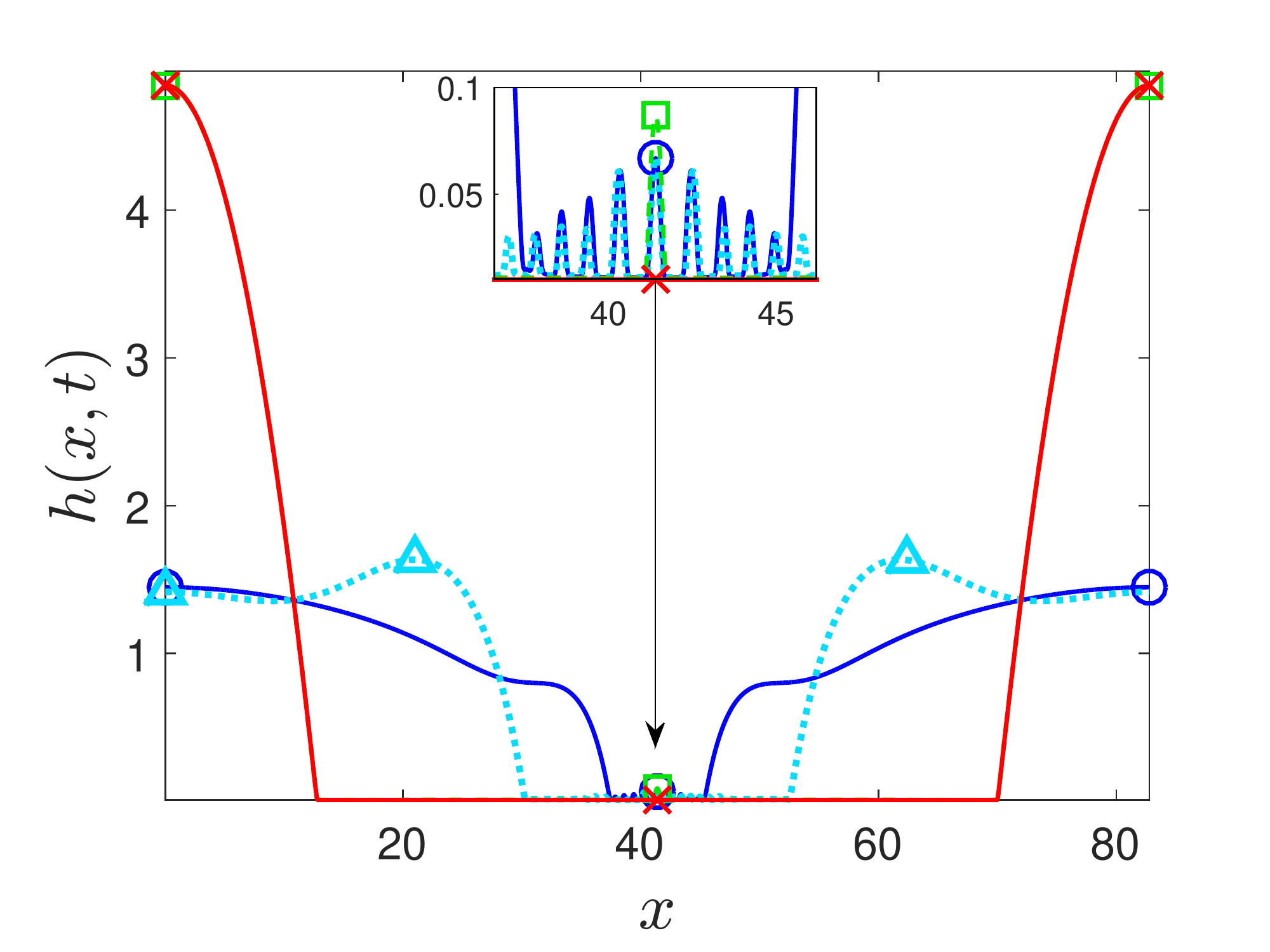}\label{fig:L1=10,L2=0,01,B=0VSB=0,1,t6}}\\
\caption{Evolution of dewetting films for $h_0=1$, $h_{\star}=0.01$, $\la_1=10$ and $\la_2=0.01$ at three selected times, for $b=0$ (blue circles), $0.001$ (cyan triangles), $0.01$ (green squares), and $0.1$ (red crosses). At time $t=2.5 \times 10^5$ in \protect\subref{fig:L1=10,L2=0,01,B=0VSB=0,1,t1}, the film with $b=0.1$ is separating, while in \protect\subref{fig:L1=10,L2=0,01,B=0VSB=0,1,t3} at time $t=3.34 \times 10^5$, and in \protect\subref{fig:L1=10,L2=0,01,B=0VSB=0,1,t6} at time $t=3.345 \times 10^5$, the films with $b=0.01$ and $b=0.1$ are already fully developed. We note that for large slip, the evolution is faster and no satellite droplets form. As $b$ is decreased, the dynamics is slower and satellites form. The insets show a close-up of the secondary droplets.}\label{fig:L1=10,L2=0,01,B=0VSB=0,1}
\end{figure}

Finally, we focus on the influence of the slip coefficient $b$. In figures \subref*{fig:L1=10,L2=0,01,B=0VSB=0,1,t1}--\subref*{fig:L1=10,L2=0,01,B=0VSB=0,1,t6}, we fix the viscoelastic parameters $\la_1=10,\la_2=0.01$, and consider $b=0$ (blue circles), $0.001$ (cyan triangles), $0.01$ (green squares), and $0.1$ (red crosses). The results show that the dynamics with a non-zero slip coefficient is faster. In fact, we can see in figure \subref*{fig:L1=10,L2=0,01,B=0VSB=0,1,t1}, that the film with $b=0.1$ is already separating at time $t=2.5 \times 10^5$, whereas the other films are still in the initial phase in which the perturbation has not grown significantly. Moreover, we see in figure \subref*{fig:L1=10,L2=0,01,B=0VSB=0,1,t3} at time $t=3.34 \times 10^5$, and in figure \subref*{fig:L1=10,L2=0,01,B=0VSB=0,1,t6} at time $t=3.345 \times 10^5$, that the films with $b=0.01$, and $b=0.1$ are already fully developed, whereas the ones with $b=0$, and $b=0.001$ are still retracting. We show that not only slip has an influence on the dynamics of the evolution, but it also has two main effects on the resulting morphologies of the interface: First, by raising the height of the retracting rims in the early stage of the evolution (we note in figure \subref*{fig:L1=10,L2=0,01,B=0VSB=0,1,t3} the dip in the interface of the receding rim for the film with $b=0.001$ in contrast to the one with no-slip); Second, by preventing the formation of the secondary droplets in the final configuration. In fact, multiple satellite droplets form in the cases with $b=0$, and $b=0.001$, while only one secondary droplet remains present when $b=0.01$, and none when $b=0.1$.

\begin{figure}[h!t1]
\captionsetup{type=figure}
\centering
\subfloat[]{\includegraphics[scale=0.265,valign=t,trim=0.07in 0in 0.35in 0in,clip=true]{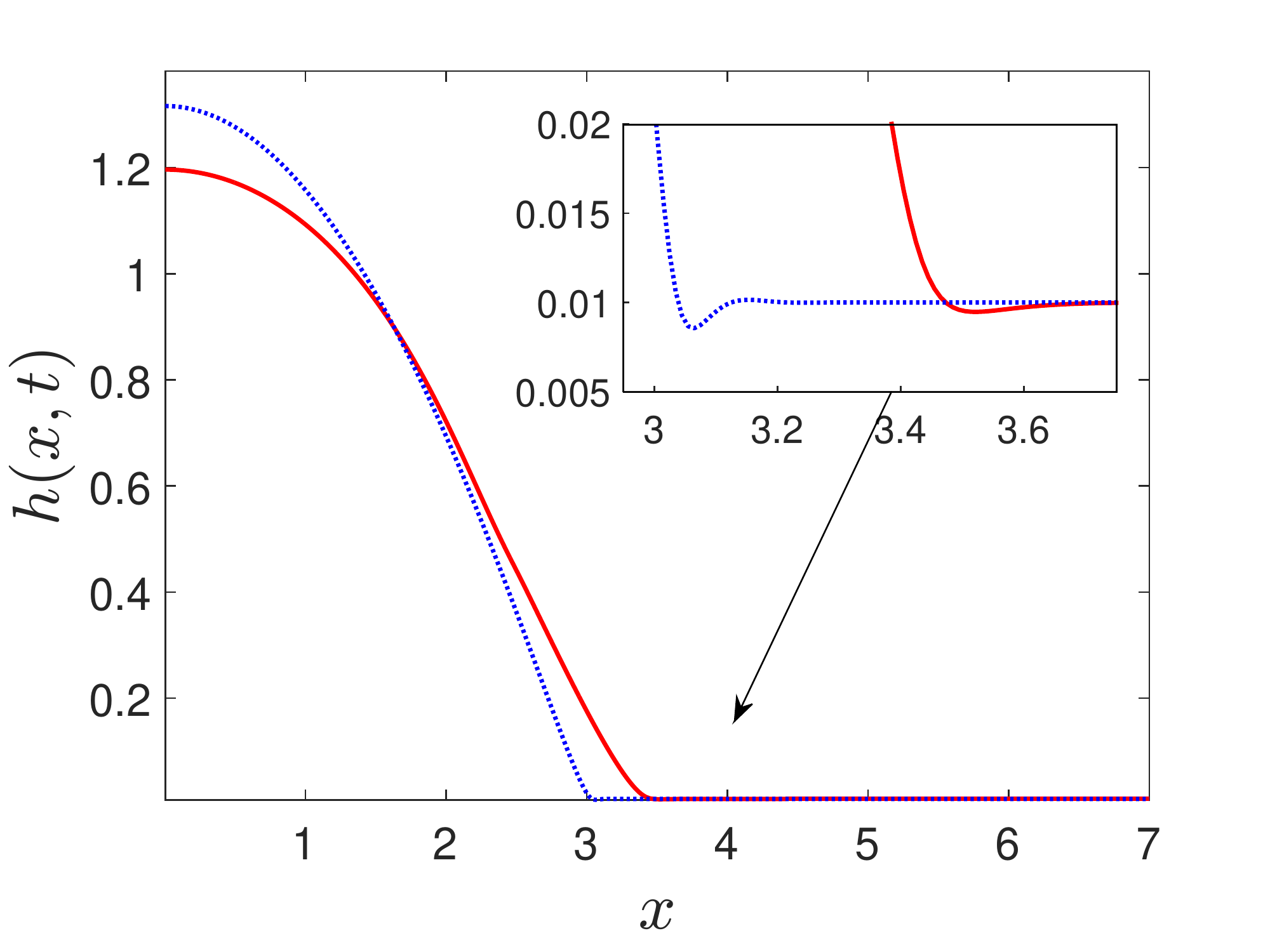}\label{fig:DropletL1=15,L2=0,01,VSL1=L2=0,hstar=0.01,t3}}
\subfloat[]{\includegraphics[scale=0.265,valign=t,trim=0.07in 0in 0.35in 0in,clip=true]{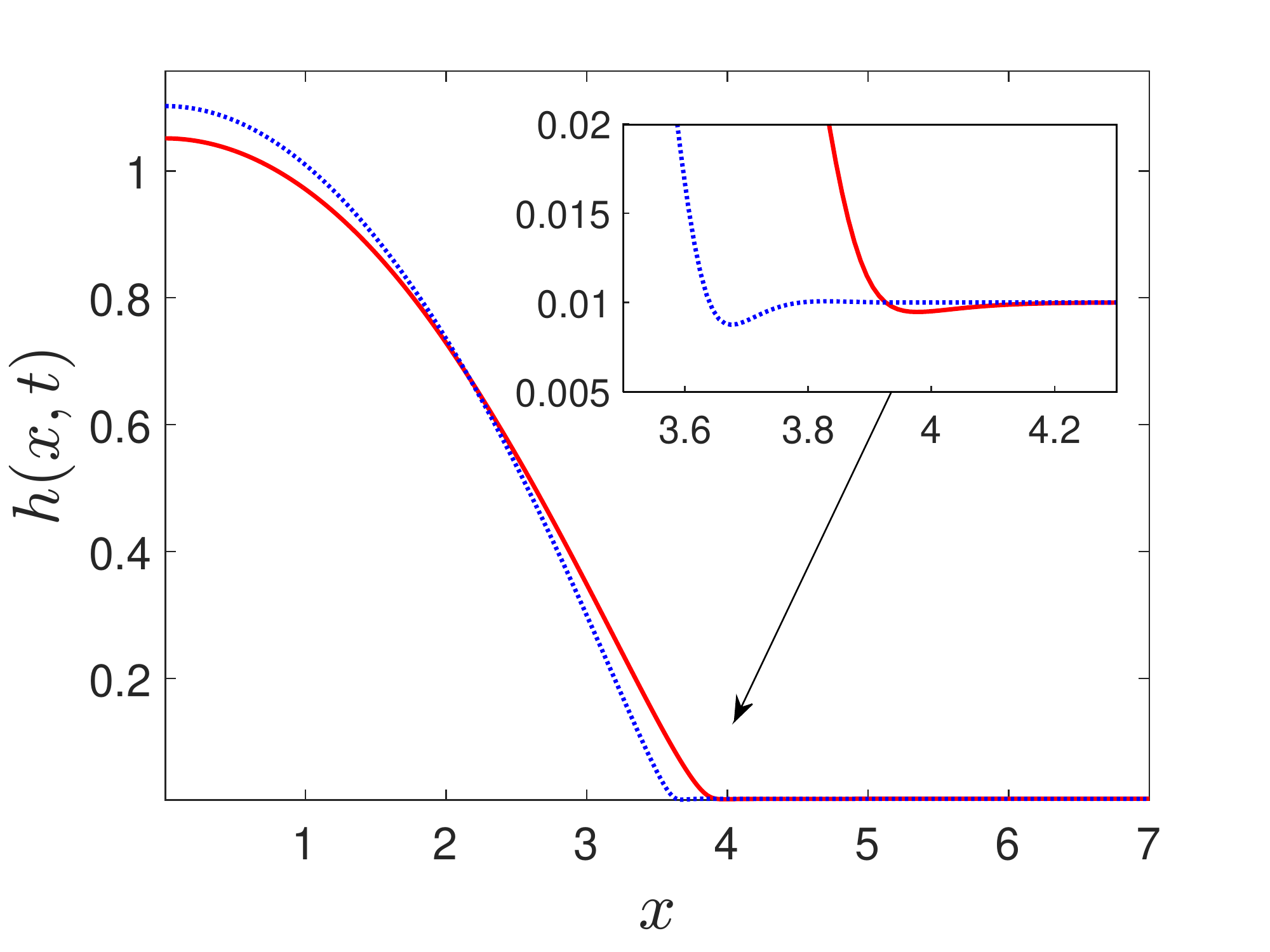}\label{fig:DropletL1=15,L2=0,01,VSL1=L2=0,hstar=0.01,t11}}
\subfloat[]{\includegraphics[scale=0.265,valign=t,trim=0.07in 0in 0.35in 0in,clip=true]{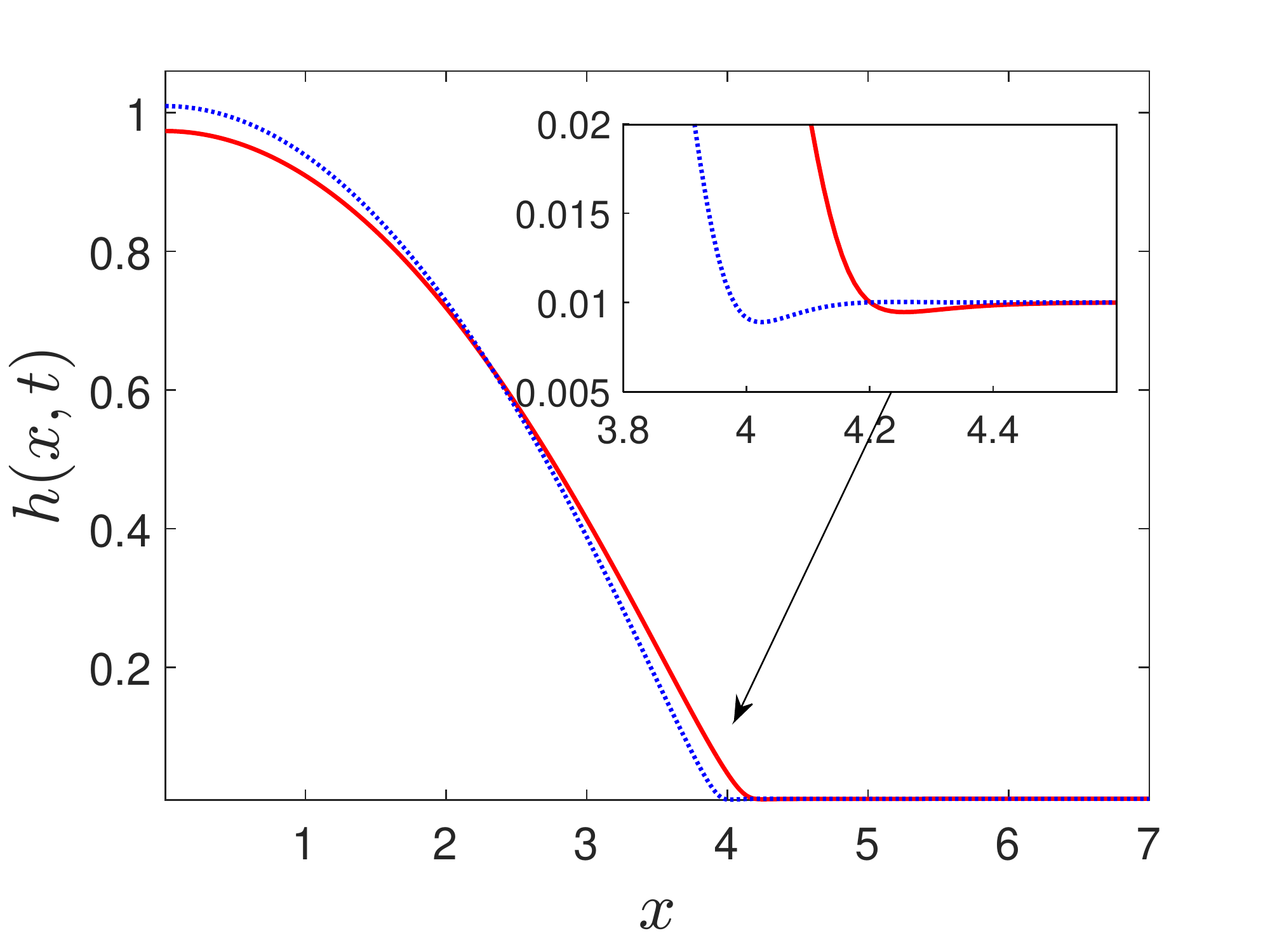}\label{fig:DropletL1=15,L2=0,01,VSL1=L2=0,hstar=0.01,t21}}\\
\subfloat[]{\includegraphics[scale=0.265,valign=t,trim=0.07in 0in 0.35in 0in,clip=true]{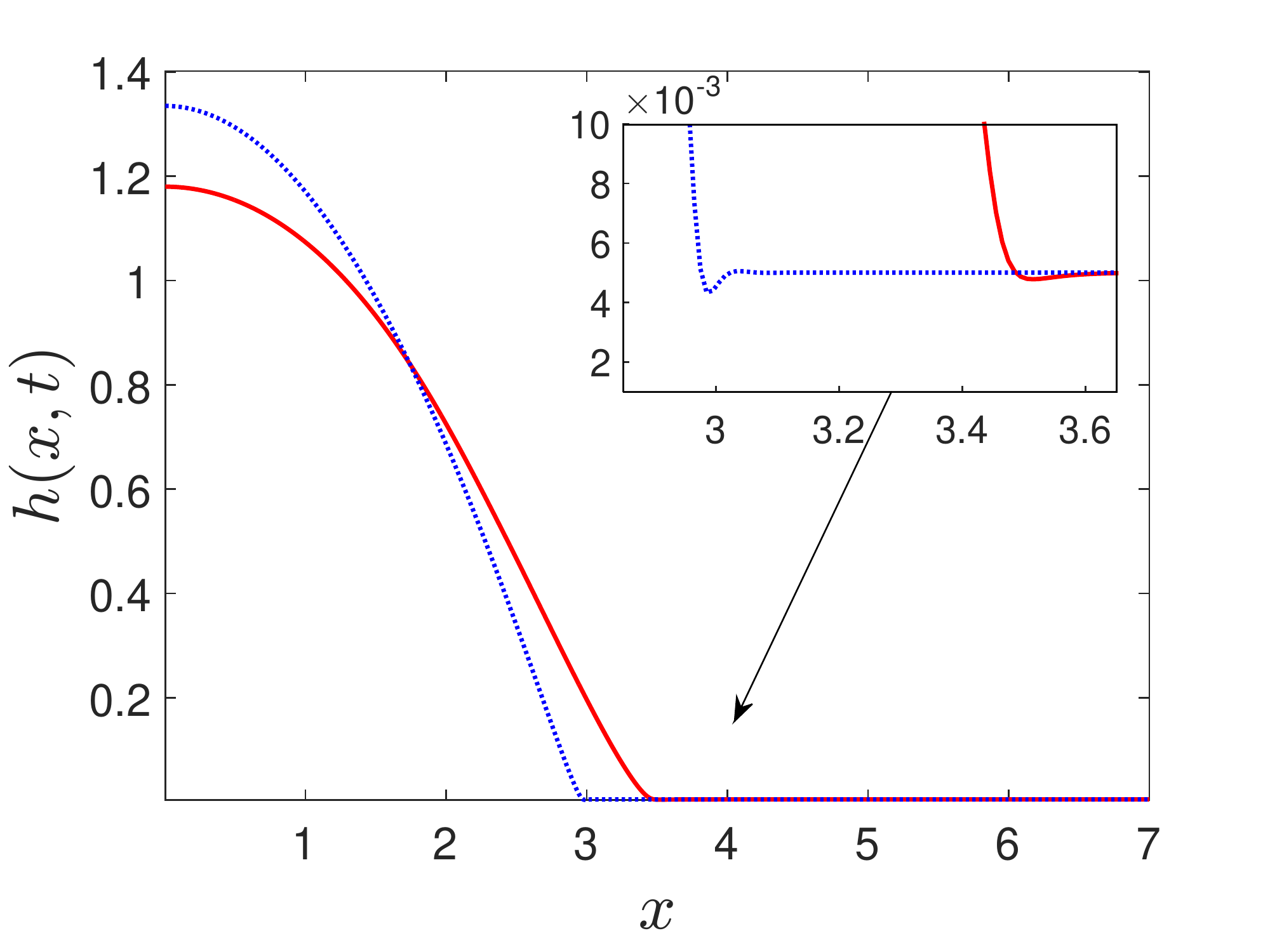}\label{fig:DropletL1=15,L2=0,01,VSL1=L2=0,hstar=0.005,t3}}
\subfloat[]{\includegraphics[scale=0.265,valign=t,trim=0.07in 0in 0.35in 0in,clip=true]{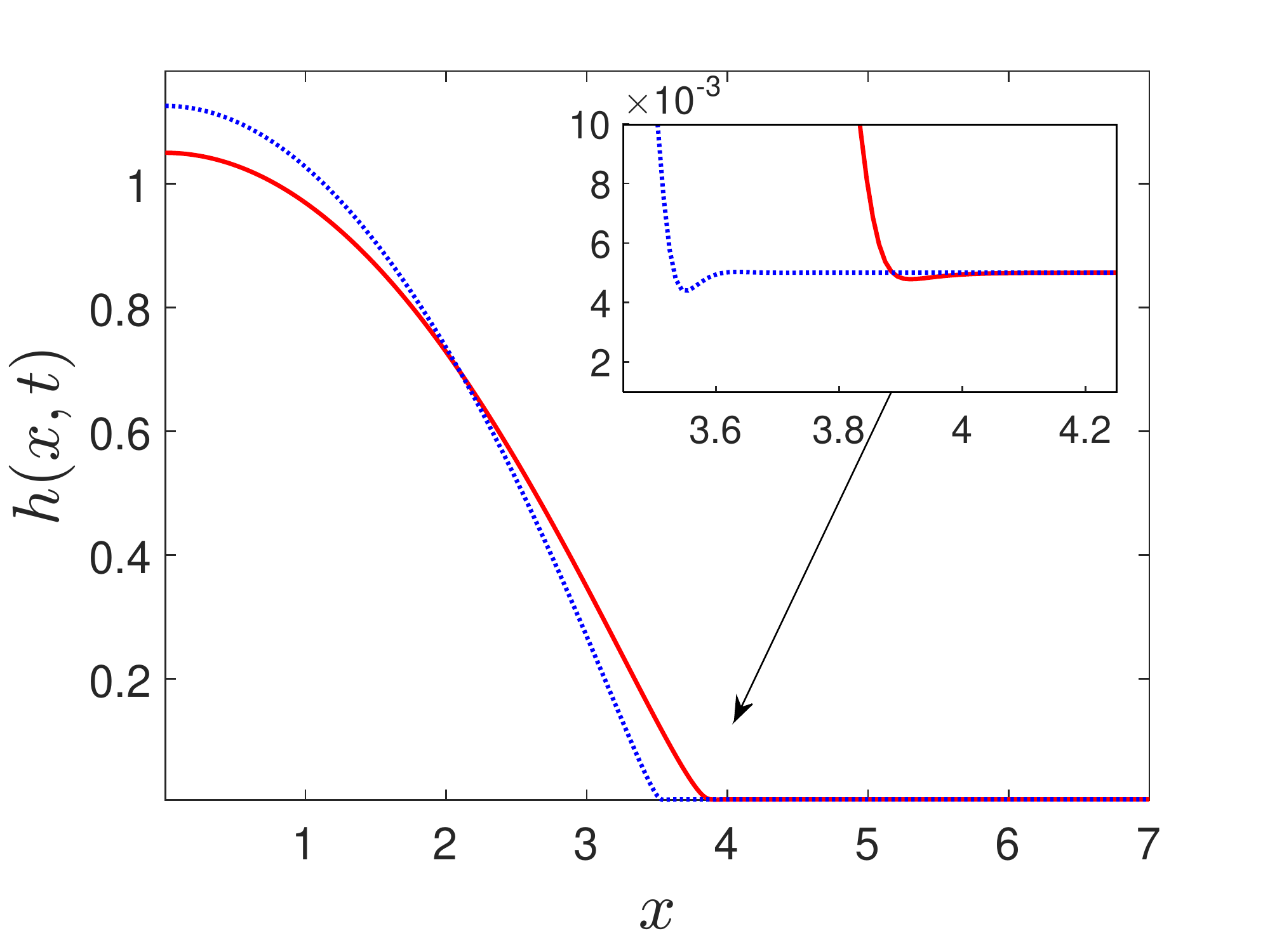}\label{fig:DropletL1=15,L2=0,01,VSL1=L2=0,hstar=0.005,t11}}
\subfloat[]{\includegraphics[scale=0.265,valign=t,trim=0.07in 0in 0.35in 0in,clip=true]{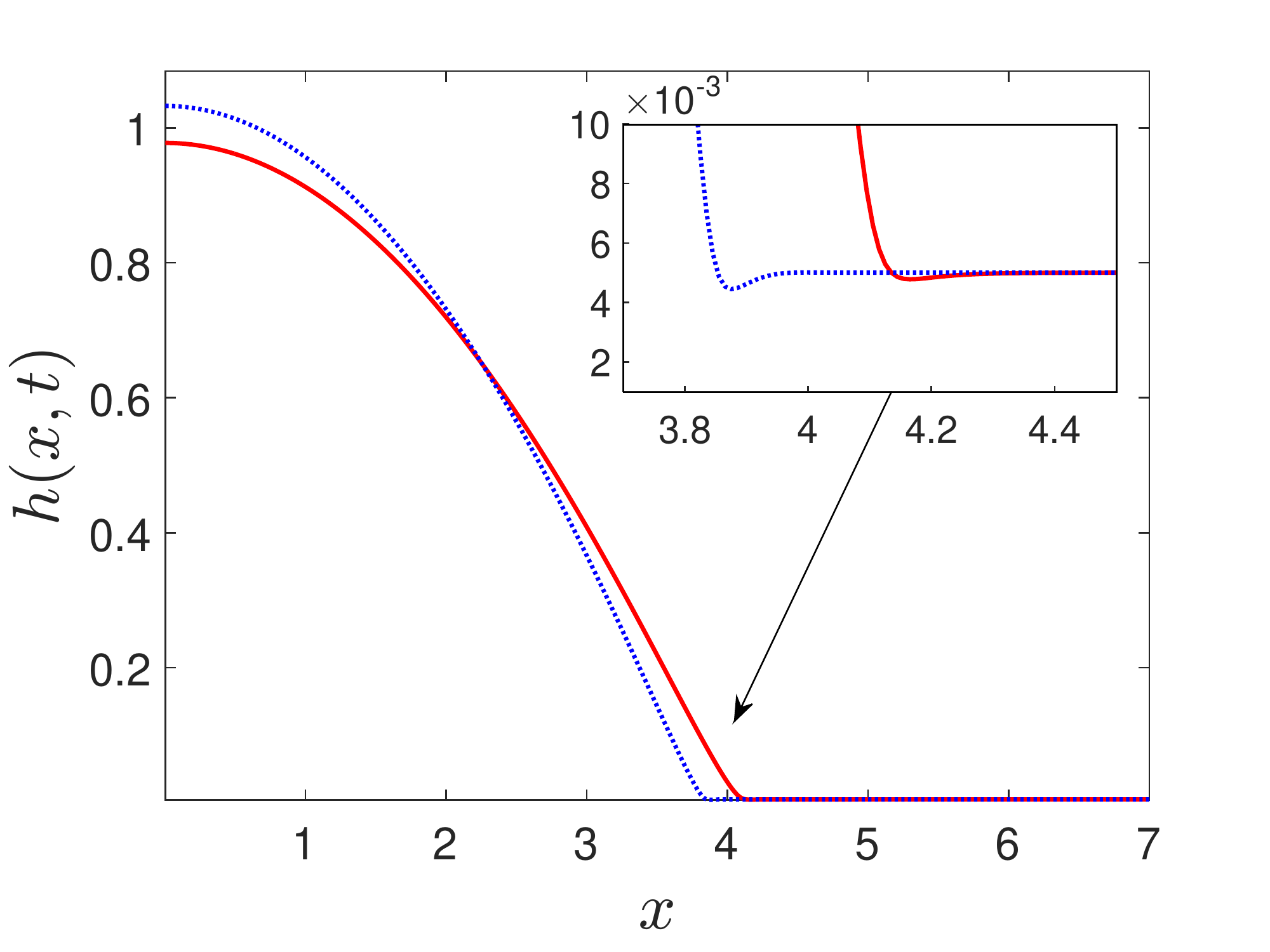}\label{fig:DropletL1=15,L2=0,01,VSL1=L2=0,hstar=0.005,t21}}
\caption{The spreading of a viscoelastic drop with $\la_1=15,\la_2=0.01$ (red solid curve) versus a Newtonian drop with $\la_1=\la_2=0$ (blue dotted curve), at $t=10,50,100$ from left to right. In \protect\subref{fig:DropletL1=15,L2=0,01,VSL1=L2=0,hstar=0.01,t3}--\protect\subref{fig:DropletL1=15,L2=0,01,VSL1=L2=0,hstar=0.01,t21} the equilibrium thickness $h_{\sta}=0.01$; in \protect\subref{fig:DropletL1=15,L2=0,01,VSL1=L2=0,hstar=0.005,t3}--\protect\subref{fig:DropletL1=15,L2=0,01,VSL1=L2=0,hstar=0.005,t21} $h_{\sta}=0.005$.
The insets show a close-up of the contact line region.}\label{fig:8}
\end{figure}

\subsection[Spreading and receding viscoelastic drops]{Spreading and receding viscoelastic drops}
\subsubsection[Spreading drops]{Spreading drops}

Next, we discuss spreading of a planar viscoelastic drop. The initial condition is a circular cap of radius $R$ and center $(0,-R \cos \theta_i)$, that lies on the substrate with an offset of thickness $h_{\sta}$. We specify the initial contact angle between the fluid interface and the solid substrate, called $\theta_i$, different from the equilibrium angle, denoted by $\theta_e$. The latter is implicitly defined by the form of the disjoining pressure given by equation (\ref{Def:VdW}). We investigate the dynamic contact angle, $\theta_D$, formed at the moving contact line, and study its relation with $\theta_e$. $\theta_D$ is calculated as the slope of the tangent line at the inflection point of the fluid interface $h(x, t)$. In the discussion that follows, we start with $\theta_i=30^{\circ}$, and let the drop relax to $\theta_e=15^{\circ}$. For all cases shown, we impose a no-slip boundary condition.

\begin{figure}[H!t1]
\captionsetup{type=figure}
\centering
\subfloat[]{\includegraphics[scale=0.35,valign=t,trim=0.15in 0in 0.3in 0in,clip=true]{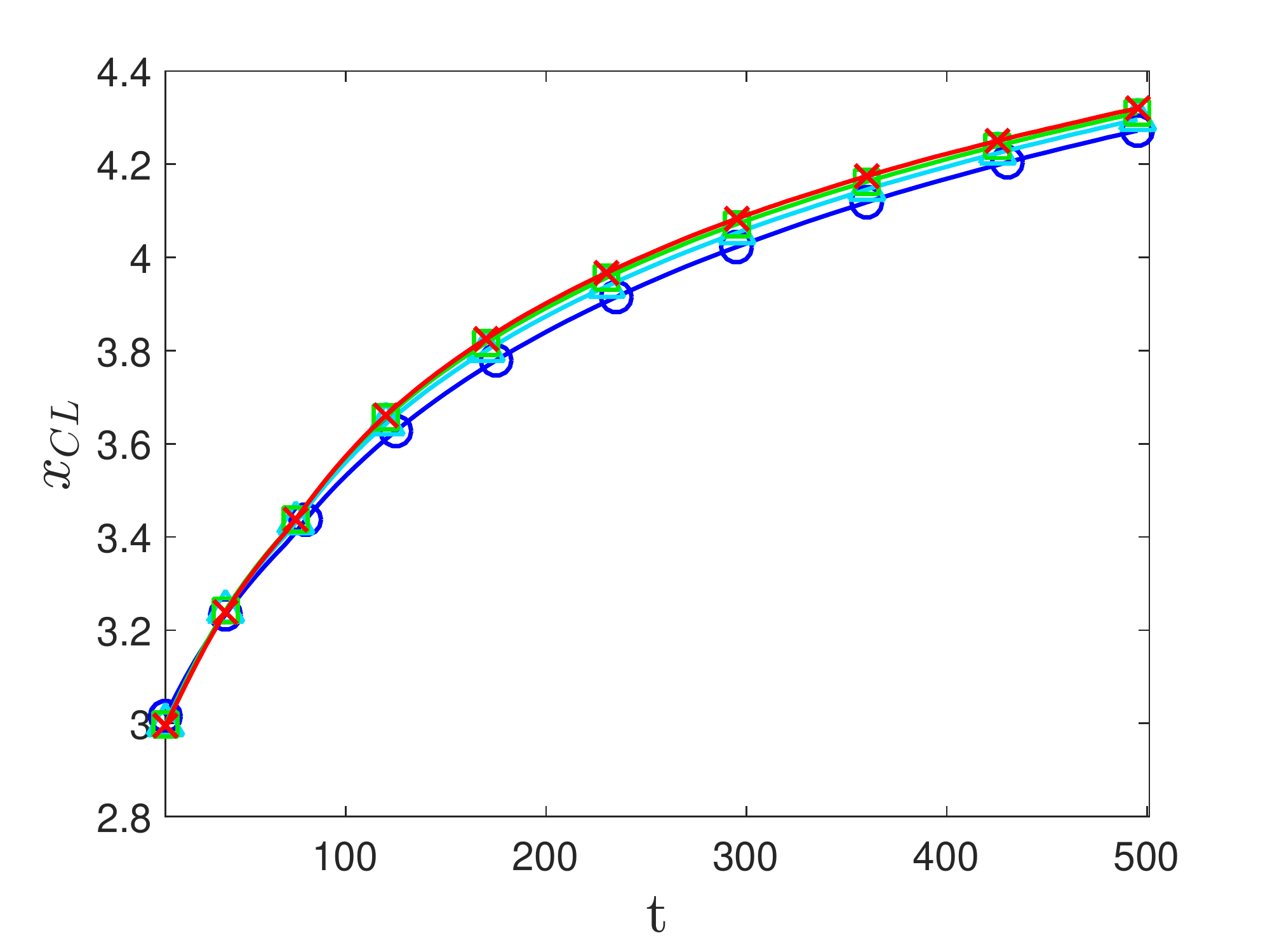}\label{fig:CLPVSTime}}
\subfloat[]{\includegraphics[scale=0.35,valign=t,trim=0.1in 0in 0.3in 0in,clip=true]{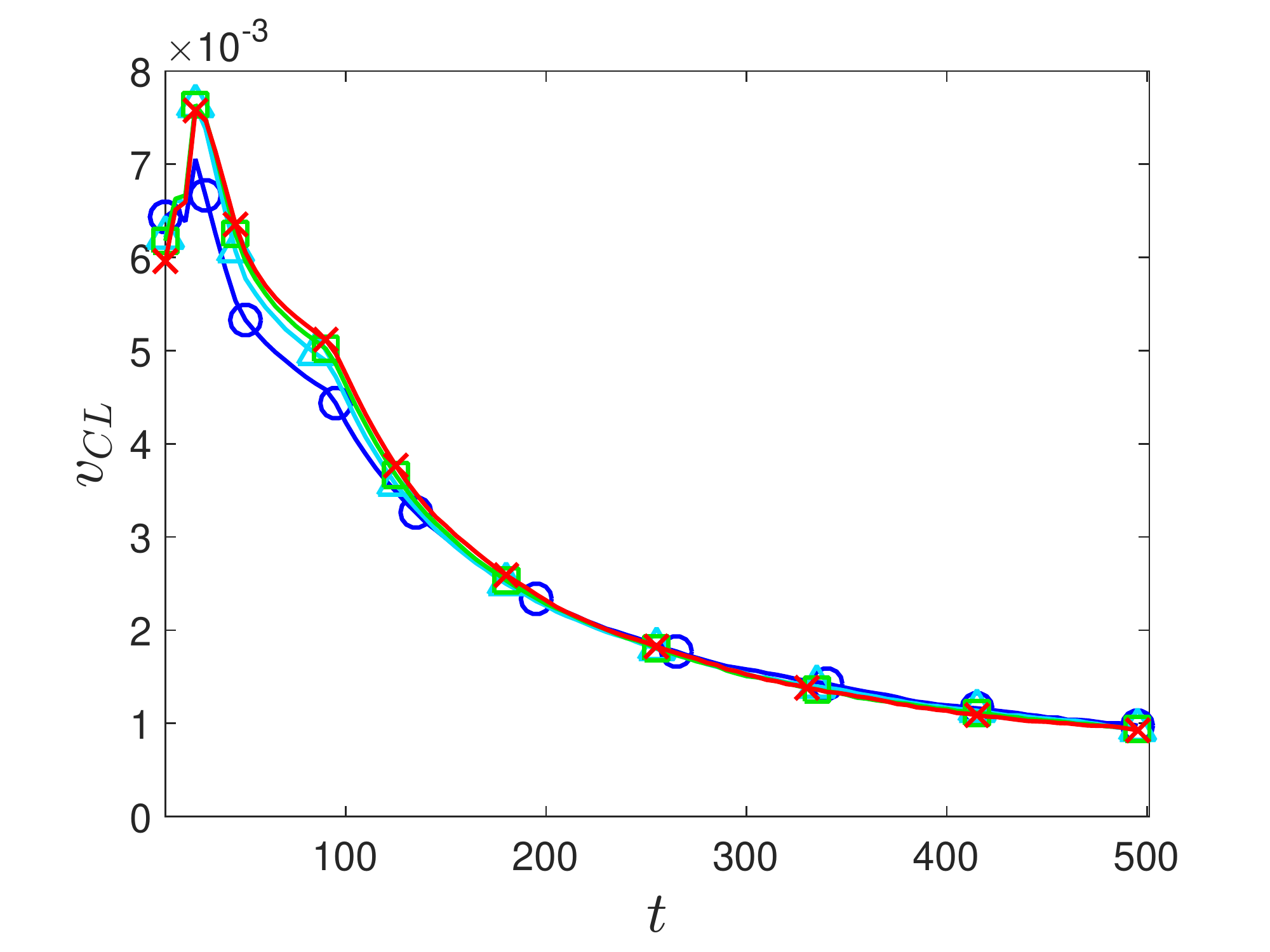}\label{fig:VelocityVSTime}}\\
\subfloat[]{\includegraphics[scale=0.35,valign=t,trim=0.03in 0in 0.3in 0in,clip=true]{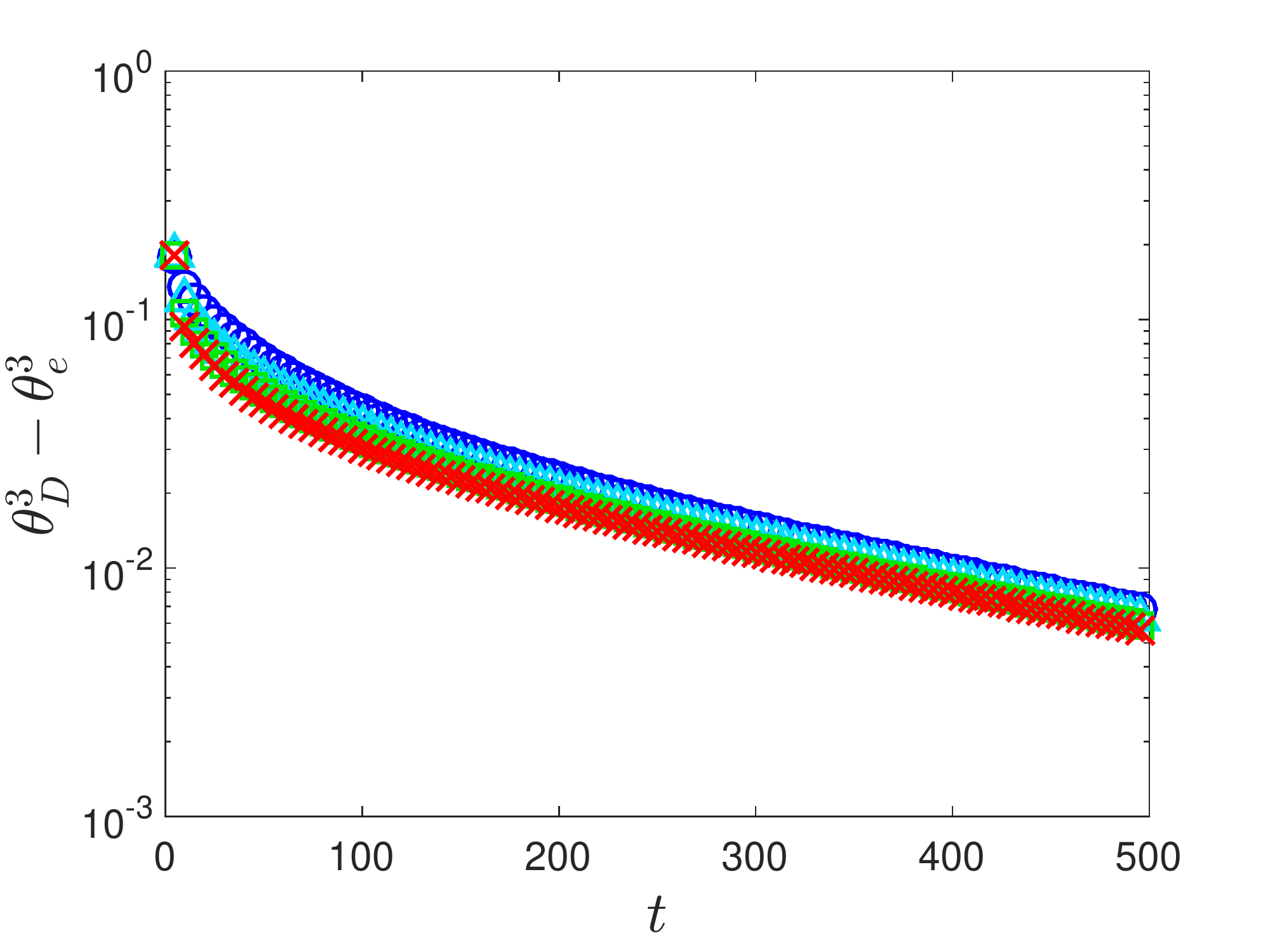}\label{fig:CubesVSTime}}
\subfloat[]{\includegraphics[scale=0.35,valign=t,trim=0.03in 0in 0.3in 0in,clip=true]{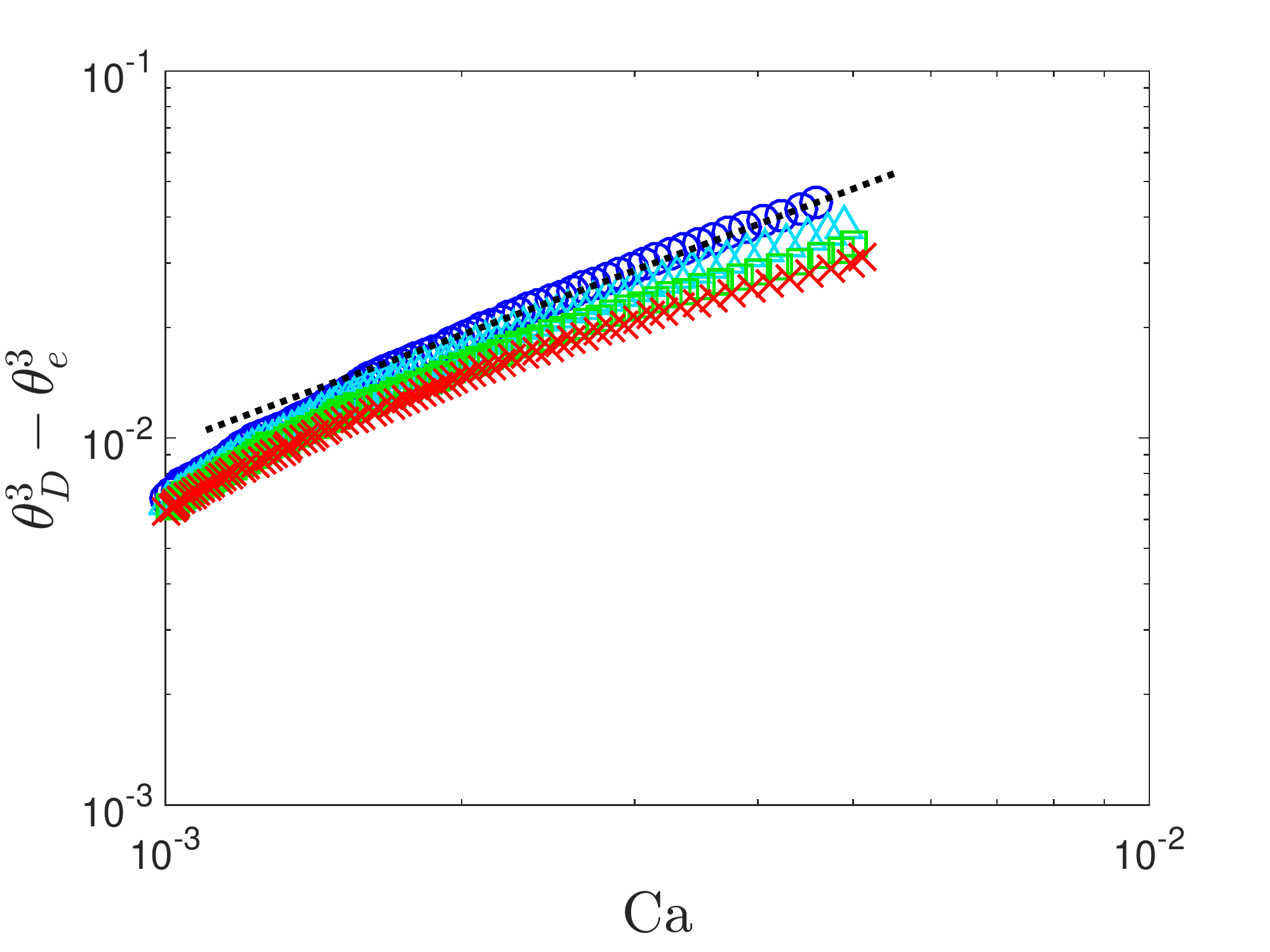}\label{fig:CubesVsCapillary}}
\caption{The spreading of a planar drop on a prewetted substrate with $h_{\star}=0.005$, from an initial angle $\theta_i=30^{\circ}$ to an equilibrium configuration with $\theta_e=15^{\circ}$ for a viscous Newtonian drop, with $\la_1=\la_2=0$ (blue circles), and viscoelastic drops, $\la_1=5$ (cyan triangles), $10$ (green squares), and $15$ (red crosses) when $\la_2=0.01$. \protect\subref{fig:CLPVSTime} Contact line position, $x_{CL}$, versus time. \protect\subref{fig:VelocityVSTime} The velocity of the contact line, $v_{CL}$, versus time. \protect\subref{fig:CubesVSTime} $\theta_D^3 - \theta_e^3$ versus time in a semilogarithmic scale. \protect\subref{fig:CubesVsCapillary} $\theta_D^3 - \theta_e^3$ versus the capillary number $Ca$ in a log-log plot. The dashed black line has a reference slope equal to one.}\label{fig:CAStudy}
\end{figure}

Figure \ref{fig:8} shows the comparison of numerical simulations of a Newtonian drop, with $\la_1=\la_2=0$ (blue dotted curve), versus a viscoelastic drop with $\la_1=15, \la_2=0.01$ (red solid curve), at three selected times ($t = 10,50,100$). Figures \subref*{fig:DropletL1=15,L2=0,01,VSL1=L2=0,hstar=0.01,t3}--\subref*{fig:DropletL1=15,L2=0,01,VSL1=L2=0,hstar=0.01,t21}, where we use $h_{\sta}=0.01$, illustrate the difference in the velocity of the contact line. In figure \subref*{fig:DropletL1=15,L2=0,01,VSL1=L2=0,hstar=0.01,t3}, we note that viscoelasticity influences predominantly the initial stage of the spreading. This behavior can be attributed to viscoelastic effects due to stretching of liquid
around the contact line region in the direction of spreading.
As the spreading velocity decreases, viscoelastic stresses relax in the contact line region, leading to the same spreading speed for both drops. In both cases, the drops relax towards the final
configuration defined by $\theta_e$ (this regime of spreading is not shown in figure \ref{fig:8}).
To shed more light on the origin of the difference in the spreading behavior, we plot in figures \subref*{fig:DropletL1=15,L2=0,01,VSL1=L2=0,hstar=0.005,t3}--\subref*{fig:DropletL1=15,L2=0,01,VSL1=L2=0,hstar=0.005,t21} the simulation results where we use $h_{\sta}=0.005$. The consideration of thinner $h_{\sta}$ is motivated by \cite{SpaidHomsy1994,SpaidHomsy}, where it was shown, within the Oldroyd-B model, that elastic effects influence the behavior of the film ridge more remarkably when $h_{\sta}$ is reduced. The comparison of figure \subref*{fig:DropletL1=15,L2=0,01,VSL1=L2=0,hstar=0.005,t3} and \subref*{fig:DropletL1=15,L2=0,01,VSL1=L2=0,hstar=0.01,t3} shows that the difference between the Newtonian and the viscoelastic drop is indeed more significant for thinner $h_{\sta}$. We attribute this difference to the dynamics of the interface at the contact line region: For the Newtonian drop, the interface of the prewetted film shows an oscillatory ``sagging'' behavior, slowing down the spreading velocity.
When decreasing $h_{\sta}$, this oscillation is more noticeable. For the viscoelastic drop, however, the prewetted film does not exhibit such pronounced oscillatory behavior, as the interface at the contact line region is smoothed by viscoelasticity. A similar oscillatory behavior is demonstrated, both analytically \cite{Rauscher2005} and experimentally \cite{SeemannEtAl}, in the far-field region of dewetting viscoelastic fronts. In particular, Seemann et al.~\cite{SeemannEtAl} also show that viscoelastic effects tend to stabilize the observed undulations, in agreement with our results. In addition, our simulations suggest that viscoelasticity enhances spreading, consistently with findings in \cite{Izbassarov,SpaidHomsy}. We note that, Spaid and Homsy \cite{SpaidHomsy} observe that the viscoelastic fluid interface tends to be stabilized primarily because of changes of transport of momentum in the spreading direction, and finite restoring forces that are present when a viscoelastic fluid is stretched. More recently, Izbassarov and Muradoglu \cite{Izbassarov} consistently demonstrate that the enhancement of the spreading of viscoelastic drops is mainly due to the stretched polymer chains that exert an extensional stress, pushing the contact line, and thus increasing the spreading rate.

We next investigate the influence of the viscoelasticity on the dynamic advancing contact angle. In figure \ref{fig:CAStudy}, we consider three viscoelastic spreading drops with a fixed retardation time, $\la_2 = 0.01$, and three different relaxation times, $\la_1=5$ (cyan triangles), $10$ (green squares), and $15$ (red crosses), and compare them to the Newtonian drop with $\la_1=\la_2=0$ (blue circles). All drops spread on a prewetted substrate with thickness $h_{\star}=0.005$. In figure \subref*{fig:CLPVSTime}, we show the front contact line, $x_{CL}$, determined as the $x$-coordinate of inflection point of $h(x,t)$. Figure \subref*{fig:VelocityVSTime} shows that viscoelastic drops initially move faster than the Newtonian counterpart. Eventually, both viscoelastic and Newtonian drops reach the same speed towards the equilibrium configuration.
We also note that increasing $\la_1$ enhances the velocity of the contact line, $v_{CL}$.
Figure \subref*{fig:CubesVSTime} shows the difference of the cubes of the dynamic and equilibrium contact angles, $\theta_D^3 - \theta_e^3$, versus time in a semilogarithmic plot. As shown, the quantity $\theta_D^3 - \theta_e^3$ is smaller for viscoelastic drops with a higher relaxation time $\la_1$, due to the fact that the viscoelastic drop contact line displaces faster from the initial configuration compared to the Newtonian one, as discussed above. Finally, figure \subref*{fig:CubesVsCapillary} shows $\theta_D^3 - \theta_e^3$ versus the capillary number
$Ca=v_{CL}$ (given our choice for scales), both in logarithmic scales.
The direct proportionality between these quantities is known as the Cox-Voinov law \cite{Cox1986,Voinov}, that we consider in the general form $\theta_D^3 - \theta_e^3 \propto Ca^{\beta}$ (consistently with \cite{Kyle2}). In figure \subref*{fig:CubesVsCapillary}, we show that this law holds for the Newtonian fluid, where the best linear fit of the data (denoted by blue circles) has unit slope (i.e.~$\beta=1$); while lower values of $\beta$ are visible for the viscoelastic counterparts. These findings are consistent with recent experimental
\cite{WeiEtAl} and computational \cite{WangEtAl} results, showing that the viscoelasticity enhances contact line motion,
and that there is a slight variation in the slope of the linear dependence on the capillary number in the Cox-Voinov law, due to viscoelasticity. Furthermore, we have verified (figures not shown for brevity) that different values of the precursor film thickness do not significantly alter the dynamic contact angle, and hence the results of the comparison with the Cox-Voinov law.

\begin{figure}[H!t1]
\captionsetup{type=figure}
\centering
\subfloat[]{\includegraphics[scale=0.265,valign=t,trim=0.07in 0in 0.35in 0in,clip=true]{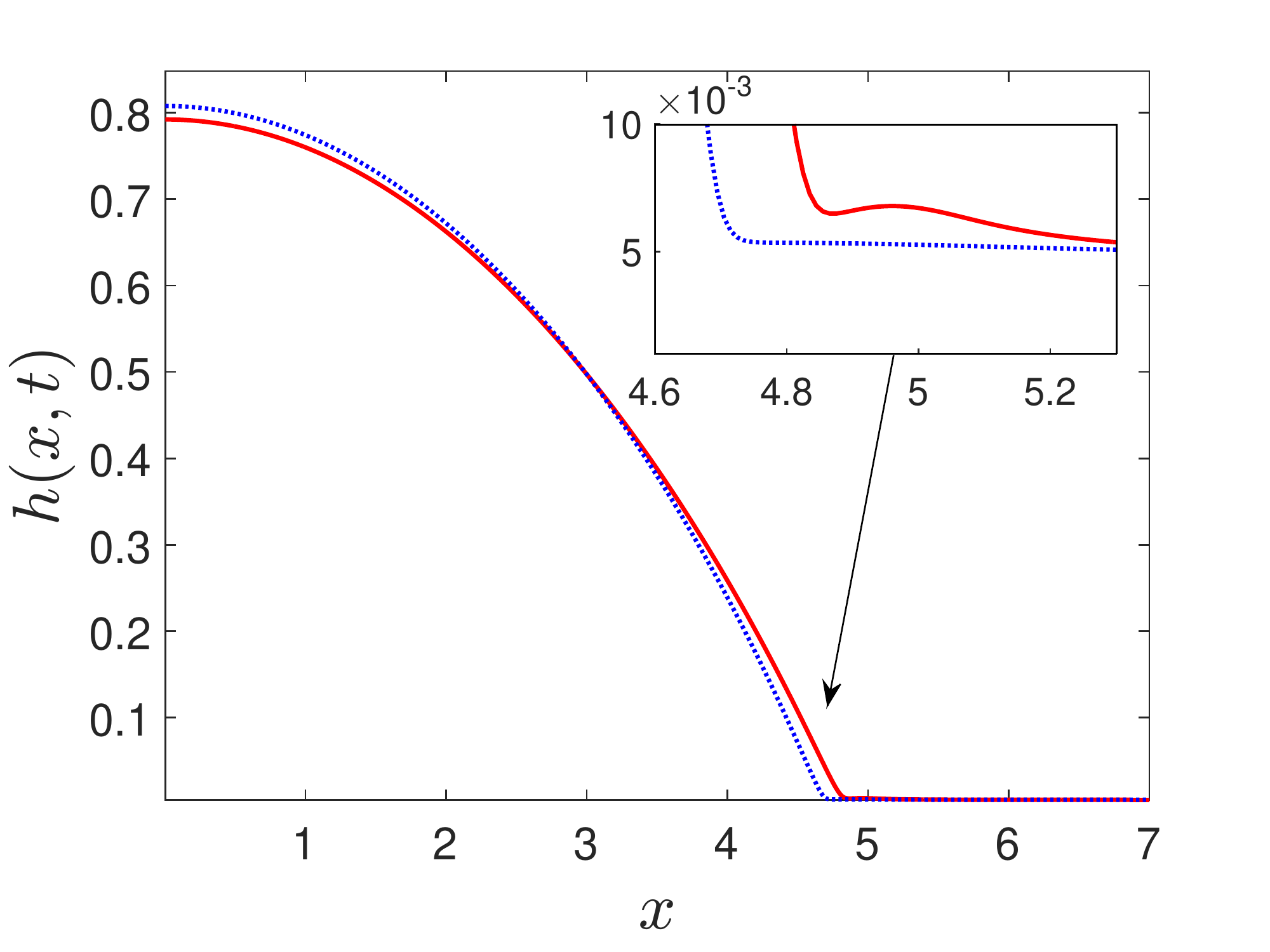}\label{fig:RetractingDropletL1=15,L2=0,01,VSL1=L2=0,hstar=0.005,t3}}
\subfloat[]{\includegraphics[scale=0.265,valign=t,trim=0.07in 0in 0.35in 0in,clip=true]{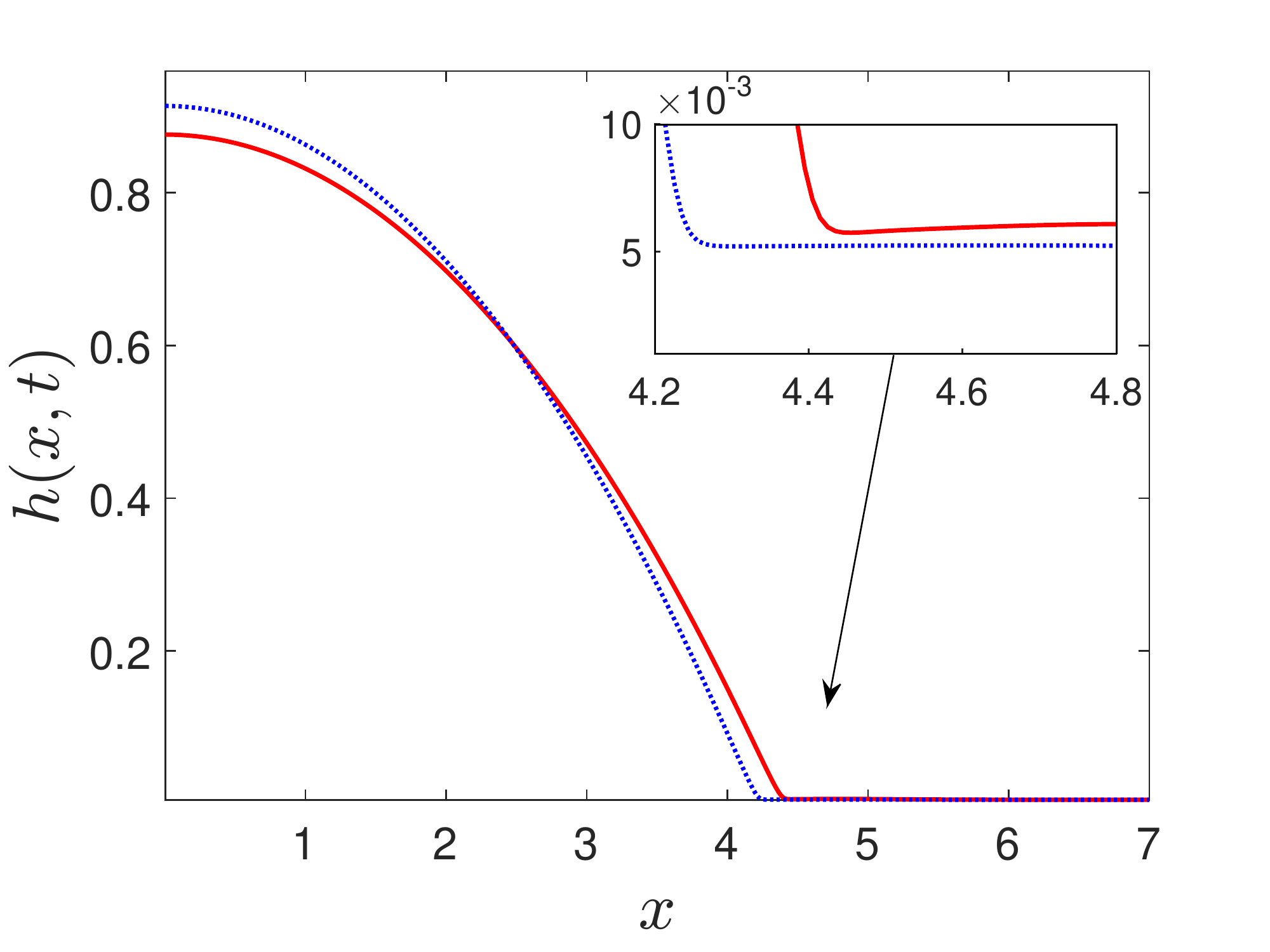}\label{fig:RetractingDropletL1=15,L2=0,01,VSL1=L2=0,hstar=0.005,t8}}
\subfloat[]{\includegraphics[scale=0.265,valign=t,trim=0.07in 0in 0.35in 0in,clip=true]{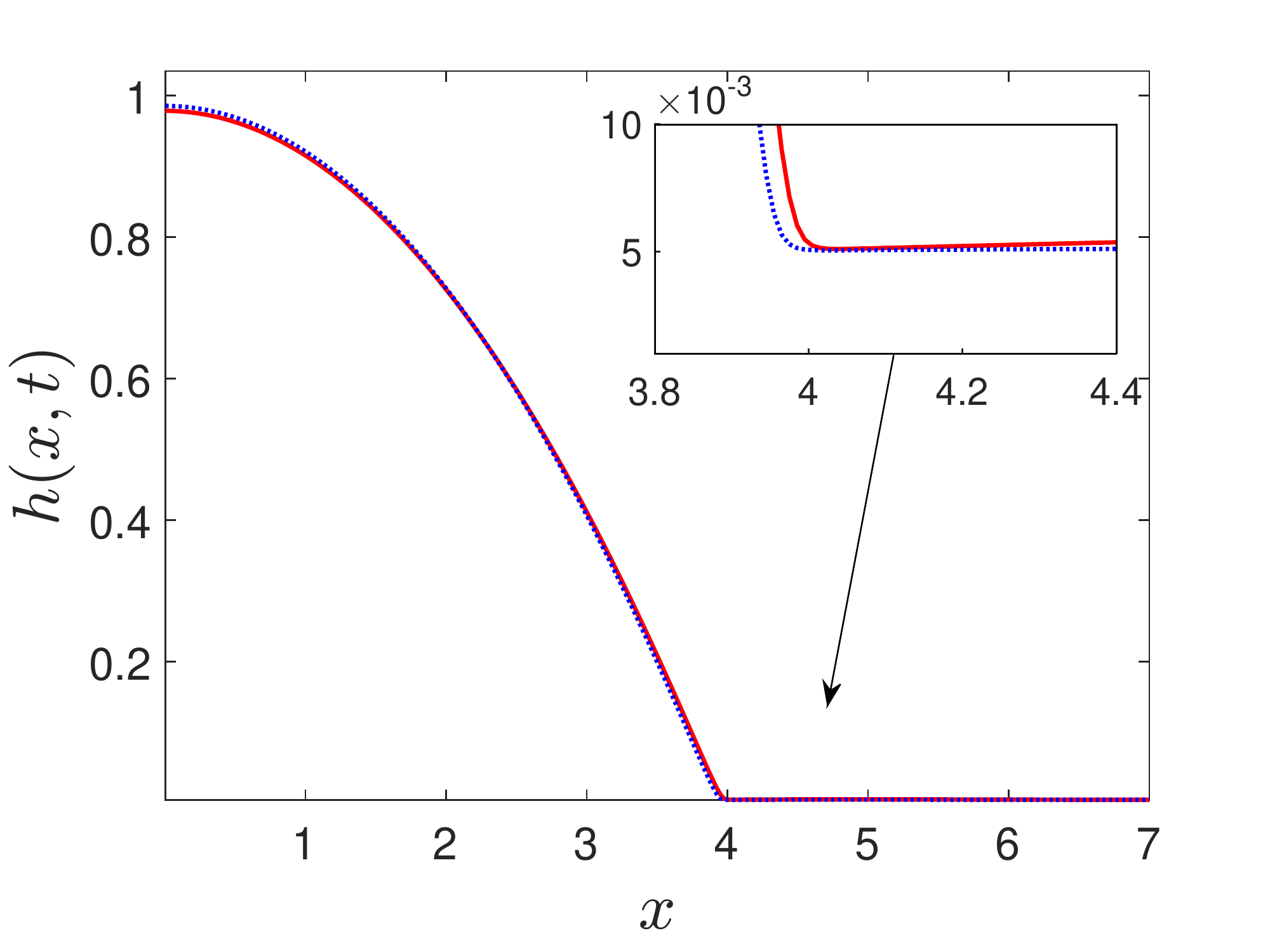}\label{fig:RetractingDropletL1=15,L2=0,01,VSL1=L2=0,hstar=0.005,t21}}
\caption{The retraction of a viscoelastic drop with $\la_1=15,\la_2=0.01$ (red solid curve) versus a Newtonian drop $\la_1=\la_2=0$ (blue dotted curve), at $t=100,350,1000$ from left to right; equilibrium film thickness $h_{\sta}=0.005$.}\label{fig:RetractingL1=10,L2=0,01}
\end{figure}

\subsubsection[Receding drops]{Receding drops}

\begin{figure}[H!t1]
\captionsetup{type=figure}
\centering
\subfloat[]{\includegraphics[scale=0.35,valign=t,trim=0.15in 0in 0.3in 0in,clip=true]{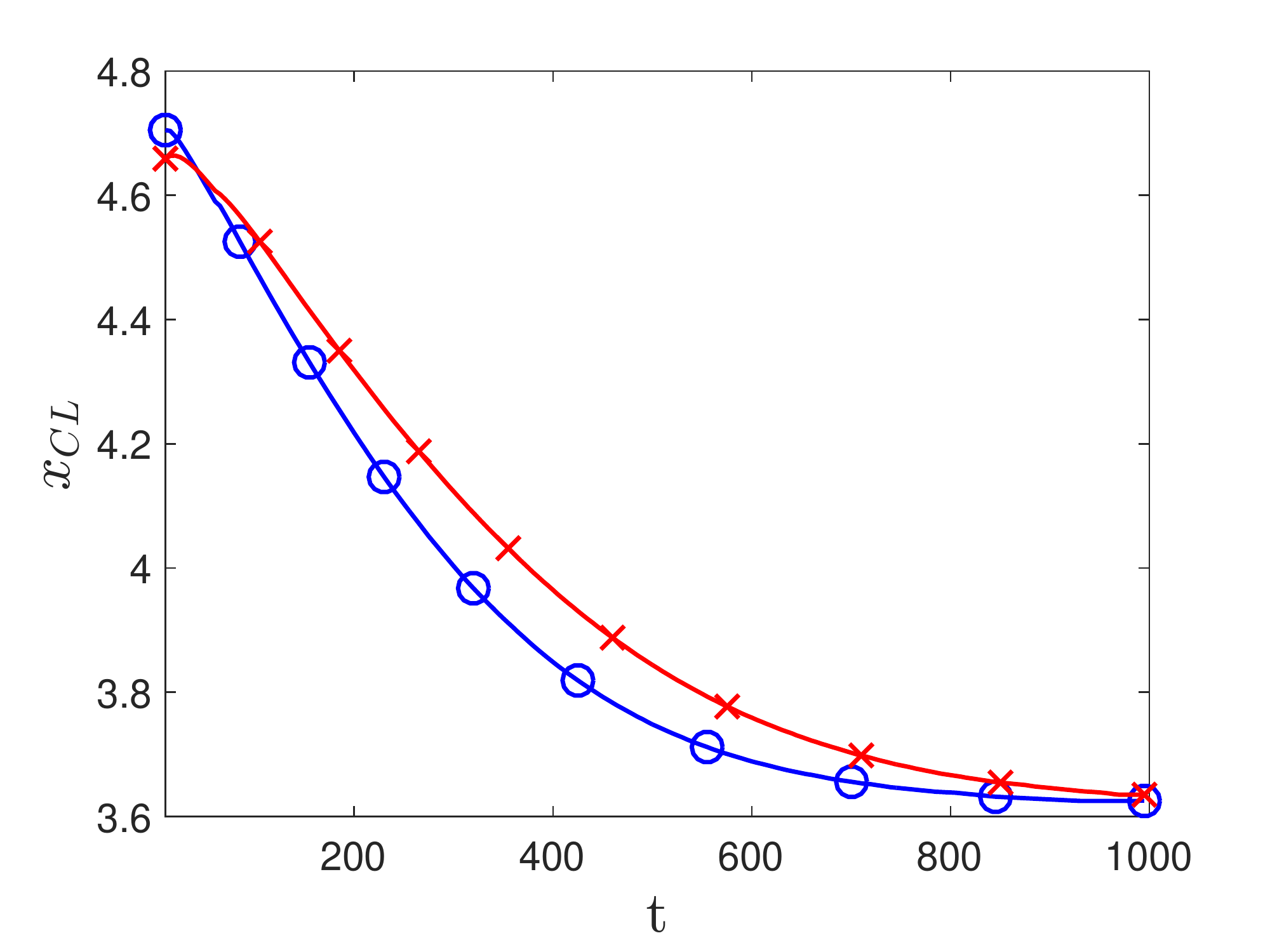}\label{fig:CLPVSTimeRetracting}}
\subfloat[]{\includegraphics[scale=0.35,valign=t,trim=0.15in 0in 0.3in 0in,clip=true]{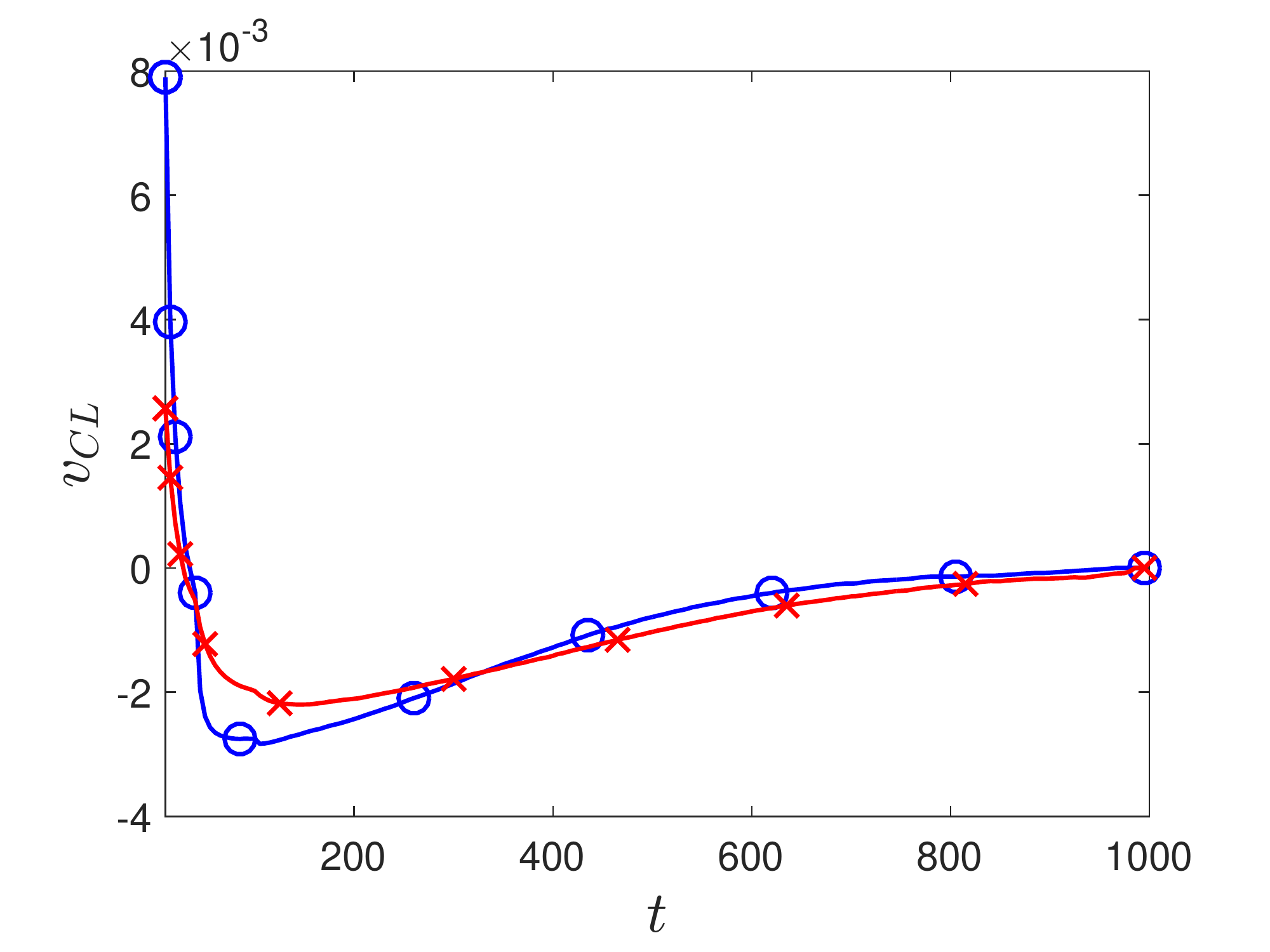}\label{fig:VelocityVSTimeRetracting}}\\
\subfloat[]{\includegraphics[scale=0.35,valign=t,trim=0.0in 0in 0.3in 0in,clip=true]{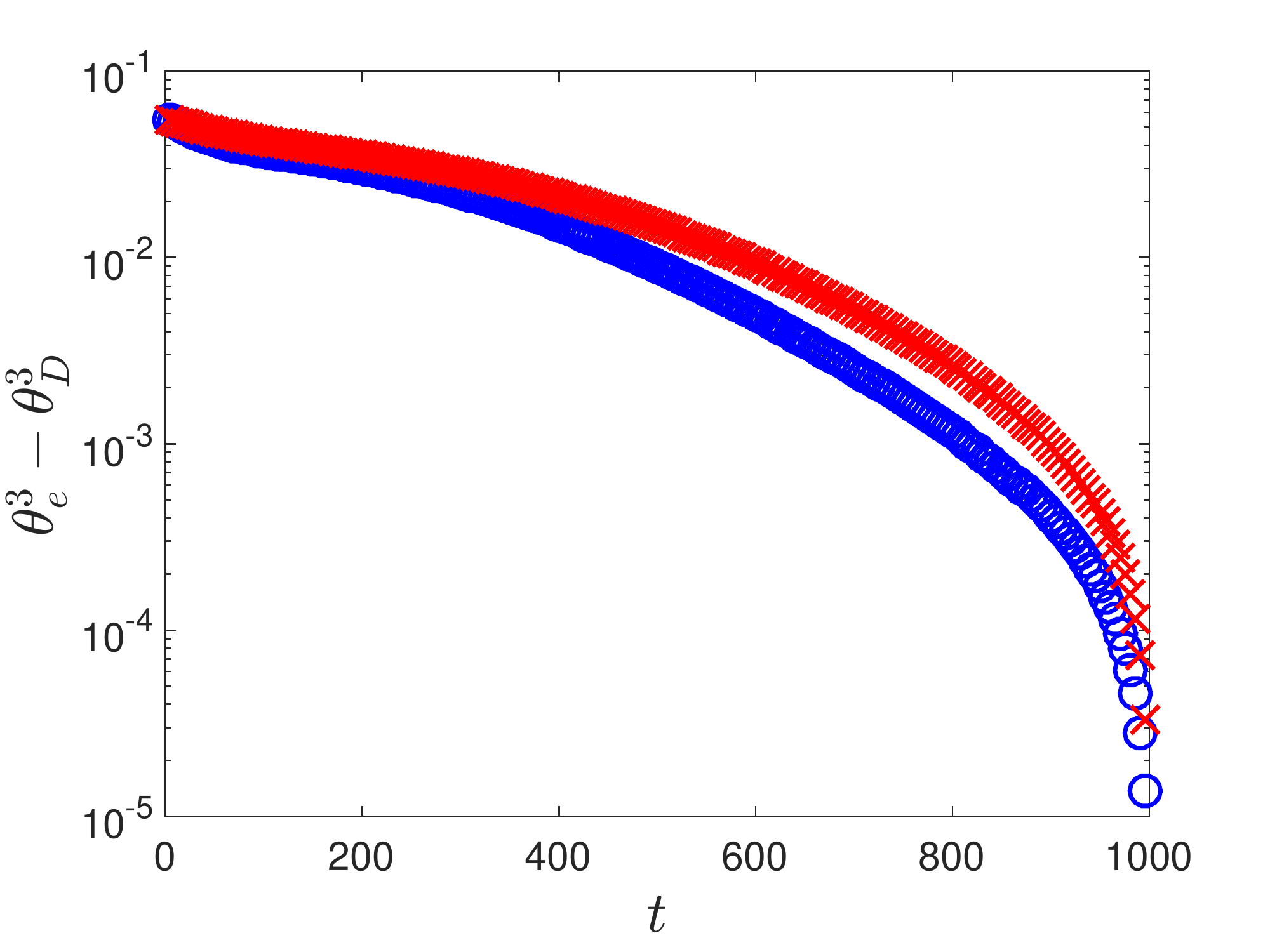}\label{fig:CubesVSTimeRetracting}}
\subfloat[]{\includegraphics[scale=0.35,valign=t,trim=0.0in 0in 0.3in 0in,clip=true]{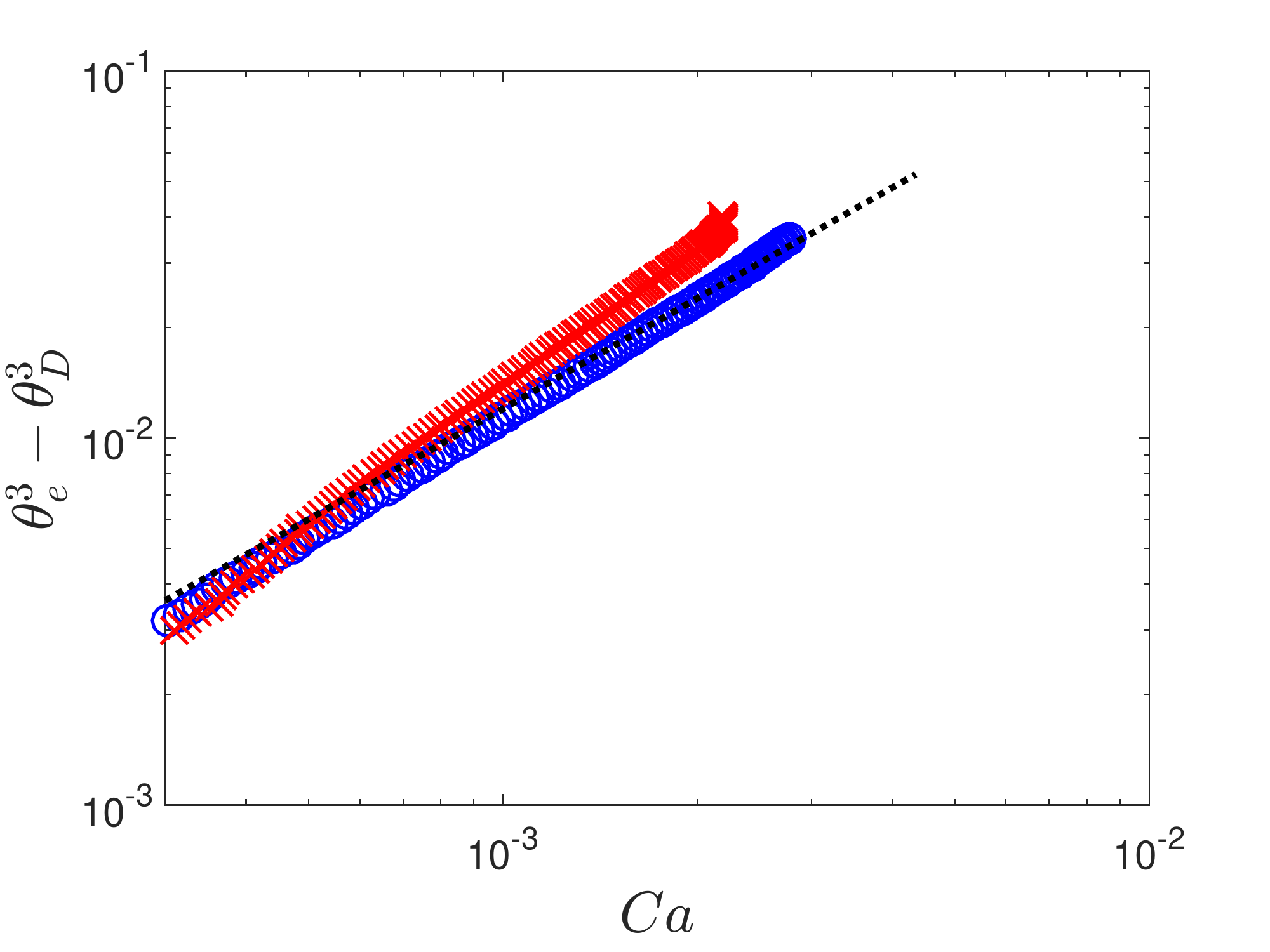}\label{fig:CubesVsCapillaryRetracting}}
\caption{The retraction of a planar drop on a prewetted substrate with $h_{\star}=0.005$, from an initial angle $\theta_i=15^{\circ}$ to an equilibrium configuration with $\theta_e=30^{\circ}$ for a viscoelastic drop with $\la_1=15,\la_2=0.01$ (red crosses), versus a Newtonian drop with $\la_1=\la_2=0$ (blue circles). \protect\subref{fig:CLPVSTimeRetracting} The contact line position, $x_{CL}$, versus time. \protect\subref{fig:VelocityVSTimeRetracting} The speed of the contact line, $v_{CL}$, versus time. \protect\subref{fig:CubesVSTimeRetracting} $\theta_e^3 - \theta_D^3$ versus time. \protect\subref{fig:CubesVsCapillaryRetracting} $\theta_e^3 - \theta_D^3$ versus the capillary number $Ca$. The dashed black line has a reference slope equal to one.}\label{fig:CAStudyRetracting}
\end{figure}

Similarly to the spreading case, we carry out investigations of the dynamic contact angle for receding drops. We use the same geometrical framework as the one for a spreading drop, but we impose $\theta_i=15^{\circ}$ and $\theta_e=30^{\circ}$. Figures \subref*{fig:RetractingDropletL1=15,L2=0,01,VSL1=L2=0,hstar=0.005,t3}--\protect\subref{fig:RetractingDropletL1=15,L2=0,01,VSL1=L2=0,hstar=0.005,t21} show the comparison of the evolution of two retracting drops at three selected times: a Newtonian drop with $\la_1=\la_2=0$ (blue dotted curve), and a viscoelastic one with $\la_1=15,\la_2=0.01$ (red solid curve); for both simulations, we impose an equilibrium film thickness $h_{\sta}=0.005$. Again, the two drops exhibit discrepancy in their evolution. In figure \subref*{fig:RetractingDropletL1=15,L2=0,01,VSL1=L2=0,hstar=0.005,t3}, at time $t=100$, we see that the Newtonian drop has receded more than the viscoelastic one. This happens because, contrary to the spreading problem, the viscoelastic fluid interface initially shows more bending (visible in the inset of the figure \subref*{fig:RetractingDropletL1=15,L2=0,01,VSL1=L2=0,hstar=0.005,t3}), indicating that there is resistance to the force that drives the retraction of the drop. In figure \subref*{fig:RetractingDropletL1=15,L2=0,01,VSL1=L2=0,hstar=0.005,t8}, at time $t=350$, this oscillation in the viscoelastic interface is flattened, and this allows for faster motion of the contact line for viscoelastic drop. As in the spreading case, eventually, both drops reach the same speed, and attain the same final configuration at time $t=1000$, shown in figure \subref*{fig:RetractingDropletL1=15,L2=0,01,VSL1=L2=0,hstar=0.005,t21}. The slower retraction of a viscoelastic drop is also investigated in \cite{BartoloEtAl}, where it is demonstrated that this behavior is due to the elastic effects near the moving contact line.

Finally, we present the results of the dynamic contact angle for the receding drops. Figure \subref*{fig:CLPVSTimeRetracting} shows that the Newtonian drop with $\la_1=\la_2=0$ (denoted by blue circles) recedes faster than the viscoelastic drop with $\la_1=15,\la_2=0.01$ (denoted by red crosses). Figure \subref*{fig:VelocityVSTimeRetracting} shows the retraction velocities of the two drops as a function of time. The Newtonian drop initially recedes faster than the viscoelastic one, and eventually they reach the same speed towards the final configuration. Figure \subref*{fig:CubesVSTimeRetracting} shows $\theta_e^3- \theta_D^3$ versus time, in a semilogarithmic plot. The quantity $\theta_e^3- \theta_D^3$ is higher for the viscoelastic drop, since it retracts slower. Figure \subref*{fig:CubesVsCapillaryRetracting} shows $\theta_e^3 - \theta_D^3$ versus $Ca$, both in logarithmic scales. Differently from the spreading case, the best linear fit of the receding viscoelastic data has a slope higher than the one corresponding to the Newtonian data ($\beta=1$).

\section[Conclusions]{Conclusions}\label{Sec4}

We numerically solve the nonlinear equation that governs the fluid interface of a dewetting thin film of viscoelastic fluid on a solid substrate. The governing equation is obtained as the long-wave approximation of the Navier-Stokes equations with Jeffreys type constitutive equation to describe the non-Newtonian nature of the viscoelastic fluid. The van der Waals interaction force drives the instabilities of the liquid interface and causes the film to break up, forming holes bounded by retracting rims. We investigate how physical parameters involved, such as the relaxation and retardation characteristic times of the viscoelastic fluid, and the slip coefficient, affect the dynamics and the final configuration of the fluid. In the linear regime, our results are in agreement with the linear stability analysis. Consistently with previous studies, we find that viscoelastic parameters and the slippage coefficient do not influence either the wavenumber corresponding to the maximum growth rate or the critical one, but influence the maximum growth rate. In particular, an increase of the relaxation time, $\la_1$, or the slip length, $b$, leads to an increase of the maximum growth rate. Conversely, increasing the retardation time, $\la_2$, leads to a decrease of maximum growth rate.

The simulations of the dewetting of thin viscoelastic films in the nonlinear regime reveal novel complex morphologies that depend on the viscoelasticity. The results show that for small values of $\la_1$, a single secondary droplet can be formed, while for large values, multiple secondary droplets can emerge. We note that the emergence of these small length scales can be related to the relaxation time $\la_1$. Here, we not only provide, for the first time, a study of these developing length scales, but also report on the migration and merging of the secondary droplets due to viscoelastic effects. Simulation results also show that the inclusion of $\la_2$ provides a numerical advantage by stabilizing the computations at high values of $\la_1$. In addition, the influence of the slip coefficient on the dynamics and final configurations is also addressed. Future work shall consider extension of these results to three spatial dimensions.

In the final part of this work, we investigate the dynamic contact angle of viscoelastic drops. Our numerical simulations show that the viscoelastic advancing front moves faster at early times, and that eventually it behaves as the Newtonian counterpart for large times. Our simulations suggest that the enhancement of the speed of the viscoelastic spreading drop is due to the smoothness of its interface in the prewetted region. The analysis of the dynamic contact angle also allows us to verify the Cox-Voinov law for the viscous Newtonian case; while we show small deviations from this law for viscoelastic drops. For receding viscoelastic drops, we show that the speed of the contact line is instead decreased, when compared to a Newtonian one. Again, we explain this behavior by the viscoelastic force at the contact line region resisting the receding motion of the contact line. Although our study is limited to the Jeffreys linear viscoelastic model, we hope that it will serve as a basis for further analysis of other viscoelastic models.

\section*{Acknowledgements}
This work was partially supported by the NSF grants No. DMS-$1320037$ (S. A.), No. CBET-$995904$ (L. K.), and No. CBET-$1604351$ (L. K., S. A.).

\numberwithin{equation}{section}
\section*{Appendix A: Derivation}
\renewcommand{\theequation}{A\arabic{equation}}
\setcounter{equation}{0}  

\noindent In this appendix, we outline the derivation of the long-wave formulation, equations (\ref{GovEq2D}) and (\ref{Eq:Q&R}), for the evolution of the interface of a thin viscoelastic film. The system of equations (\ref{GovMomentum}) and (\ref{Incompressability}) is subject to boundary conditions at the free surface, represented parametrically by the function $f(x,y,t)=y-h(x,t)=0$, and boundary conditions at the $x$-axis ($y=0$), as introduced in \S~\ref{Sec1}, and illustrated in figure \ref{fig:Setup}. The stress balance at the interface (where the top fluid is passive, as in our study) is expressed as
\begin{equation}\label{Eq:LaplacePressBC}
\left( \tau -(p+\Pi) I \right) \cdot \mathbf{n} =\sigma \kappa  \mathbf{n} \, ,
\end{equation}

\noindent where $I$ denotes the identity matrix. In the absence of motion, this condition describes the jump in the pressure across the interface with outward unit normal $\mathbf{n}$, and a local curvature $\kappa= - \nabla \cdot \mathbf{n}$, due to the surface tension $\sigma$. We define the two mutually orthogonal vectors $\mathbf{n}$, $\mathbf{t}$ as
\begin{equation}\label{nt}
\mathbf{n}=\frac{1}{{\left(h_x^2 +1\right)}^{1/2}}\left(-h_x, 1 \right), \qquad \,\mathbf{t}=\frac{1}{{\left(h_x^2 +1\right)}^{1/2}}\left(1, h_x \right) \, .
\end{equation}

\noindent The kinematic boundary condition is given by $f_t + \textbf{u} \cdot \nabla f = 0$, where substituting $f(x,y,t)=y-h(x,t)$ gives
\begin{equation}\label{ht}
\frac{\partial h}{\partial t} (x,t) = - \frac{\partial}{\partial x} \int_{0}^{h(x,t)} u(x,y) dy \, .
\end{equation}

\noindent The boundary conditions at the solid substrate are described by the non-penetration condition for the normal component of the velocity and the Navier slip boundary condition for the tangential one
\begin{equation}\label{Eq:BottomBC}
v=0, \quad u= \frac{b}{\eta}\tau_{12} \, ,
\end{equation}

\noindent respectively, where $b \geq 0$ is the slip length ($b=0$ implies no-slip).

The equilibrium contact angle, $\theta_e$, formed between the fluid interface $y=h(x,t)$ and the solid substrate can be included directly in the disjoining pressure term $\Pi(h)$, leading to
\begin{equation}
\label{Def:VdW}
\Pi(h)= \frac{\sigma(1 - \cos \theta_e)}{M h_{\sta}} \left[ {\left( \frac{h_{\star}}{h}  \right)}^{m_1} - {\left( \frac{h_{\star}}{h}\right)}^{m_2} \right] \, ,
\end{equation}

\noindent with $M = (m_1-m_2)/[(m_2-1)(m_1-1)]$, where $m_1$ and $m_2$ are constants such that $m_1>m_2>1$, (in this work we chose $m_1=3$ and $m_2=2$, as also widely used in the literature, for instance by the authors in \cite{Teletzke,DiezKondic2007,Ivana}, but other values can be modeled as verified in \cite{Kyle}).

Next, we nondimensionalize using commonly implemented scaling appropriate to long-wave formulation
\begin{align}
&x=Lx^* \, , \; (y,h,h_{\sta},b)=H(y^*,h^*,h_{\sta}^*,b^*) \, , \; (p,\Pi)= P(p^*,\Pi^*) \, , \label{NonDim1}\\
&u=Uu^* \, , \; v=\varepsilon U v^* \, , \; (t,\la_1,\la_2)=T( t^*,\la_1^*,\la_2^*) \, , \; \sigma = \frac{U \eta}{\vare^3} \sigma^{*}\, , \label{NonDim2}
\end{align}
\begin{equation}\label{NonDimTensor}
\left(
\begin{array}{cc}
\tau_{11}&\tau_{12}\\
\tau_{21}&\tau_{22}\\
\end{array}
\right)
=\frac{\eta}{T}
\left(
\begin{array}{cc}
\tau_{11}^*&\frac{\tau_{12}^*}{\vare}\\
\frac{\tau_{21}^*}{\vare}&\tau_{22}^*\\
\end{array}
\right) \, ,
\end{equation}

\noindent where ${H}/{L}= \varepsilon \ll 1$ is the small parameter. To balance pressure, viscous and capillary forces, the pressure is scaled with $P= {\eta}/({T \varepsilon^2})$, and the time with $T = L/ U$. Following the derivation in \cite{Rauscher2005}, we can scale the surface tension $\sigma^*$ to one by an \textit{ad hoc} choice of the length scale. We note that the Weissenberg number, $Wi$, given the choice for scales, is $Wi= \la_1 U / L= \la_1/T=\la_1^*$. To avoid cumbersome notation, we drop the superscript `$^*$' and we consider from now on all quantities to be dimensionless.

The incompressibility condition, equation (\ref{Incompressability}), is invariant under rescalings, while the dimensionless form of equation (\ref{GovMomentum}) for the $x$ and $y$ component is, respectively,
\begin{subequations}\label{Eq:x-y-compMomentum}
\begin{align}
\varepsilon^2 Re \frac{du}{dt}&= \varepsilon^2 \frac{\partial \tau_{11}}{\partial x} +\frac{\partial \tau_{21}}{\partial y} - \frac{\partial p}{\partial x} \, , \label{Eq:x-compMomentum}\\
\varepsilon^4 Re \frac{dv}{dt}&= \varepsilon^2 \left( \frac{\partial \tau_{12}}{\partial x}  + \frac{\partial \tau_{22}}{\partial y}
 \right) - \frac{\partial p}{\partial y} \, , \label{Eq:y-compMomentum}
\end{align}
\end{subequations}

\noindent where $Re= {\rho U L}/{\eta}$ is the Reynolds number, assumed to be $O(1 / \varepsilon)$ or smaller. The dimensionless components of the stress tensor given by the Jeffreys model, equation (\ref{Jeffreys}), satisfy
\begin{subequations}\label{nondimJeffrey}
\begin{align}
\tau_{11} + \lambda_1 \frac{\partial \tau_{11}}{\partial t} &= 2 \left( \frac{\partial u}{\partial x} + \lambda_2 \frac{\partial }{\partial t} \left(\frac{\partial u }{\partial x}\right) \right) \, , \\
\tau_{22} + \lambda_1 \frac{\partial \tau_{22}}{\partial t} &= 2 \left( \frac{\partial v}{\partial y} + \lambda_2 \frac{\partial }{\partial t} \left(\frac{\partial v }{\partial y}\right) \right) \, , \\
\tau_{12} + \lambda_1 \frac{\partial \tau_{12}}{\partial t} &= \frac{\partial u}{ \partial y} + \lambda_2 \frac{\partial}{\partial t}\left( \frac{\partial u}{\partial y } \right) + \varepsilon^2 \left( \frac{\partial v}{ \partial x} + \lambda_2 \frac{\partial}{\partial t}\left( \frac{\partial v}{\partial x } \right)\right)\, . \label{nondimJeffrey3}
\end{align}
\end{subequations}

\noindent The kinematic boundary condition (\ref{ht}) is invariant under rescaling, while the non-penetration condition and the Navier slip boundary condition for the velocity components parallel to the substrate (\ref{Eq:BottomBC}) in dimensionless form are
\begin{equation}\label{Eq:DimlessBottomBCs}
v=0 \, ,\quad u= b \tau_{12} \, ,
\end{equation}

\noindent where in the weak slip regime $b=O(1)$. The leading-order terms in the governing equations (\ref{Eq:x-compMomentum}) and (\ref{Eq:y-compMomentum}) respectively, are
\begin{subequations}\label{Eq:Rescaled-x-y-compMomentum}
\begin{align}
\frac{\partial \tau_{21}}{\partial y} &= \frac{\partial p}{\partial x} \, , \label{Eq:Rescaled-x-compMomentum} \\
\frac{\partial p}{\partial y} &= 0 \, . \label{Eq:Rescaled-y-compMomentum}
\end{align}
\end{subequations}

\noindent The leading-order terms of the normal and tangential components of the stress balance at the free surface (\ref{Eq:LaplacePressBC})
are
\begin{align}
p &= -\frac{\partial^2 h }{\partial x^2}  - \Pi(h) \, , \label{Def:pR}\\
h_x \tau_{12} &= 0 \, , \label{Def:36}
\end{align}

\noindent where the form of $\Pi(h)$ in (\ref{Def:pR}) is given by (\ref{Def:VdW}), with all the quantities considered nondimensional. Considering in general $h_x \neq 0$, from (\ref{Def:36}) follows that $\tau_{12} = 0$, at the interface. Now, integrating (\ref{Eq:Rescaled-x-compMomentum}) with respect to $y$, from $y$ to $h(x,t)$, we obtain $\tau_{21}= (y -h) {p}_x $. Noting that the stress tensor is symmetric, and substituting $\tau_{21}$ into (\ref{nondimJeffrey3}), we obtain (up to the leading-order)

\begin{equation}\label{Eq:40}
\frac{\partial p}{\partial x} (y-h) + \lambda_1 \frac{\partial}{\partial t}\left( \frac{\partial p}{\partial x} (y-h) \right) = \frac{\partial u}{\partial y} + \lambda_2 \frac{\partial}{\partial t }\left( \frac{\partial u}{\partial y}\right) \, .
\end{equation}

\noindent Integrating (\ref{Eq:40}) from $0$ to $y=h(x,t)$ and using the corresponding boundary conditions at the substrate $y=0$, we obtain
\begin{equation}\label{Eq:41}
\left( 1 + \lambda_2 \frac{\partial }{\partial t} \right) \left( u + b h \frac{\partial p}{\partial x} \right) = \left( 1 + \lambda_1 \frac{\partial}{\partial t} \right) \left( \left( \frac{y^2}{2} - yh \right) \frac{\partial p}{\partial x} \right) \, .
\end{equation}

\noindent Integrating (\ref{Eq:41}) again from $y=0$ to $y=h(x,t)$ gives
\begin{align}\label{Eq:42}
& \int_{0}^{h(x,t)}u dy + b h^2 \frac{\partial p}{\partial x} + \lambda_{2} \frac{\partial}{\partial t} \int_{0}^{h(x,t)}u dy + \lambda_2 b \frac{\partial h^2}{\partial t} \frac{\partial p}{ \partial x} - \lambda_2 h_t u(y=h(x,t))- \lambda_2 h_t b h \frac{\partial p}{\partial x} = \nonumber \\
& \qquad - \frac{h^3}{3} \frac{\partial p}{\partial x} -\lambda_1 h^2 h_t \frac{\partial p}{\partial x} + \lambda_1 \frac{h^2}{2} h_t \frac{\partial p}{\partial x} \, .
\end{align}

\noindent Taking the spatial derivative of the latter equation and substituting it into the kinematic boundary condition (\ref{ht}), we obtain a long-wave approximation in terms of $u$ and $h(x,t)$
\begin{align}\label{Eq:43}
& h_t + \lambda_2 \left[ h_{tt} + \frac{\partial}{\partial x} ({{u}} (y=h(x,t)) h_t ) \right] = \frac{\partial}{\partial x}\left[ (1 + \lambda_1 \partial_t) \left( \frac{h^3}{3} \frac{\partial p}{\partial x} \right) + \right. \nonumber \\
& \left. + \ (1 + \lambda_2 \partial_t ) \left(bh^2 \frac{\partial p}{\partial x}\right) \right] - \frac{\partial}{\partial x} \left[ \left( \lambda_1 \frac{h^2}{2}\frac{\partial p}{\partial x} + \lambda_2 b h \frac{\partial p}{\partial x} \right) h_t \right] \, .
\end{align}

\noindent To write this in a closed form relation for $h(x,t)$, we note that equation (\ref{Eq:41}) can be written in a more compact form as a linear ordinary differential equation
\begin{equation}\label{Eq:44}
u + \la_2 \frac{\partial u}{\partial t} = - (1 + \la_2 \partial_t) bh \frac{\partial p}{\partial x} + (1 + \la_1 \partial_t) \left[ \left( \frac{y^2}{2} - hy \right) \frac{\partial p}{\partial x} \right] \, .
\end{equation}

\noindent One can simply solve this linear differential equation, obtaining
\begin{equation}\label{Eq:46}
u = \frac{1}{\la_2} \int_{- \infty}^{t} e^{- \frac{t - {t'}}{\la_2}} \widehat{f}(x,y,{t'}) d {t'}\, ,
\end{equation}

\noindent with $\widehat{f}$ equal to the right-hand side of equation (\ref{Eq:44}). Integration by parts can be performed to recast (\ref{Eq:46}) at $y=h(x,t)$, and one finally obtains equations (\ref{GovEq2D}) and (\ref{Eq:Q&R}).

\numberwithin{equation}{section}
\section*{Appendix B: Numerical Discretization}
\renewcommand{\theequation}{A\arabic{equation}}
\setcounter{equation}{0}  

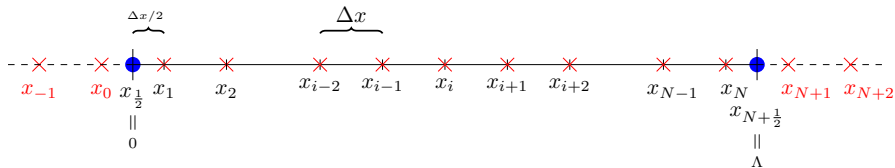
\begin{figure}\centering
\resizebox{0.8\textwidth}{!}{
\begin{tikzpicture}
\tikzstyle{ground}=[fill,pattern=north east lines,draw=none,minimum width=0.3,minimum height=0.6]

\draw (-2,0) node[cross=3.5pt,red]{};
\draw[red] (-2,-0.2) node[anchor=north]{$x_{-1}$};
\draw (-1,0) node[cross=3.5pt,red]{};
\draw[red] (-1,-0.2) node[anchor=north]{$x_{0}$};
\fill[blue] (-1/2,0) circle(3.5pt);
\fill[blue] (9.5,0) circle(3.5pt);
\draw[dashed,black] (-5/2,0) -- (-0.5,0);
\draw[black] (-0.5,0) -- (9.5,0) ;
\draw (-1/2,1/4) -- (-1/2,-1/4) node[black,anchor=north] {$\equalto{x_{\frac{1}{2}}}{ 0}$};

\draw (0,0) node[cross=3.5pt,red]{};
\draw (0,-0.2) node[black,anchor=north]{$x_1$};
\draw (0,1/12) -- (0,-1/12)node[black,anchor=north] {};
\draw (1,0) node[cross=3.5pt,red]{};
\draw (1,-0.2) node[black,anchor=north]{$x_2$};
\draw (1,1/12) -- (1,-1/12)node[black,anchor=north] {};

\draw [black,
    thick,
    decoration={
        brace,
        raise=0.5cm
    },
    decorate
] (2.5,0) -- (3.5,0)
node [black,pos=0.5,anchor=north,yshift=1.05cm] {$\Delta x$};

\draw [black,
    thick,
    decoration={
        brace,
        raise=0.5cm
    },
    decorate
] (-0.5,0) -- (0,0)
node [black,pos=0.4,anchor=north,yshift=1cm] {\tiny{$\Delta x /2 $}};

\draw (2.5,1/12) -- (2.5,-1/12)node[black,anchor=north] {};
\draw (2.5,-0.1) node[black,anchor=north]{$x_{i-2}$};
\draw (2.5,0) node[cross=3.5pt,red]{};

\draw (3.5,1/12) -- (3.5,-1/12)node[black,anchor=north] {};
\draw (3.5,-0.1) node[black,anchor=north]{$x_{i-1}$};
\draw (3.5,0) node[cross=3.5pt,red]{};

\draw (4.5,1/12) -- (4.5,-1/12)node[black,anchor=north] {};
\draw (4.5,-0.1) node[black,anchor=north]{$x_{i}$};
\draw (4.5,0) node[cross=3.5pt,red]{};

\draw (5.5,1/12) -- (5.5,-1/12)node[black,anchor=north] {};
\draw (5.5,-0.1) node[black,anchor=north]{$x_{i+1}$};
\draw (5.5,0) node[cross=3.5pt,red]{};

\draw (6.5,1/12) -- (6.5,-1/12)node[black,anchor=north] {};
\draw (6.5,-0.1) node[black,anchor=north]{$x_{i+2}$};
\draw (6.5,0) node[cross=3.5pt,red]{};

\draw (8,1/12) -- (8,-1/12)node[black,anchor=north,yshift=-0.3cm] {};
\draw (8,0) node[cross=3.5pt,red]{};
\draw (8.15,-0.2) node[black,anchor=north]{$x_{N-1}$};
\draw (9,1/12) -- (9,-1/12)node[black,anchor=north] {};
\draw (9,0) node[cross=3.5pt,red]{};
\draw (9.15,-0.2) node[black,anchor=north]{$x_{N}$};

\draw (9.5,1/4) -- (9.5,-1/4)node[black,anchor=north,yshift=-0.3cm] {$\equalto{x_{N+\frac{1}{2}}}{ \Lambda}$};

\draw[black,dashed] (9.5,0) -- (11.5,0);
\draw (10,0) node[cross=3.5pt,red]{};
\draw[red] (10.3,-0.2) node[anchor=north]{$x_{N+1}$};
\draw (11,0) node[cross=3.5pt,red]{};
\draw[red] (11.3,-0.2) node[anchor=north]{$x_{N+2}$};
\end{tikzpicture}
}
\caption{Discretization of the spatial domain.}
\label{fig:NumericalDiscretization}
\end{figure}

The spatial domain $[0, \Lambda]$ is discretized by uniformly spaced grid points, that constitute a vertex-centered grid (see figure \ref{fig:NumericalDiscretization}). Following the natural order from left to right, adjacent vertices are associated to the indices $i-1,i,i+1$, respectively. Thus, we let $x_{i}=x_0+i \Delta x ,\; i=1,2, \ldots, \, N$ (where $N=\Lambda / \Delta x$, and $\Delta x$ is the grid size), so that the endpoints of the physical domain, $0$ and $\Lambda$, correspond to the $x_{1}- \frac{\Delta x}{2}$ and $x_{N}+\frac{\Delta x}{2}$ cell-centers, respectively. Similarly, we discretize the time domain and denote by $h_i^n$ the approximation to the solution at the point ($x_i, n\Delta t$), where $n=0,1, \ldots$ indicates the number of time steps, and $\Delta t$ is the temporal step size.

In order to approximate the fourth order spatial derivative in equation (\ref{GovEq2D}), we need at least a $5$-point stencil to obtain second order accuracy. We define the first and third derivatives at the cell-centers so that the second and fourth order derivatives are centered at the grid points (see \cite{Kondic2003} for a detailed description).

We recall that the class of $\theta$-schemes for the finite difference discretization of the time derivative, can be written as
\begin{equation}\label{Eq:ThetaScheme}
\frac{h_i^{n+1} -h_i^n}{\triangle t} = -\left[\theta F_i^n + (1- \theta )F_i^{n+1}\right] , \; \quad i=1,2, \ldots, \, N,
\end{equation}

\noindent where $0 \leq \theta \leq 1$ and the nonlinear function $F_i$ is related to the spatial discretization of equations (\ref{GovEq2D}) and (\ref{Eq:Q&R}). Here, $\theta = 0$ leads to the explicit forward Euler scheme, $\theta = 1$ to the implicit backward Euler scheme, and $\theta = 1/2$ to the implicit second order Crank-Nicolson scheme. We use the latter, similarly to \cite{DiezKondic2007,DiezKondic2002}, leading to a system of $N$ nonlinear algebraic equations for $h_i^{n+1}, \; i=1,2, \ldots, \, N$. Following the procedure outlined in \cite{Kondic2003}, we linearize the nonlinear terms with Newton's method by expanding $h_i^{n+1}= h_i^{\dagger} + \xi_i$, and $F_i^{n+1}=F_i^{\dagger} + ( {\partial F_i^{\dagger}}/{\partial \xi_j})\xi_j$, for $i=1 ,2\; \ldots , \; N, \; j=1 ,2\; \ldots , \; N$; where $h^{\dagger}$ is a guess for the solution (commonly the previous time step solution $h^n$), $\xi$ is the correction term, and the notation $F_i^{\dagger}$ indicates that $F_i$ is calculated using $h_i^{\dagger}$. Once the linearized system is solved for the correction term, the guess for the solution is updated, and this iterative scheme is repeated until the convergence criterion is met (up to a desired tolerance).

To solve the discrete equations efficiently, we use an adaptive time step $\Delta t$. In fact, $\Delta t$ is increased to accelerate the time integration at stages where the solution does not vary rapidly. On the other hand, $\Delta t$ is decreased when the solution shows a high variation, where the Newton's method requires more than a few steps to converge. The behavior of the solution is discussed in detail in \S~\ref{sec:NumericalResults}, where we present our numerical results.

At the endpoints of the domain we impose the $h_x=h_{xxx}=0$ boundary conditions. The condition $h_x=0$ gives the value of $h$ at the two ghost points $x_{0}$ and $x_{N+1}$ outside the physical domain, i.e.~$h_{0}=h_{1}$ and $h_{N+1}=h_{N}$; the condition $h_{xxx}=0$ specifies the two ghost points $x_{-1}$ and $x_{N+2}$, i.e.~$h_{-1}=h_2$ and $h_{N+2}=h_{N-1}$.

\end{document}